\documentclass[%
reprint,
superscriptaddress,
amsmath,amssymb,
aps,
]{revtex4-1}

\usepackage{graphicx}% Include figure files
\usepackage{dcolumn}% Align table columns on decimal point
\usepackage{bm}% bold math
\usepackage{color}% color
\begin{document}
	
%\preprint{APS/123-QED}
	
\title{Quantum nondemolition measurement of mechanical motion quanta}% Force line breaks with \\
	
\author{Luca Dellantonio}
\affiliation{The Niels Bohr Institute, University of Copenhagen, Blegdamsvej 17, DK-2100 Copenhagen \O, Denmark}
\affiliation{Center for Hybrid Quantum Networks (Hy-Q), The Niels Bohr Institute, University of Copenhagen, Blegdamsvej 17, DK-2100 Copenhagen \O, Denmark}
\author{Oleksandr Kyriienko}%
\affiliation{The Niels Bohr Institute, University of Copenhagen, Blegdamsvej 17, DK-2100 Copenhagen \O, Denmark}
\affiliation{NORDITA, KTH Royal Institute of Technology and Stockholm University, Roslagstullsbacken 23, SE-106 91 Stockholm, Sweden}
\author{Florian Marquardt}
\affiliation{Institute for Theoretical Physics, University Erlangen-N\"{u}rnberg, Staudstra\ss e 7, 91058 Erlangen, Germany}
\affiliation{Max Planck Institute for the Science of Light, G\"{u}nther-Scharowsky-Stra\ss e 1, 91058 Erlangen, Germany}
\author{Anders S. S\o rensen}
\affiliation{The Niels Bohr Institute, University of Copenhagen, Blegdamsvej 17, DK-2100 Copenhagen \O, Denmark}
\affiliation{Center for Hybrid Quantum Networks (Hy-Q), The Niels Bohr Institute, University of Copenhagen, Blegdamsvej 17, DK-2100 Copenhagen \O, Denmark}

\date{\today}% It is always \today, today,
	
\begin{abstract}
The fields of opto- and electromechanics have facilitated numerous advances in the areas of precision measurement and sensing, ultimately driving the studies of mechanical systems into the quantum regime. To date, however, the quantization of the mechanical motion and the associated quantum jumps between phonon states remains elusive. For optomechanical systems, the coupling to the environment was shown to preclude the detection of the mechanical mode occupation, unless strong single photon optomechanical coupling is achieved. Here, we propose and analyse an electromechanical setup, which allows us to overcome this limitation and resolve the energy levels of a mechanical oscillator. We find that the heating of the membrane, caused by the interaction with the environment and unwanted couplings, can be suppressed for carefully designed electromechanical systems. The results suggest that phonon number measurement is within reach for modern electromechanical setups.

%http://www.nature.com/nphys/authors/submit/index.html#References
		
%\begin{description}
%\end{description}
		
\end{abstract}
	
\maketitle

%INTRODUCTION
Energy quantization is one of the hallmarks of quantum mechanics. First theorized for light by Einstein and Planck, it was found to be ubiquitous in nature and represents a cornerstone of modern physics. It has been observed in various microscopic systems starting from nuclei, atoms, and molecules, to larger mesoscopic condensed matter systems such as superconductors \cite{EnQSCsystem}. For macroscopic systems, however, the observation of energy quantization is hindered by the smallness of the Planck constant. Thus, although being a milestone of contemporary physics, up to date the discrete energy spectrum of mechanical resonators has never been seen directly. 
	
Extreme progress in studying mechanical systems has been achieved in experiments exploiting radiation pressure. This is the core of optomechanics \cite{MarquardtReview}, where photons and phonons of the optical and mechanical subsystems interact with each other. A similar type of coupling can be realized in the microwave domain with electrical circuits, leading to the field of electromechanics \cite{Teufel2011sc,Cooling2,Cooling3,PhononQNDmw,Wollman2015,Sillanpaa2017}. The numerous advances of optomechanics and electromechanics include ground state cooling \cite{Cooling1,Cooling2,AmCooling1,Cooling3,Cooling6}, ultra precise sensing \cite{Teufel2009,Sensing1,Sensing2,ExpNumb6}, generation of squeezed light and mechanical states \cite{SqLight1,SqLight2,SqLight3,Wollman2015,Sillanpaa2017}, back action cancellation \cite{QB-EvasionNBI,QB-Evasion2}, and detection of gravitational waves \cite{Abbott2016}. In all of these systems, however, the operation in the single photon/phonon regime is challenging due to the small value of the bare coupling \cite{Teufel2011sc,Groblacher2009}. Instead, experiments exploit an enhanced linearized effective coupling induced by a large driving field. This severely limits the nature of the interactions \cite{LloydQC} and possible quantum effects. In particular, it precludes the observation of the energy quantization in mechanical resonators.

Quantization of mechanical energy can be observed by a quantum nondemolition (QND) measurement \cite{BraginskyRev,ClerkRev} of an oscillator's phonon number operator $\hat{n}_{b}$. Here, QND means that the interaction, which couples the mechanical system with the measurement apparatus, does not affect the observable we are interested in. This is achieved if the total Hamiltonian commutes with $\hat{n}_{b}$, and the influence of the environment is minimized.
%%%
\begin{figure}[htbp]
	\centering
	\includegraphics[width=8 cm]{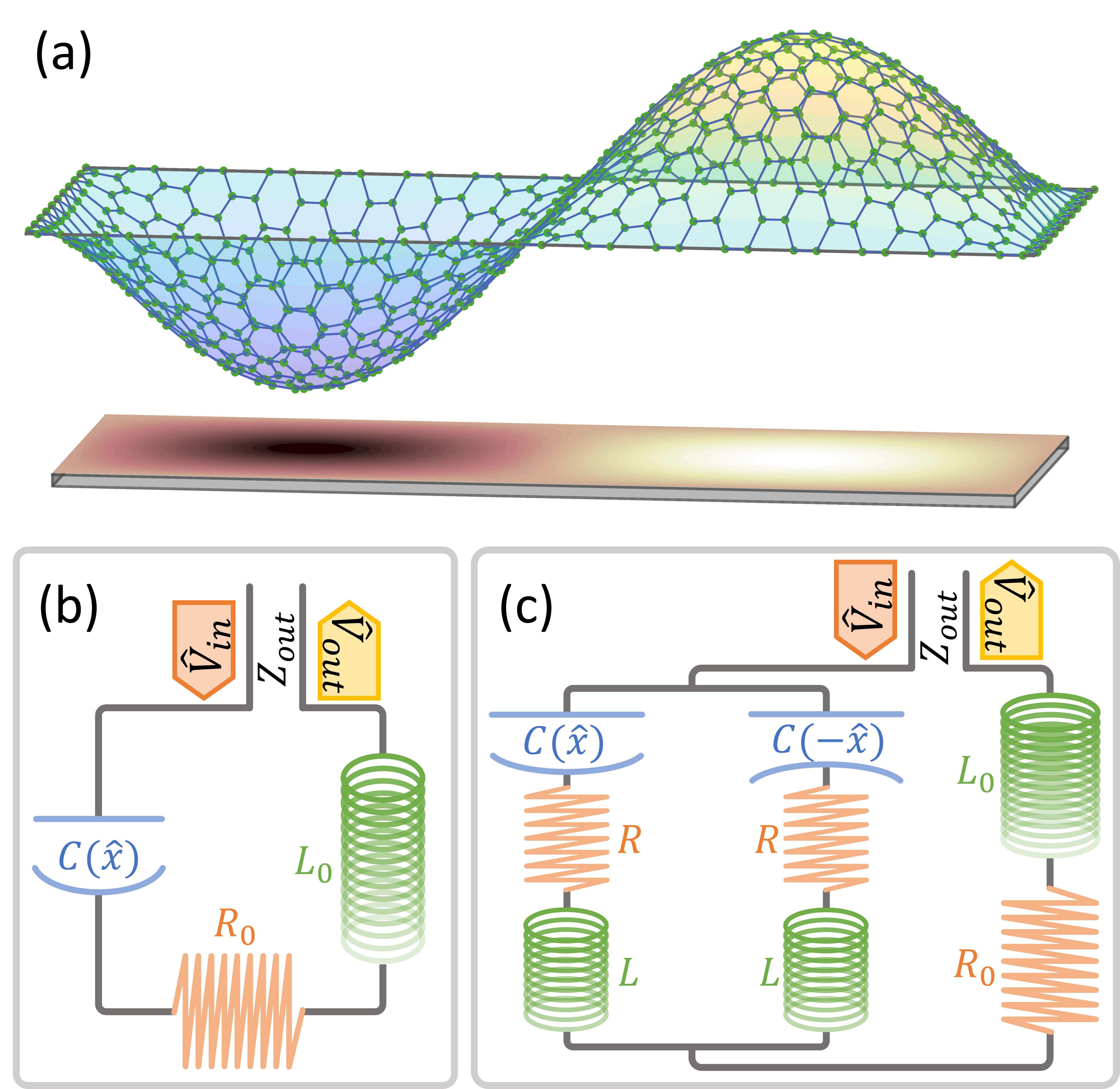}
	\caption[Fig1]{\textbf{(a)} Sketch of a capacitor with an oscillating plate, here represented by a graphene membrane [top]. We consider an antisymmetric $(2,1)$ mechanical mode. \textbf{(b)} $RLC$ oscillator formed by the inductance $L_{0}$, resistance $R_{0}$, and position--dependent capacitance $C(\hat{x})$. The circuit is driven by the input voltage $\hat{V}_{in}$ through a transmission line of impedance $Z_{out}$. $\hat{V}_{out}$ is the reflected signal. \textbf{(c)} Model circuit for an $RLC$ system where the capacitor has the same form as in \textbf{(a)}. The membrane has a vanishing linear coupling to the symmetric electrical mode used for probing the system. The antisymmetric mode, residing in the small loop containing parasitic inductances $L$ and resistances $R$, describes the redistribution of charge on the capacitor.}
	\label{Fig:Fig1}
\end{figure}
%%%
Considering the electromechanical setups in Fig.~\ref{Fig:Fig1}, we show that QND detection is feasible for a capacitor in which one of the electrodes is a light micromechanical oscillator. By choosing an antisymmetric mode for the oscillator, the interaction between the electrical and mechanical subsystems is quadratic in the displacement. Along with the suppression of the linear coupling, this ensures the QND nature of the measurement, as originally proposed in Refs. \cite{Marquardt,MarquardtFollow} for an optomechanical system. In that system, however, it was shown in Refs. \cite{YanbeiChen,Clerk2016} that the combination of unwanted losses and the coupling to an orthogonal electromagnetic mode spoils the interaction, unless strong single photon coupling is achieved. Here, we show that for the considered electromechanical setup the equivalent orthogonal mode can have dramatically different properties, allowing for the phonon QND detection. We derive general conditions under which the QND measurement is possible, and characterize its experimental signatures. %A particularly attractive feature of our approach is that it only depends on the ratio of the coupling constants. It is thus largely insensitive to parasitic capacitances which often limits the performance of electromechanical systems.
As compared to previous approaches to phonon QND measurement \cite{Marquardt,MarquardtFollow,MarquardtPRL,Yanay,AtPhon}, our procedure does not impose stringent requirements on the single photon optomechanical coupling, but relies on the ratio of the involved coupling constants. This makes our approach attractive even for systems where the interaction is limited, e.g., due to stray capacitances in the setup. For a measurement of the square displacement, a similar advantage was identified in Ref. \cite{Yanay}.

We first study an $RLC$ circuit with one capacitor plate being an oscillating membrane, without assuming the symmetry discussed above [Fig.~\ref{Fig:Fig1}(b)]. The mechanical motion of the plate shifts the resonance frequency of the circuit, while the electric potential exerts a force on the membrane. In order to perform a QND measurement of the phonon number, we require this interaction to be proportional to $\hat{n}_{b}$. We therefore Taylor expand the inverse of the capacitance to second order in the displacement, $1/C(\hat{x}) \simeq C^{-1}_{0} + \tilde{g}_{1} ( \hat{b}+\hat{b}^{\dagger} ) + \tilde{g}_{2} ( \hat{b}+\hat{b}^{\dagger})^{2}/2$, where we replaced the position $\hat{x}$ with the creation $\hat{b}^{\dagger}$ and annihilation $\hat{b}$ operators of the mechanical motion, and $\tilde{g}_{1,2}$ denote linear and quadratic coupling constants. Within the rotating wave approximation, $\tilde{g}_{2} ( \hat{b}+\hat{b}^{\dagger})^{2}/2 \simeq \tilde{g}_{2}\hat{n}_{b}$, leading to the desired QND interaction, while the $\tilde{g}_{1}$ term adds unwanted heating that spoils the phonon measurement.
	
\begin{figure}[thbp]
	\centering
	\includegraphics[width=8 cm]{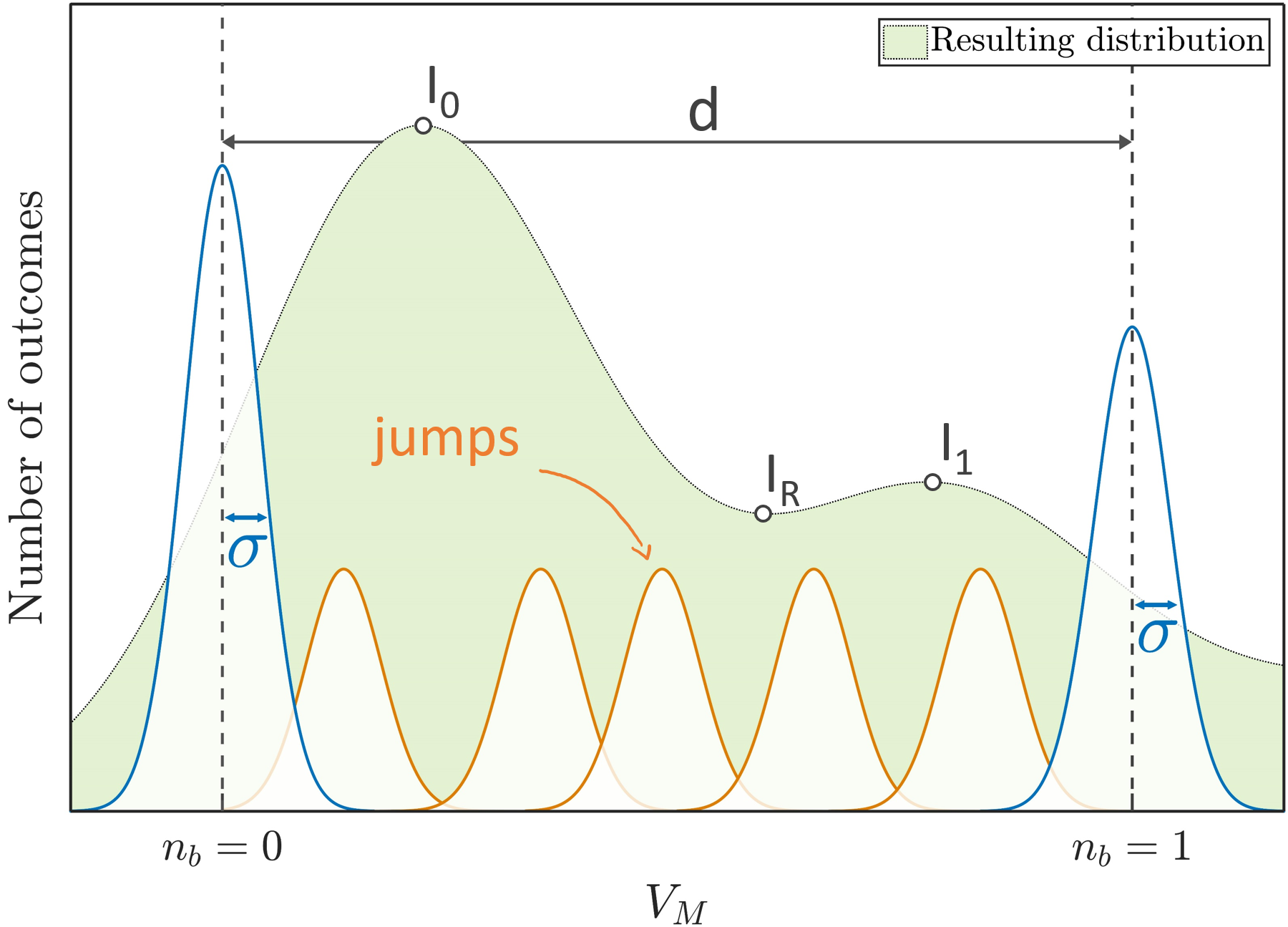}
	\caption[Fig2]{Distribution of outcomes $V_{M}$ for two different phonon numbers: $n_{b}=0$ (first peak to the left) and $n_{b}=1$ (last peak to the right). For a given value of $n_{b}$, repeated measurements are Gaussian distributed with a variance $\sigma^{2}\propto 1+2\bar{n}_{e}$ of the outgoing signal $\hat{V}_{out}$, consisting of vacuum and thermal noise. The distance $d$ between the two peaks depends on the circuit parameters and the number of incident photons, and identifies the signal to noise ratio $D=d/\sigma$. Ideally, for each shot of the measurement, the mechanics is either in its ground or first excited state. However, for $\Delta n_{b} > 0$ there will be events where the mechanical state jumps, resulting in outcomes $V_{M}$ in between the peaks relative to $n_{b}=0$ and $n_{b}=1$ (smaller peaks in the figure). This leads to the smeared distribution shown in the back. The visibility of the QND measurement is quantified by the values at the peaks and valleys, as indicated by $I_0$, $I_1$ and $I_R$ [see Eq.~\eqref{Eq:Visibility}]. The figure is for illustration only, and is not to scale.}
	\label{Fig:Fig2}
\end{figure}

The main aim of this work is to identify conditions under which the QND measurement is feasible, despite the presence of heating. We first consider the simple circuit in Fig.~\ref{Fig:Fig1}(b), and assume the incoming signal $\hat V_{in}$ to be in a coherent state resonant with the circuit. The quadratic interaction then shifts the electrical resonance frequency proportionally to the phonon number $\tilde{g}_{2}\hat{n}_{b}$. For small $\tilde{g}_{2}$, this shift leads to a phase change of the outgoing signal $\hat{V}_{out}$, that can be determined by homodyne measurement. %This shift can be determined by a homodyne measurement of the phase quadrature of $\hat{V}_{out}$. 
Different phononic states will thus lead to distinct outcomes $V_{M}$, as shown in Fig.~\ref{Fig:Fig2}. The distance $d$ between output signals for different $\hat{n}_{b}$ and the standard deviation $\sigma$ of the noise define the signal to noise ratio $D = d/\sigma$ (see Fig.~\ref{Fig:Fig2}), that needs to be maximised. 

In order to have a successful QND measurement, the phonon number $\hat{n}_b$ must be conserved. If the mechanical state jumps during a measurement, the outcome $V_{M}$ will end up in between the desired peaks. This leads to a reduced contrast, as illustrated by the distribution in the background of Fig~\ref{Fig:Fig2}. The probability for $\hat{n}_{b}$ to change is generally state--dependent, in the sense that higher Fock states are more likely to jump. A state--independent characterization of this heating, is given by the average phonons $\Delta n_b$ added to the ground state during the measurement time $T$. The jump probability for any state can then be derived from $\Delta n_{b}$ using standard results for harmonic oscillators \cite{SM}.
%An important figure of merit is therefore the distance $d$ between output signals for two different phonon numbers, compared to the standard deviation $\sigma$ of the noise. We quantify this by $D=d/\sigma$ (see Fig.~\ref{Fig:Fig2}). At the same time the phonon number must be conserved. The average number of phonons $\Delta n_{b}$ added to the ground state during the measurement time $T$ must therefore be small. 

Both $D$ and $\Delta n_{b}$ are proportional to the incoming intensity. We therefore characterize a setup by the parameter $\lambda =D^{2}/\Delta n_{b}$, where $\lambda \gg 1$ is required for successful QND detection. For the $RLC$ circuit in Fig.~\ref{Fig:Fig1}(b) we find below that
\begin{equation}\label{Eq:LambdaSimple}
\lambda = \frac{1}{2(1+2\bar{n}_{e})^{2}}\left(\frac{g_{2}}{g_{1}}\right)^{2}\left(\frac{\omega_{m}}{\gamma_{t}}\right)^{2},
\end{equation}
where $g_{1} = \tilde{g}_{1}C_{0}\omega_{s}$, $g_{2} = \tilde{g}_{2}C_{0}\omega_{s}$ and $\bar{n}_{e}$ is the thermal occupation of $R_{0}$ and $Z_{out}$ (assumed equal, $R_{0}=Z_{out}$). Here, $\omega_{m}$ and $\omega_{s}=(C_{0}L_{0})^{-1/2}\gg\omega_{m}$ are the mechanical and electrical frequencies, respectively, and $\gamma_{t}=Z_{out}/L_{0}$ corresponds to the output coupling rate. A result similar to Eq.~\eqref{Eq:LambdaSimple} is derived in Ref. \cite{QNDphonon2}. 

Despite progress in reaching the resolved sideband regime $\omega_{m}\gg\gamma_{t}$ in both opto-- and electromechanical systems, $g_{2}$ is generally much smaller than $g_{1}$, implying $\lambda \ll 1$ in Eq.~\eqref{Eq:LambdaSimple}. To circumvent this problem, we use the second fundamental mode of the membrane in the capacitor, as depicted in Fig.~\ref{Fig:Fig1}(a). The first order coefficient $\tilde{g}_{1}$ of the $1/C(\hat{x})$ expansion then vanishes, leaving $\tilde{g}_{2}$ to be the largest contribution to the electromechanical coupling. In this situation $\lambda$ seemingly grows indefinitely, the induced heating disappears, and the QND measurement of the phonon number is easily realized. In practice, however, two effects will limit the achievable value of $\lambda$. First, inaccuracies in the nanofabrication can cause misalignments and, consequently, a residual linear coupling. Second, the oscillation of the membrane induces a charge redistribution in the capacitor to maintain it at an equipotential. The associated antisymmetric electrical mode introduces an effective linear coupling, and a similar heating mechanism as the one identified in Ref. \cite{YanbeiChen} for the optomechanical setup of Refs. \cite{Marquardt,MarquardtFollow}. In these papers, the quadratic interaction results from a hybridization of two modes linearly coupled to the mechanical position, and the QND detection was found to be impossible unless the single--photon coupling $g_{1}$ exceeded the intrinsic cavity damping. In our case, the QND interaction arises directly from the Taylor expansion of the capacitance. Hence, there is no constraint tying the second--order coupling $g_{2}$ to the properties of the symmetric and antisymmetric electrical modes, which can have vastly different resonance frequencies and dampings \cite{SM}. This inhibits the mechanical heating and ultimately allows for the QND detection of the phonon number. We model the charge redistribution in the capacitor by parasitic inductances ($L$) and resistances ($R$) in the equivalent circuit of Fig.~\ref{Fig:Fig1}(c). Each of the two arms containing $R$ and $L$ represents one half of the capacitor, with opposite dependence on the membrane position, $C(\hat{x})$ and $C(-\hat{x})$. 
	
%Below, we evaluate the effect of charge redistribution and inaccuracies in the nanofabrication and thereby obtain a realistic estimate for the possible value of $\lambda$ which can be achieved in practice. 
	
%Given $\lambda$ for a specific setup, we then develop a measurement theory that optimizes the experimental outcome. In this context, the optimal measurement is a procedure that allows to distinguish different phonon number states. In fact, for a given measurement scheme, $\lambda$ sets an upper limit on what we later call the visibility: a tool that quantifies how successful the QND measurement is. In other words, having an experimental apparatus with a very large $\lambda$ does not necessarily allow to detect the energy quantization of the mechanical eigenstates, \textit{unless} we intelligently plan the employed measurement scheme. The acquisition time, the number of photons sent in the setup, and the initial state of the membrane are all variables to be optimized. We perform optimization analytically for the specific case in which the membrane is kept cooled to a fixed average phonon number during the whole measurement. These results can then be used in numerical optimizations for other measurement schemes. 
%Then the results can be employed for a numerical optimization of the visibility for other experimental procedure, like -- for instance -- initial cooling of the membrane followed by data acquisition during the mechanical thermalisation. 
	
\paragraph{``Single-arm'' RLC circuit}
In the following, we derive Eq.~\eqref{Eq:LambdaSimple} for the $RLC$ circuit in Fig.~\ref{Fig:Fig1}(b). The methods sketched here will then be generalised for the ``double-arm'' circuit in Fig.~\ref{Fig:Fig1}(c). Using the standard approach \cite{devoret1995quantum}, we write the circuit Hamiltonian as $\hat{\mathcal{H}}(\hat{x}) = \hat{\Phi}^{2}/\left[2 L_{0}\right] + \hat{Q}^{2}/\left[2C(\hat{x})\right] $, where the conjugate variables $\hat{Q}$ and $\hat{\Phi}$ are the charge and magnetic flux, respectively. We can expand $\hat{\mathcal{H}}(\hat{x})$ in the mechanical position $\hat{x}\propto \hat{b}+\hat{b}^{\dagger}$, in order to obtain the circuit Hamiltonian $\hat{\mathcal{H}}_{e} = \hat{\mathcal{H}}(\hat{x}=0)$ and the coupling Hamiltonian $\hat{\mathcal{H}}_{em} = g_{1} \omega_{s} L_{0} \hat{Q}^{2}(\hat{b}+\hat{b}^{\dagger})/2+ g_{2} \omega_{s} L_{0} \hat{Q}^{2} (\hat{n}_{b}+\hat{b}\hat{b}/2+\hat{b}^{\dagger}\hat{b}^{\dagger}/2)$. The total Hamiltonian $\hat{\mathcal{H}}_{tot} = \hat{\mathcal{H}}_{e} + \hat{\mathcal{H}}_{em} + \hat{\mathcal{H}}_{m}$ is therefore the sum of the circuit, interaction, and the mechanical Hamiltonian $\hat{\mathcal{H}}_{m} = \hbar \omega_{m} \hat{b}^{\dagger}\hat{b}$. 

Next, we describe the environmental effects corresponding to decay and heating of the modes. Associating each resistor $R_{i}$ with its own Johnson--Nyquist noise $\hat{V}_{Ri}$, we find the equations of motion of the composite system
\begin{subequations}\label{Eq:EOMsimple}
	\begin{align}
	\dot{\hat{Q}} & =  \frac{\hat{\Phi}}{L_{0}} \label{Eq:EOMqS} \\
	\begin{split}
	\dot{\hat{\Phi}} & =  -\frac{\hat{Q}}{C_{0}}-(\gamma_{t}+\gamma_{r})\hat{\Phi}-g_{1} \omega_{s}L_{0} \hat{Q}\left(\hat{b}+\hat{b}^{\dagger}\right) \\ & -g_{2}\omega_{s}L_{0}\hat{Q}\left(\hat{n}_{b} +\frac{\hat{b}\hat{b}+\hat{b}^{\dagger}\hat{b}^{\dagger}}{2} \right)+2\left(\hat{V}_{in}+\hat{V}_{R0}\right) \label{Eq:EOMphiS}
	\end{split} \\
	\begin{split}
	\dot{\hat{b}} & =  -i \omega_{m}\hat{b}-g_{1}\frac{i\omega_{s}L_{0}\hat{Q}^{2}}{2\hbar} -g_{2}\frac{i\omega_{s}L_{0}\hat{Q}^{2}}{2\hbar}\left( \hat{b}+\hat{b}^{\dagger} \right) \\ & - \frac{\gamma_{b}}{2}\hat{b}+i \frac{x_{0}}{\hbar}\hat{F}_{b}, \label{Eq:EOMbS}
	\end{split}
	\end{align}
\end{subequations}
where $\gamma_{r} = R_{0}/L_{0}$, $\gamma_{b}$ is the intrinsic mechanical damping rate with associated noise $\hat{F}_{b}$, and $x_{0}=\sqrt{\hbar/(2 m \omega_{m})}$ is the amplitude of the zero--point motion for a membrane of mass $m$. From now on, we consider optimally loaded setups with $\gamma_{r}=\gamma_{t}$. Eqs.~\eqref{Eq:EOMsimple} fully characterize the dynamics of the system, and represent the starting point for our detailed analysis.

The feedback of the membrane's motion on the electrical circuit is described by Eq.~\eqref{Eq:EOMphiS}. Driving the system at the electrical resonance frequency $\omega_{s}$, the terms proportional to $g_{1}(\hat{b}+\hat{b}^{\dagger})$ and $g_{2}(\hat{b}\hat{b}+\hat{b}^{\dagger}\hat{b}^{\dagger})$ give rise to sidebands at frequencies $\omega_{s}\pm\omega_{m}$ and $\omega_{s}\pm 2\omega_{m}$, respectively, whereas $g_{2}\hat{n}_{b}$ induces a phonon--dependent frequency shift of the microwave cavity. Since homodyne detection is only sensitive to signals at the measured frequency, the sidebands are removed in the outcome $V_{M}$, which is defined as the phase quadrature of $\hat{V}_{out}=\hat{V}_{in}-\gamma_{t}\hat{\Phi}$. This allows us to neglect oscillating terms in the calculation of $V_{M}$ \cite{Sesta}. The only contribution to $V_{M}$ is therefore the phonon--dependent frequency shift, that allows us to resolve the mechanical state. On the contrary, the electrically induced mechanical heating only involves the sidebands $\omega_{s}\pm\omega_{m}$ and $\omega_{s}\pm 2\omega_{m}$, being unaffected by the term $g_{2} \hat{n}_{b}$ in the Hamiltonian. For the RLC circuit in Fig.~\ref{Fig:Fig1}(b), the heating is dominated by the linear term, since $g_{1}\gg g_{2}$, and we shall neglect $g_{2}$ for the calculation of $\Delta n_{b}$ below.
%%%

Below, we quantify the heating of the membrane and the phonon--dependent $LC$--frequency shift. We first assume that the mechanical state does not jump during the measurement. Then, the equations of motion of the two subsystems decouple and we find $D^{2}=g_{2}^{2} \lvert\alpha\rvert^{2} /[4(1+2\bar{n}_{e})\gamma_{t}^{2}]$, where the number of photons $\lvert\alpha\rvert^{2}$ sent into the circuit within the measurement time $T$ sets the measurement strength. As discussed above, $\Delta n_{b}$ is the average phonon number at the end of the measurement $\Delta n_{b} = \langle \hat{n}_{b}(T)\rangle$, with the mechanics initially in its ground state. For $T$ much shorter than the mechanical lifetime $\gamma_{b}^{-1}$, $\Delta n_{b}$ can be linearised to find the rate at which the membrane heats up. For the $RLC$ circuit in Fig.~\ref{Fig:Fig1}(b), we find $\Delta n_{b}=(1+2\bar{n}_{e})g_{1}^{2}\lvert\alpha\rvert^{2}/(2 \omega_{m}^{2})$. The parameter $\lambda$ given in Eq.~\eqref{Eq:LambdaSimple} is then found as the ratio $\lambda =D^{2}/\Delta n_{b}$. For details, see \cite{SM}.
	
\paragraph{``Double arm'' circuit}
With the overall linear coupling vanishing, the parameter $\lambda$ will be limited by fabrication imperfections and coupling to the antisymmetric mode. To model these phenomena we consider the circuit in Fig.~\ref{Fig:Fig1}(c), where the antisymmetric mode resides inside the small loop containing the two capacitors, and the symmetric one probes the system. We derive $g_{1}$ and $g_{2}$ from the expansion of \textit{each} of the two capacitors: $1/C\left(\pm \hat{x}\right) \simeq C^{-1}_{0} \pm \tilde{g}_{1} ( \hat{b}+\hat{b}^{\dagger} ) + \tilde{g}_{2} \hat{n}_{b}$, so that in the absence of fabrication imperfections the \textit{total} capacitor $C_{tot} = C\left( \hat{x}\right)+C\left(- \hat{x}\right)$ is not linearly coupled to the symmetric mode. The coefficients $g_{1}$ and $g_{2}$ are related to their tilde counterparts in the same way as before, and the parameters $D^{2}$ and $\Delta n_{b}$ are evaluated in a similar fashion as we did for the $RLC$ circuit. Since we quantify two sources of heating, it is convenient to write $\lambda =(\lambda_{b}^{-1}+\lambda_{p}^{-1})^{-1}$, where $\lambda_{b}$ takes into account heating from charge redistribution, and $\lambda_{p}$ describes the influence of fabrication imperfections \cite{Prima}. With the details presented in Methods and Supplementary \cite{SM}, we find
\begin{subequations}\label{Eq:LambdaCompl}
	\begin{align}
	\lambda_{b} = & \frac{2}{\left( 1+2\bar{n}_{e} \right)^{2}}\left( \frac{g_{2}}{g_{1}} \right)^{2} \left( \frac{\omega_{s}}{\gamma_{t}} \right)^{2}\frac{Z_{out}}{R}, \label{Eq:lambdaBCompl}\\ 
	\lambda_{p} = & \frac{2}{\left( 1+2\bar{n}_{e} \right)^{2}} \left( \frac{g_{2}}{g_{1}} \right)^{2}\left( \frac{g_{1}}{g_{r}} \right)^{2} \left(\frac{\omega_{m}}{\gamma_{t}}\right)^{2}, \label{Eq:lambdaPCompl}
	\end{align}
\end{subequations}
where $\omega_{s} = \left[C_{0}(L+2 L_{0})\right]^{-1/2}$ is the frequency of the symmetric mode, $\gamma_{t} = [2 Z_{out}]/[L+2L_{0}]$ is the decay to the transmission line and $g_{r}=2C_{0}x_{0}\omega_{s}\partial_{x}C_{tot}^{-1}(x)$ is the residual linear coupling induced by fabrication imperfection. We use the same notation introduced for the $RLC$ circuit to allow a direct comparison. Eqs.~\eqref{Eq:LambdaCompl} express the gain of our approach to QND detection. First, Eq.~\eqref{Eq:lambdaPCompl} quantifies the advantage of symmetry: $\lambda$ dramatically improves compared to Eq.~\eqref{Eq:LambdaSimple} by having a small residual linear coupling $g_{r}\ll g_{1} $. Second, Eq.~\eqref{Eq:lambdaBCompl} is multiplied by the factor $(\omega_{s}/\omega_{m})^{2}$ with respect to Eq.~\eqref{Eq:LambdaSimple}. For microwave readout of a MHz oscillator, this factor can be substantial. Furthermore, the mechnical oscillator is now only susceptible to the noise associated with charge redistribution on the capacitor, and not to the resistance in the inductor. This gives an additional improvement if $R < Z_{out}$.	
	
To describe a realistic situation, we numerically simulate the case in which the parasitic resistances $R$, inductances $L$ and the two bare capacitances $C_{0}$ differ from each other. In Fig.~\ref{Fig:Fig5}, we test the system with these asymmetries and the physical parameters given below. In the left plot, the role of a residual linear coupling $g_{r}$ is investigated. In the right one, we consider unbalanced resistances $R\pm\delta R$, inductances $L\pm\delta L$, and capacitances $C_{0}\pm\delta C$. The results show that our analytical predictions accurately describe a system with non-zero $g_{r}$ and $\delta C$. Furthermore, the numerical points confirm that $\delta R$ and $\delta L$ enter as higher order perturbations. In fact, we generally find that Eqs.~\eqref{Eq:LambdaCompl} are accurate for relatively large perturbations (up to $25\%$).
\begin{figure}[htbp]
	\centering
	\includegraphics[width=8.5 cm]{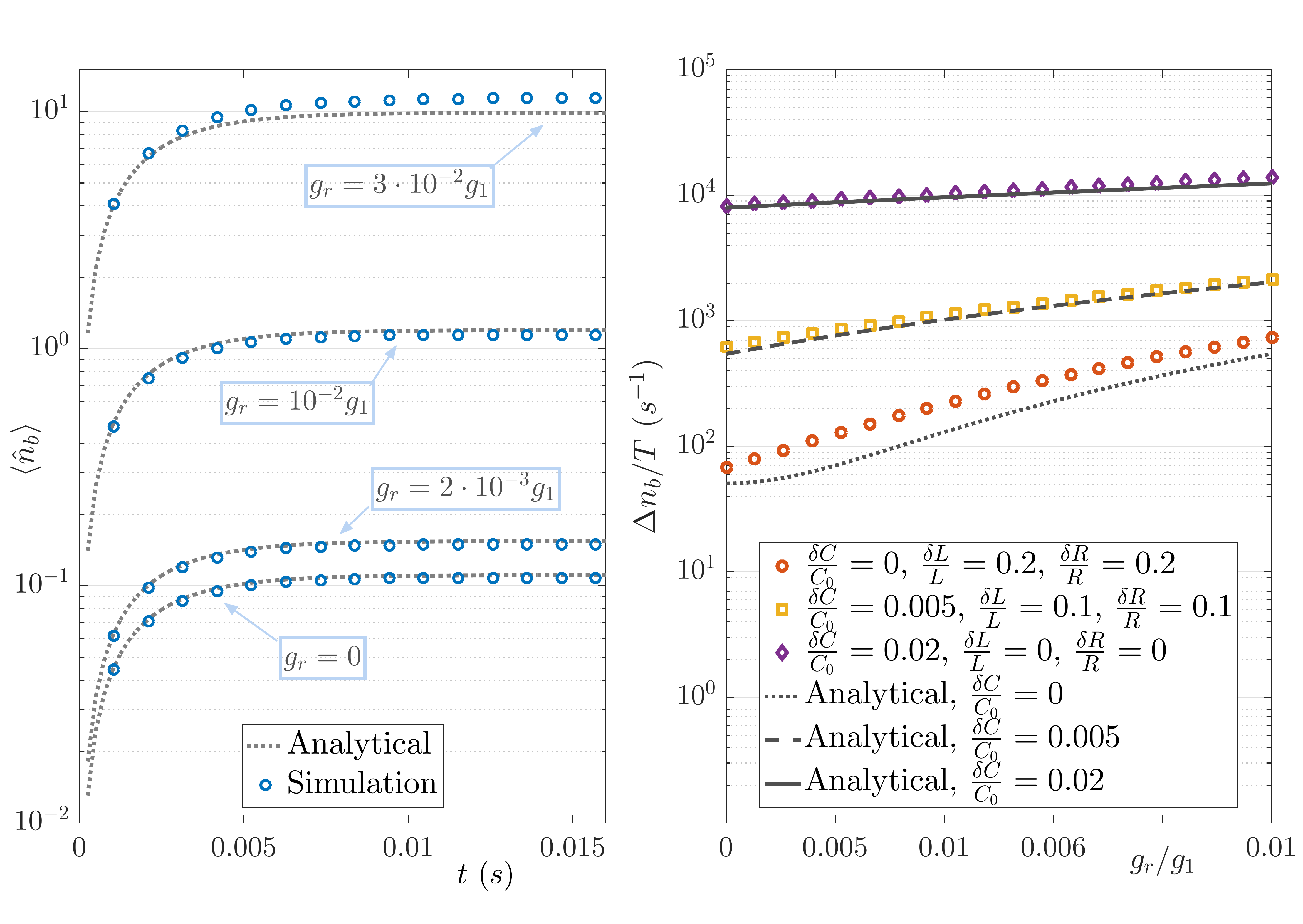}
	\caption[Fig5]{\textbf{Left}: Average phonon number $n_{b}(t)$ as a function of time. We present a comparison between the analytical curves (grey, dotted lines) and the full simulations of the system (blue dots). From the bottom to the top we set $g_{r}/g_{1}$ to be $0$, $2\cdot 10^{-3}$, $10^{-2}$ and $3\cdot 10^{-2}$. We use $\delta R = \delta L = \delta C = 0$. \textbf{Right}: heating rate $\Delta n_{b}/T$ as a function of the normalised residual linear coupling $g_{r}/g_{1}$. Here we analyse the system in presence of asymmetries in the parasitic elements of the circuit. The three dark grey lines are the analytical predictions for $\delta C/C_{0}$ being equal to $0$ (dotted), $0.005$ (dashed) and $0.02$ (full). The circles, squares and diamonds are the simulated results for the values $\delta R/R$, $\delta L /L$, and $\delta C/C$ reported in the legend. We assume $L/L_{0}=10^{-2}$, $R/Z_{out} = 10^{-1}$, $\omega_{s}=(2\pi)7$ GHz, $\omega_{m} = (2\pi)80$ MHz, $\gamma_{r}\simeq\gamma_{t}=(2\pi)0.15$ MHz, $\gamma_{b} = (2\pi)80$ Hz, $g_{1}=(2\pi)7$ kHz, $\bar{n}_{e} = \bar{n}_{m} = 0$, and an incident photon flux $\lvert\tilde{\alpha}\rvert^{2}=1.15\cdot 10^{15}$ s$^{-1}$.}
	\label{Fig:Fig5}
\end{figure}
	
Inspired by recent experiments \cite{ExpNumb1,ExpNumb3,ExpNumb4,ExpNumb5,ExpNumb6}, we estimate the value of $\lambda$, which can be reached in state--of--the--art setups. We consider a rectangular monolayer graphene membrane of length $1$ $\mu$m and width $0.3$ $\mu$m, with a mechanical frequency of $\omega_{m}= (2 \pi) 80$ MHz and a quality factor $Q=10^6$. It is suspended $d_{0}=10$ nm above a conducting plate, forming the capacitor [see sketch in Fig.~\ref{Fig:Fig1}(a)]. Assuming that the membrane is clamped to the substrate along its boundaries, we identify the ratio of the coupling coefficients for each capacitor $C(\pm\hat{x})$ in Fig. \ref{Fig:Fig1}(c) to be $g_{2}/g_{1} = \pi^{2} x_{0}/(8 d_{0})$ \cite{EmilThesis}. Considering that for these geometries stray capacitances $C_{s}$ are typically preponderant with respect to $C_{0}$, we take $g_{1}\simeq (2\pi) 7$ kHz and $g_{2}\simeq(2\pi) 1$ Hz, corresponding to $C_{s} \simeq 100 C_{0}$. For comparison, a value of $C_{s}=50$ $f$F is obtained in Ref. \cite{ExpNumb1}, for a graphene membrane about two and a half times the size considered here. This stray capacitance would be $376$ times $C_{0} \simeq 13$ $f$F. Assuming a reduction of $C_{s}$ due to the smaller dimensions, we take $C_{s} = 100 C_{0}$.

\begin{figure}[htbp]
	\centering
	\includegraphics[width=8.5 cm]{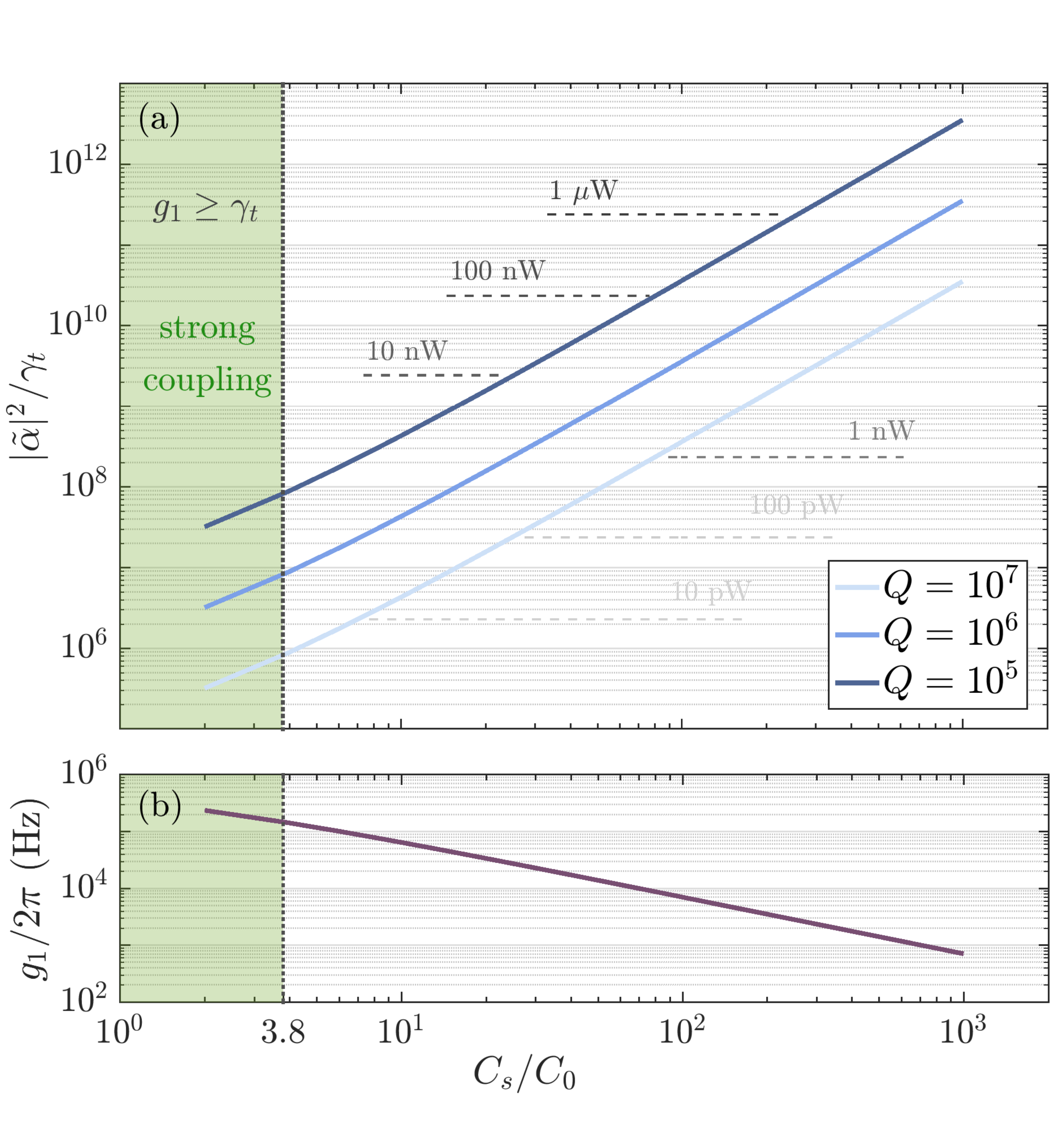}
	\caption[Fig4]{\textbf{(a)}: average intracavity photons $\lvert\tilde{\alpha}\rvert^{2}/\gamma_{t}$ required for the QND measurement, as a function of the relative value of the stray capacitance $C_{s}/C_{0}$. The three lines correspond to different values of the mechanical quality factor, as indicated in the legend. We assume $\Delta n_{b} = 0.3$ and equal contributions from the mechanical and electrically induced reservoirs $\bar{n}_{m}=\bar{N}_{ \rm eff}/2 = 3$. As a reference, the grey dashed lines indicate the associated powers of the probe. \textbf{(b)}: linear coupling $g_{1}$ as a function of $C_{s}/C_{0}$. For both figures, the shadowed region indicates the strong coupling $g_{1}\geq \gamma_{t}$, where QND detection is feasible with other approaches \cite{Marquardt,Yanay,AtPhon}.}
	\label{Fig:Fig4}
\end{figure}

With an electrical reservoir at zero temperature $\bar{n}_{e}\simeq0$ (valid for milliKelvin experiments), an electrical frequency $\omega_{s} = (2 \pi) 7$ GHz, and decay rate $\gamma_{t}=(2 \pi) 150$ kHz, we get $\lambda_{b} = 105 \times Z_{out}/R$ and $\lambda_{p} = 0.014 \times (g_{1}/g_{r})^{2}$. Since the graphene coupling can be tuned via electric fields \cite{ExpNumb2,GrapheneTuning1,GrapheneTuning2}, we assume $g_{1}/g_{r}\sim 100$, which fixes $\lambda$ between $60$ ($R=Z_{out}$) and $122$ ($R= Z_{out}/10$), mostly restricted by $\lambda_{p}$. This limit is well above the threshold for having a good visibility of the phonon number states (see below), and can be further improved by either increasing the sideband resolution $\omega_{m}/\gamma_{t}$, the electrical frequency $\omega_{s}$, or by reducing the size of the membrane. In Fig.~\ref{Fig:Fig4}(b), we show the linear coupling $g_{1}$ as a function of the stray capacitance. For small values of $C_{s}$, we reach the strong coupling regime, where $g_{1} \geq \gamma_{t}$. In the realistic scenario described above, where $C_{s}\gg C_{0}$, our scheme still allows for phonon QND measurement even for $g_1,g_2\ll\gamma_t$. This is in contrast to the optomechanical regime, where strong coupling $g_1>\gamma_t$ is required \cite{YanbeiChen}. Regardless of how much $C_s$ reduces the coupling constants, it is in principle always possible to compensate by using stronger power.

\begin{figure}[htbp]
	\centering
	\includegraphics[width=8.5 cm]{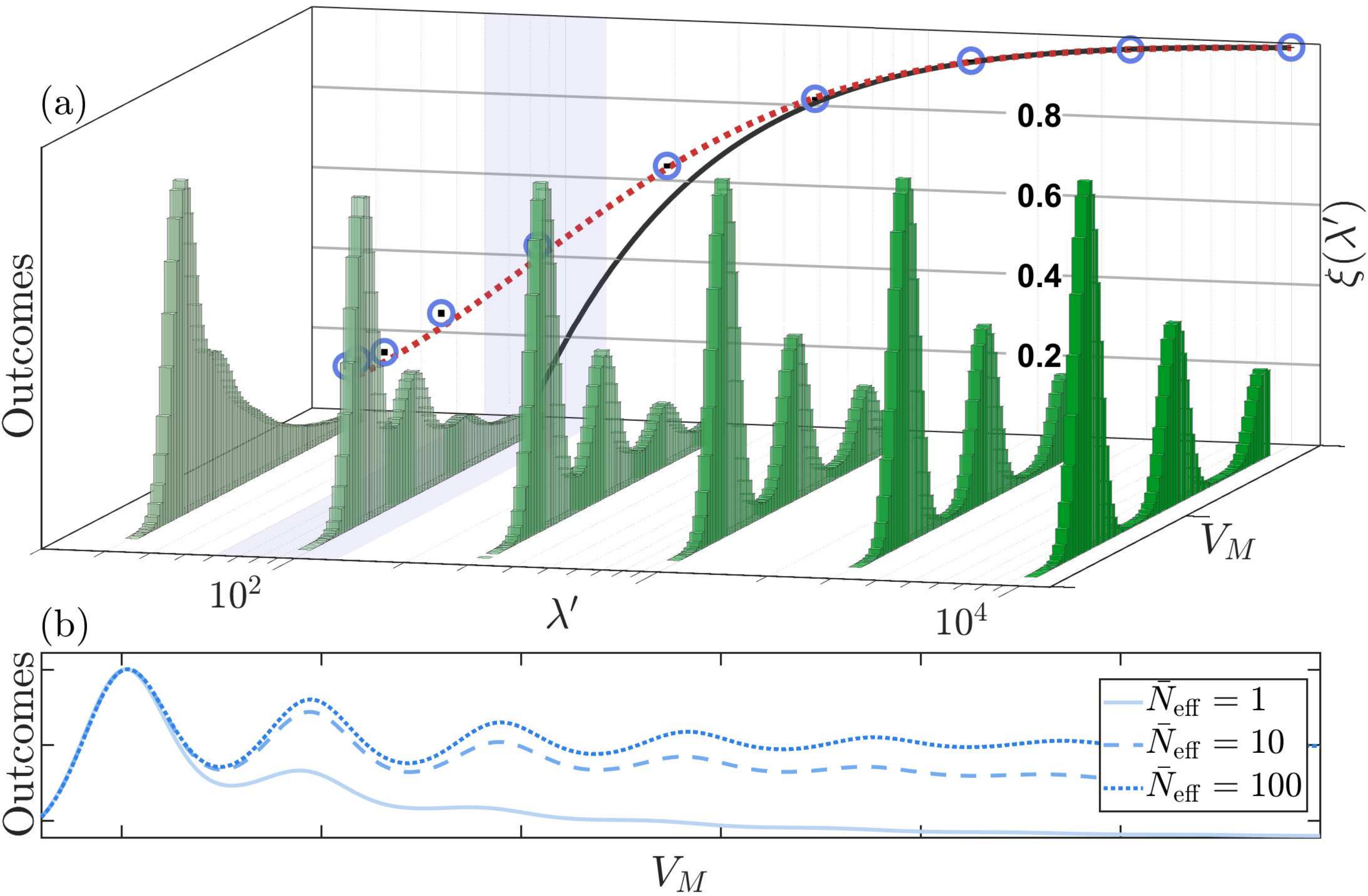}
	\caption[Fig3]{\textbf{(a), 3D plot}: histograms of outcomes for different $\lambda'$ (from left to right, $\lambda' = 32$, $10^{2}$, $3 \cdot 10^{2}$, $10^{3}$, $3 \cdot 10^{3}$ and $10^{4}$). The optimal values of $\Delta n_{b}$ are (from left to right) $0.43$, $0.27$, $0.12$, $0.05$, $2\cdot 10^{-3}$ and $8\cdot 10^{-4}$, and have been determined by a numerical optimization. The shadowed region corresponds to the estimated visibility for state--of--the--art technology, $\lambda \simeq 60-130$. \textbf{2D plot (back)}: maximum visibility $\xi$ for different values of the parameter $\lambda'$. The blue circles have been evaluated numerically from the histograms in the 3D plot (and others). The error bars of the Monte Carlo simulation (black lines inside) have been determined assuming Poissonian statistics in each bin, and are negligible on this scale. The red dotted curve comes from our model for the visibility, and the black solid curve is the simplified expression presented in Eq.~\eqref{Eq:VisApp}. We consider $\bar{N}_{\text{\rm eff}}=1$. See Methods for more details. \textbf{(b)}: expected outcomes for $\lambda' = 75$ and $\bar{N}_{\rm eff}$ being $1$ (full), $10$ (dashed) and $100$ (dotted line). The parameter $\Delta n_{b}$ has been optimized to achieve maximum visibility for each value of $\bar{N}_{\rm eff}$.}
	\label{Fig:Fig3}
\end{figure}

\paragraph{Measurement}
We now evaluate how well a given value of $\lambda$ allows for the QND detection of the phonon number. To this end we consider a situation where the system is continuosly probed and measured. The output is then turned into discrete results by averaging over a suitable time $T$, and a histogram is constructed from the measured values $V_{M}$. We assume that the heating of the continuous QND probing is in equilibrium with the mechanical damping and the associated reservoir. In this case, one also needs to consider the thermal bath of the membrane. In addition to $\Delta n_{b}$ determined above, the total heating out of the ground state is thus $\Delta n_{b}+\gamma_b \bar{n} _{m} T$. This additional term leads to a redefinition of the parameter $\lambda$ to
\begin{equation}
\lambda'=\lambda \frac{\Delta n_{b}}{\Delta n_{b}+\gamma_b \bar{n} _{m} T},
\end{equation}
and the equilibrium average mechanical occupation, resulting from both the mechanical reservoir and the QND probe, becomes
\begin{equation}
\bar N_{{\rm eff}} \simeq \bar n_m \frac{\lambda}{\lambda-\lambda'}. 
\end{equation}
The phonon QND measurement is then characterized by $\lambda'$, which is desirable to have as close as possible to its maximum $\lambda$. This can be achieved by choosing a sufficiently strong probing power and a short measurement time $T$, such that the mechanical heating can be neglected. This leads to a large $\bar N_{{\rm eff}}$, that does not significantly change the contrast of the QND measurement [see Eq.~\eqref{Eq:VisApp} and Fig.~\ref{Fig:Fig3}(b)], but increases the time for acquiring significant statistics (the mechanical system spends less time in each Fock state).

Given $\lambda'$, we now want to optimize all remaining parameters of the system, to be able to discern the ground and first excited states with the largest contrast. We simulate the mechanical system with the quantum--jump method, and pick Gaussian distributed random values for the electrical vacuum and thermal noise. From this, we make the histogram of the resulting output voltages $V_{M}$ presented in Fig.~\ref{Fig:Fig3}(a), where the induced heating $\Delta n_{b}$ is optimized numerically. For the optimization we consider the visibility
\begin{equation}\label{Eq:Visibility}
\xi=\frac{\frac{1}{2}\left(I_{0}+I_{1}\right)-I_{R}}{\frac{1}{2}\left(I_{0}+I_{1}\right)+I_{R}},
\end{equation}
where $I_{0}$ and $I_{1}$ are the heights of the peaks corresponding to $n_{b}=0$ and $n_{b}=1$ phonons, while $I_{R}$ is the lowest height in between $I_{0}$ and $I_{1}$ (see Fig.~\ref{Fig:Fig2}). 

Additionally, we make an analytical model where we allow for one jump during each measurement period. We can extract the asymptotic behaviour of the visibility
\begin{equation}\label{Eq:VisApp}
\xi\left( \lambda',\bar{N}_{ \rm eff} \right)=1-8\frac{3+5 \bar{N}_{\rm eff}}{1+2\bar{N}_{\rm eff}}\frac{\sqrt{\pi\log \lambda'}}{\lambda'},
\end{equation}
reflecting the compromise between the contributions to $I_{R}$ from the noise $\propto \exp(-D^{2}/8)$ and from the jumps during the measurements $\propto \Delta n_{b}$. 

The results of simulations and model are shown in Fig.~\ref{Fig:Fig3}(a). The blue points are the numerical optimization, which are in good agreement with the analytical result (red, dotted line). Notice that for small values of $\lambda'$, the optimal $\Delta n_{b}$ is sufficiently high to allow multiple jumps during the measurement time $T$, leading to minor discrepancies. The black, solid line is Eq.~\eqref{Eq:VisApp}, and the shadowed region corresponds to the predicted values of $\lambda$ for the parameters introduced above. Qualitatively, clear signatures of the mechanical energy quantization are present for $\lambda' \gtrsim 40$, where the visibility exceeds $20 \%$.
%%%
%%%
%To evaluate the perfomance for a given $\lambda'$ and $\bar N_{{\rm eff}}$ we ..... and now continue with what we have about Monte Carlo. Only I think that we should switch to $\lambda'$. We can then keep everything until we say that we have visibility better than 20 \%

For the experimental parameters considered above, the maximum attainable value of $\lambda'$ is $\lambda = 122$ (for $R=Z_{out}/10$), and is achieved with a strong probe such that $\bar{N}_{ \rm eff} \gg \bar{n}_{m}$. The incident power and the measurement time $T$ provide a handle to optimize the performance for given experimental conditions. Qualitatively, a short value of $T$ minimizes the effects of the mechanical heating, and makes $\lambda' \simeq \lambda$. On the other hand, the required power to reach such a regime can be troublesome \cite{TemperatureRising}, and we may need to integrate for too long time to have sufficient statistics (since $\bar{N}_{\rm eff} \gg 1$). This last problem can be solved by adding an electrical cooling, red--detuned by $\omega_{m}\gg\gamma_{t}$ from the QND probe. This cooling would not affect the parameter $\lambda'$, since it does not heat up the system, but only reduces $\bar{N}_{\rm eff}$. The visibility $\xi$ thus remains almost unaltered [see Eq.~\eqref{Eq:VisApp} and Fig.~\ref{Fig:Fig3}(b)], but the probability to find the membrane in low excited states is increased, reducing the experimental time.

As an example, assume that the heating from the electrical feedback and the mechanical bath are equal, such that $\lambda'=\lambda/2=61$. Considering a cryogenic temperature of $14$ mK \cite{ExpNumb4}, the average mechanical occupation is $\bar{n}_{m}\simeq 3$, implying $\bar{N}_{ \rm eff} = 6$. The optimal $\Delta n_{b}$ is then $0.3$, and can be obtained with a driving power of $16$ nW and a measurement time of $0.1$ ms for a mechanical quality factor $Q=10^6$ and a stray capacitance $C_{s}=100 C_{0}$. For other values of $Q$ and $C_{s}$, the driving power can be varied to fulfil the constraint $\bar{N}_{\rm eff}=2\bar{n}_{m}$, as shown in Fig.~\ref{Fig:Fig4}(a). The incident field is rather intense, which may cause additional heating to the system. In the setup of Ref. \cite{TemperatureRising}, such additional heating has been observed above an intracavity photon number of $10^8$. For comparison, in Fig.~\ref{Fig:Fig4}(a) we show the intracavity photon number $ \lvert\tilde{\alpha}\rvert^2/\gamma_t$ for our system, where $\lvert\tilde{\alpha}\rvert^2 = \lvert \alpha \rvert^{2}/T$ is the photon flux. Depending on the parameters, we see that $ \lvert\tilde{\alpha}\rvert^{2}/\gamma_t$ will be similar or higher than $10^{8}$ for $C_s\gtrsim 100 C_0$. These devices cannot, however, be compared directly. Nevertheless, since Ref. \cite{TemperatureRising} indicates that the source of this heating is electrical, we believe that it would be strongly suppressed for the QND measurement considered here. Since the linear coupling is almost cancelled by symmetry, the resulting heating rate is likely reduced by a factor $(g_r/g_1)^2\simeq 10^{-4} $. In absence of this suppression, conducting our experiment in a pulsed regime may substantially reduce other heating mechanisms \cite{SM}.

\paragraph{Conclusions and Outlook}

We have revisited the challenge of performing phonon QND measurement. Employing symmetry to inhibit the linear coupling, the detrimental heating is suppressed while retaining the desired quadratic coupling. Contrary to the optomechanical case \cite{YanbeiChen}, the residual coupling to the antisymmetric mode is strongly suppressed by its higher frequency and reduced resistance. A particularly attractive feature of the current approach is that it is only sensitive to the ratio $g_{2}/g_{1}$, and not to their absolute values. Stray capacitances, which reduce the electromechanical couplings, can thus be compensated using stronger input fields. 

These attractive features put QND detection within reach of presently available technology. A successful realization of a QND detection will not only represent a demonstration of genuine non-classical behaviour of mechanical systems, but also extend the interactions available in electro/opto--mechanics to non--Gaussian operations \cite{Braunstein2005}. This will considerably expand the realm of effects that can be studied with these systems, and facilitate their application for quantum information processing \cite{LloydQC}.

As an outlook, it is desirable to extend this work to the optomechanical case. The electromechanical systems considered here can be described with Kirchoff's laws, that give rigorous results within a well defined model. The physical mechanisms behind the heating are identified to be the Johnson-Nyquist noises associated to the resistors, and fabrication imperfections. For comparison, the exact description of dissipation in a multi--mode optomechanical system may be more involved. Nevertheless, the results presented here could be useful for guiding the intuition towards QND detection in the optical regime. As a further extension, it would be interesting to investigate the effect of squeezing. By reducing the vacuum noise, squeezing can lead to a direct improvement in $\lambda$, thus reducing the physical requirements for the QND detection.
	
\begin{acknowledgements}
	We gratefully acknowledge funding from the European Union Seventh Framework Programme through the ERC Grant QIOS, the European HOT network, and the Danish Council for Independent Research (DFF). We thank Emil Zeuthen and Albert Schliesser for fruitful discussions. 
\end{acknowledgements}

\section*{Author contributions}
O.K., A.S. and F.M. conceived the study. L.D. derived the main results, did the numerical calculations and wrote the first draft. All authors contributed to the manuscript. A.S supervised the project.

\section*{Methods}
\begin{footnotesize}
	
Below, we explain the crucial steps to derive the results proposed above. For further details, see \cite{SM}.
	
\paragraph{The double arm circuit} 
The Hamiltonian for the system in Fig.~\ref{Fig:Fig1}(c) is given by 
\begin{equation}
\begin{split}
\hat{\mathcal{H}} =& \hbar \omega_{m} \hat{b}^{\dagger} \hat{b} + \frac{\hat{\Phi}_{a}^{2}}{4L}+\frac{\hat{Q}_{a}^{2}}{C_{0}}+\frac{\hat{\Phi}_{s}^{2}}{L+2L_{0}}+\frac{\hat{Q}_{s}^{2}}{4C_{0}} \\ & + \frac{g_{1}}{C_{0} \omega_{s}}\hat{Q}_{a}\hat{Q}_{s}\left( \hat{b}+\hat{b}^{\dagger} \right) + \frac{g_{2}}{C_{0} \omega_{s}}\hat{Q}_{a}^{2}\hat{b}^{\dagger}\hat{b} + \frac{g_{2}}{4 C_{0} \omega_{s}}\hat{Q}_{s}^{2}\hat{b}^{\dagger}\hat{b},
\end{split}
\label{Eq:HamQuantCompleteBal}
\end{equation}
where subscripts ``$a$'' and ``$s$'' indicate the asymmetric and the symmetric electrical fields, respectively. From Eq.~\eqref{Eq:HamQuantCompleteBal} and using Kirchoff's laws, it is possible to determine the equations of motions, including noises and decays. The normalized distance $D^{2} = d^{2}/\sigma^{2}$ is obtained assuming the phonon number to be constant within $T$ -- i.e.: setting $g_{1}=0$ -- so that the asymmetric and symmetric fields decouple. Looking at the phase quadrature of the reflected signal $\hat{V}_{out}=\hat{V}_{in}-\gamma_{t}\hat{\Phi}_{s}$, we determine $d$. The noise $\sigma$ is the sum of vacuum noise from the input coherent field, and the Johnson Nyquist noises of the resistors. 

As discussed above, the heating $\Delta n_{b} = \langle \hat{n}_{b}(T)\rangle$ has two contributions: asymmetries leading to a non--vanishing linear coupling $g_{r}$, and the charge redistribution. The first, is found by assuming $R\ll R_{0}$ and $L\ll L_{0}$, such that the circuit in Fig.~\ref{Fig:Fig1}(c) is equivalent to the one in Fig.~\ref{Fig:Fig1}(b), for which we already know $\Delta n_{b}$. The contribution from charge redistribution is determined from the Hamiltonian in Eq.~\eqref{Eq:HamQuantCompleteBal} neglecting the quadratic interaction, that does not alter the phonon number. The strongly driven symmetric electrical field is then substituted with its steady state, obtained assuming a constant photon flux. The time evolution of $\langle \hat{n}_{b}\rangle$ is finally found by looking at the equations of motion for the asymmetric field and the mechanical creation/annihilation operators. With the amplitude of the symmetric mode replaced by its steady state, these equations are now linear in the annihilation (creation) operators $\hat{b}$ ($\hat{b}^{\dagger}$) and can be solved by standard optomechanics techniques.

\paragraph{Asymmetric circuit}
To obtain Fig.~\ref{Fig:Fig5}, we analyse the system in presence of asymmetries. First, we derive the generalization of the Hamiltonian in Eq.~\eqref{Eq:HamQuantCompleteBal} with unequal rest capacitors, resistors, inductors and linear couplings. Differently from above, we linearise the symmetric/asymmetric electrical fields around their mean values ($\hat{Q}_{a/s}\rightarrow \langle\hat{Q}_{a/s}\rangle + \hat{\delta Q}_{a/s}$ and $\hat{\phi}_{a/s}\rightarrow \langle\hat{\phi}_{a/s}\rangle + \hat{\delta\phi}_{a/s}$), and the mechanical creation/annihilation operators ($\hat{b}^{(\dagger)}\rightarrow \langle \hat{b}^{(\dagger)} \rangle + \hat{\delta b}^{(\dagger)}$). Here, besides the usual oscillatory behaviour of the mechanical operators $\langle \hat{b}^{(\dagger)} \rangle$, the amplitude is generally time dependent \cite{Linearization}. This can be understood by looking at Eq.~\eqref{Eq:HamQuantCompleteBal}; since both the electrical fields have now non-zero average, the three body interaction $\propto \hat{Q}_{a} \hat{Q}_{s} (\hat{b}+\hat{b}^{\dagger})$ is equivalent to a force directly driving the mechanical system. Once solutions for the averages are found, it is possible to determine the variations, and finally the time evolution of the phonon number. 

\paragraph{Optimization of the visibility}
To obtain Fig.~\ref{Fig:Fig3} we rely on both an analytical and a numerical optimization of the visibility $\xi$. To determine the red, dotted curve, we assume that the initial mechanical state is thermal, such that the occupations of the Fock states can be found. Given $\Delta n_{b}$, the probability to jump once either up or down during the measurement time $T$ is a Poissonian process. The probability distribution function for the outcomes $V_{M}$ can then be obtained and maximised, by varying $\Delta n_{b}$. The histograms and the blue points are derived with Monte--Carlo simulations, where the time evolution of single mechanical trajectories are replicated with the stochastic wave-function method \cite{Dalibard1992}. Importantly, every measurement interval of duration $T$ has been discretized, to allow for multiple jumps. The parameter $\Delta n_{b}$ is then varied to find the best visibility $\xi$.

\end{footnotesize}


\begin{thebibliography}{49}%
	\makeatletter
	\providecommand \@ifxundefined [1]{%
		\@ifx{#1\undefined}
	}%
	\providecommand \@ifnum [1]{%
		\ifnum #1\expandafter \@firstoftwo
		\else \expandafter \@secondoftwo
		\fi
	}%
	\providecommand \@ifx [1]{%
		\ifx #1\expandafter \@firstoftwo
		\else \expandafter \@secondoftwo
		\fi
	}%
	\providecommand \natexlab [1]{#1}%
	\providecommand \enquote  [1]{``#1''}%
	\providecommand \bibnamefont  [1]{#1}%
	\providecommand \bibfnamefont [1]{#1}%
	\providecommand \citenamefont [1]{#1}%
	\providecommand \href@noop [0]{\@secondoftwo}%
	\providecommand \href [0]{\begingroup \@sanitize@url \@href}%
	\providecommand \@href[1]{\@@startlink{#1}\@@href}%
	\providecommand \@@href[1]{\endgroup#1\@@endlink}%
	\providecommand \@sanitize@url [0]{\catcode `\\12\catcode `\$12\catcode
		`\&12\catcode `\#12\catcode `\^12\catcode `\_12\catcode `\%12\relax}%
	\providecommand \@@startlink[1]{}%
	\providecommand \@@endlink[0]{}%
	\providecommand \url  [0]{\begingroup\@sanitize@url \@url }%
	\providecommand \@url [1]{\endgroup\@href {#1}{\urlprefix }}%
	\providecommand \urlprefix  [0]{URL }%
	\providecommand \Eprint [0]{\href }%
	\providecommand \doibase [0]{http://dx.doi.org/}%
	\providecommand \selectlanguage [0]{\@gobble}%
	\providecommand \bibinfo  [0]{\@secondoftwo}%
	\providecommand \bibfield  [0]{\@secondoftwo}%
	\providecommand \translation [1]{[#1]}%
	\providecommand \BibitemOpen [0]{}%
	\providecommand \bibitemStop [0]{}%
	\providecommand \bibitemNoStop [0]{.\EOS\space}%
	\providecommand \EOS [0]{\spacefactor3000\relax}%
	\providecommand \BibitemShut  [1]{\csname bibitem#1\endcsname}%
	\let\auto@bib@innerbib\@empty
	%</preamble>
	\bibitem [{\citenamefont {Martinis}\ \emph {et~al.}(1985)\citenamefont
		{Martinis}, \citenamefont {Devoret},\ and\ \citenamefont
		{Clarke}}]{EnQSCsystem}%
	\BibitemOpen
	\bibfield  {author} {\bibinfo {author} {\bibfnamefont {J.~M.}\ \bibnamefont
			{Martinis}}, \bibinfo {author} {\bibfnamefont {M.~H.}\ \bibnamefont
			{Devoret}}, \ and\ \bibinfo {author} {\bibfnamefont {J.}~\bibnamefont
			{Clarke}},\ }\href {\doibase 10.1103/PhysRevLett.55.1543} {\bibfield
		{journal} {\bibinfo  {journal} {Phys. Rev. Lett.}\ }\textbf {\bibinfo
			{volume} {55}},\ \bibinfo {pages} {1543} (\bibinfo {year}
		{1985})}\BibitemShut {NoStop}%
	\bibitem [{\citenamefont {Aspelmeyer}\ \emph {et~al.}(2014)\citenamefont
		{Aspelmeyer}, \citenamefont {Kippenberg},\ and\ \citenamefont
		{Marquardt}}]{MarquardtReview}%
	\BibitemOpen
	\bibfield  {author} {\bibinfo {author} {\bibfnamefont {M.}~\bibnamefont
			{Aspelmeyer}}, \bibinfo {author} {\bibfnamefont {T.~J.}\ \bibnamefont
			{Kippenberg}}, \ and\ \bibinfo {author} {\bibfnamefont {F.}~\bibnamefont
			{Marquardt}},\ }\href@noop {} {\bibfield  {journal} {\bibinfo  {journal}
			{Reviews of Modern Physics}\ }\textbf {\bibinfo {volume} {86}},\ \bibinfo
		{pages} {1391} (\bibinfo {year} {2014})}\BibitemShut {NoStop}%
	\bibitem [{\citenamefont {Teufel}\ \emph
		{et~al.}(2011{\natexlab{a}})\citenamefont {Teufel}, \citenamefont {Li},
		\citenamefont {Allman}, \citenamefont {Cicak}, \citenamefont {Sirois},
		\citenamefont {Whittaker},\ and\ \citenamefont {Simmonds}}]{Teufel2011sc}%
	\BibitemOpen
	\bibfield  {author} {\bibinfo {author} {\bibfnamefont {J.}~\bibnamefont
			{Teufel}}, \bibinfo {author} {\bibfnamefont {D.}~\bibnamefont {Li}}, \bibinfo
		{author} {\bibfnamefont {M.}~\bibnamefont {Allman}}, \bibinfo {author}
		{\bibfnamefont {K.}~\bibnamefont {Cicak}}, \bibinfo {author} {\bibfnamefont
			{A.}~\bibnamefont {Sirois}}, \bibinfo {author} {\bibfnamefont
			{J.}~\bibnamefont {Whittaker}}, \ and\ \bibinfo {author} {\bibfnamefont
			{R.}~\bibnamefont {Simmonds}},\ }\href@noop {} {\bibfield  {journal}
		{\bibinfo  {journal} {Nature}\ }\textbf {\bibinfo {volume} {471}},\ \bibinfo
		{pages} {204} (\bibinfo {year} {2011}{\natexlab{a}})}\BibitemShut {NoStop}%
	\bibitem [{\citenamefont {Teufel}\ \emph
		{et~al.}(2011{\natexlab{b}})\citenamefont {Teufel}, \citenamefont {Donner},
		\citenamefont {Li}, \citenamefont {Harlow}, \citenamefont {Allman},
		\citenamefont {Cicak}, \citenamefont {Sirois}, \citenamefont {Whittaker},
		\citenamefont {Lehnert},\ and\ \citenamefont {Simmonds}}]{Cooling2}%
	\BibitemOpen
	\bibfield  {author} {\bibinfo {author} {\bibfnamefont {J.~D.}\ \bibnamefont
			{Teufel}}, \bibinfo {author} {\bibfnamefont {T.}~\bibnamefont {Donner}},
		\bibinfo {author} {\bibfnamefont {D.}~\bibnamefont {Li}}, \bibinfo {author}
		{\bibfnamefont {J.~W.}\ \bibnamefont {Harlow}}, \bibinfo {author}
		{\bibfnamefont {M.~S.}\ \bibnamefont {Allman}}, \bibinfo {author}
		{\bibfnamefont {K.}~\bibnamefont {Cicak}}, \bibinfo {author} {\bibfnamefont
			{A.~J.}\ \bibnamefont {Sirois}}, \bibinfo {author} {\bibfnamefont {J.~D.}\
			\bibnamefont {Whittaker}}, \bibinfo {author} {\bibfnamefont {K.~W.}\
			\bibnamefont {Lehnert}}, \ and\ \bibinfo {author} {\bibfnamefont {R.~W.}\
			\bibnamefont {Simmonds}},\ }\href {\doibase 10.1038/nature10261} {\bibfield
		{journal} {\bibinfo  {journal} {Nature}\ }\textbf {\bibinfo {volume} {475}},\
		\bibinfo {pages} {359} (\bibinfo {year} {2011}{\natexlab{b}})}\BibitemShut
	{NoStop}%
	\bibitem [{\citenamefont {O’Connell}\ \emph {et~al.}(2010)\citenamefont
		{O’Connell}, \citenamefont {Hofheinz}, \citenamefont {Ansmann},
		\citenamefont {Bialczak}, \citenamefont {Lenander}, \citenamefont {Lucero},
		\citenamefont {Neeley}, \citenamefont {Sank}, \citenamefont {Wang},
		\citenamefont {Weides}, \citenamefont {Wenner}, \citenamefont {Martinis},\
		and\ \citenamefont {Cleland}}]{Cooling3}%
	\BibitemOpen
	\bibfield  {author} {\bibinfo {author} {\bibfnamefont {A.~D.}\ \bibnamefont
			{O’Connell}}, \bibinfo {author} {\bibfnamefont {M.}~\bibnamefont
			{Hofheinz}}, \bibinfo {author} {\bibfnamefont {M.}~\bibnamefont {Ansmann}},
		\bibinfo {author} {\bibfnamefont {R.~C.}\ \bibnamefont {Bialczak}}, \bibinfo
		{author} {\bibfnamefont {M.}~\bibnamefont {Lenander}}, \bibinfo {author}
		{\bibfnamefont {E.}~\bibnamefont {Lucero}}, \bibinfo {author} {\bibfnamefont
			{M.}~\bibnamefont {Neeley}}, \bibinfo {author} {\bibfnamefont
			{D.}~\bibnamefont {Sank}}, \bibinfo {author} {\bibfnamefont {H.}~\bibnamefont
			{Wang}}, \bibinfo {author} {\bibfnamefont {M.}~\bibnamefont {Weides}},
		\bibinfo {author} {\bibfnamefont {J.}~\bibnamefont {Wenner}}, \bibinfo
		{author} {\bibfnamefont {J.~M.}\ \bibnamefont {Martinis}}, \ and\ \bibinfo
		{author} {\bibfnamefont {A.~N.}\ \bibnamefont {Cleland}},\ }\href {\doibase
		10.1038/nature08967} {\bibfield  {journal} {\bibinfo  {journal} {Nature}\
		}\textbf {\bibinfo {volume} {464}},\ \bibinfo {pages} {697} (\bibinfo {year}
		{2010})}\BibitemShut {NoStop}%
	\bibitem [{\citenamefont {Heinrich}\ and\ \citenamefont
		{Marquardt}(2011)}]{PhononQNDmw}%
	\BibitemOpen
	\bibfield  {author} {\bibinfo {author} {\bibfnamefont {G.}~\bibnamefont
			{Heinrich}}\ and\ \bibinfo {author} {\bibfnamefont {F.}~\bibnamefont
			{Marquardt}},\ }\href {http://stacks.iop.org/0295-5075/93/i=1/a=18003}
	{\bibfield  {journal} {\bibinfo  {journal} {EPL (Europhysics Letters)}\
		}\textbf {\bibinfo {volume} {93}},\ \bibinfo {pages} {18003} (\bibinfo {year}
		{2011})}\BibitemShut {NoStop}%
	\bibitem [{\citenamefont {Wollman}\ \emph {et~al.}(2015)\citenamefont
		{Wollman}, \citenamefont {Lei}, \citenamefont {Weinstein}, \citenamefont
		{Suh}, \citenamefont {Kronwald}, \citenamefont {Marquardt}, \citenamefont
		{Clerk},\ and\ \citenamefont {Schwab}}]{Wollman2015}%
	\BibitemOpen
	\bibfield  {author} {\bibinfo {author} {\bibfnamefont {E.~E.}\ \bibnamefont
			{Wollman}}, \bibinfo {author} {\bibfnamefont {C.}~\bibnamefont {Lei}},
		\bibinfo {author} {\bibfnamefont {A.}~\bibnamefont {Weinstein}}, \bibinfo
		{author} {\bibfnamefont {J.}~\bibnamefont {Suh}}, \bibinfo {author}
		{\bibfnamefont {A.}~\bibnamefont {Kronwald}}, \bibinfo {author}
		{\bibfnamefont {F.}~\bibnamefont {Marquardt}}, \bibinfo {author}
		{\bibfnamefont {A.}~\bibnamefont {Clerk}}, \ and\ \bibinfo {author}
		{\bibfnamefont {K.}~\bibnamefont {Schwab}},\ }\href@noop {} {\bibfield
		{journal} {\bibinfo  {journal} {Science}\ }\textbf {\bibinfo {volume}
			{349}},\ \bibinfo {pages} {952} (\bibinfo {year} {2015})}\BibitemShut
	{NoStop}%
	\bibitem [{\citenamefont {Ockeloen-Korppi}\ \emph {et~al.}(2017)\citenamefont
		{Ockeloen-Korppi}, \citenamefont {Damsk\"agg}, \citenamefont {Pirkkalainen},
		\citenamefont {Heikkil\"a}, \citenamefont {Massel},\ and\ \citenamefont
		{Sillanp\"a\"a}}]{Sillanpaa2017}%
	\BibitemOpen
	\bibfield  {author} {\bibinfo {author} {\bibfnamefont {C.~F.}\ \bibnamefont
			{Ockeloen-Korppi}}, \bibinfo {author} {\bibfnamefont {E.}~\bibnamefont
			{Damsk\"agg}}, \bibinfo {author} {\bibfnamefont {J.-M.}\ \bibnamefont
			{Pirkkalainen}}, \bibinfo {author} {\bibfnamefont {T.~T.}\ \bibnamefont
			{Heikkil\"a}}, \bibinfo {author} {\bibfnamefont {F.}~\bibnamefont {Massel}},
		\ and\ \bibinfo {author} {\bibfnamefont {M.~A.}\ \bibnamefont
			{Sillanp\"a\"a}},\ }\href {\doibase 10.1103/PhysRevLett.118.103601}
	{\bibfield  {journal} {\bibinfo  {journal} {Phys. Rev. Lett.}\ }\textbf
		{\bibinfo {volume} {118}},\ \bibinfo {pages} {103601} (\bibinfo {year}
		{2017})}\BibitemShut {NoStop}%
	\bibitem [{\citenamefont {Chan}\ \emph {et~al.}(2011)\citenamefont {Chan},
		\citenamefont {Alegre}, \citenamefont {Safavi-Naeini}, \citenamefont {Hill},
		\citenamefont {Krause}, \citenamefont {Groblacher}, \citenamefont
		{Aspelmeyer},\ and\ \citenamefont {Painter}}]{Cooling1}%
	\BibitemOpen
	\bibfield  {author} {\bibinfo {author} {\bibfnamefont {J.}~\bibnamefont
			{Chan}}, \bibinfo {author} {\bibfnamefont {T.~P.~M.}\ \bibnamefont {Alegre}},
		\bibinfo {author} {\bibfnamefont {A.~H.}\ \bibnamefont {Safavi-Naeini}},
		\bibinfo {author} {\bibfnamefont {J.~T.}\ \bibnamefont {Hill}}, \bibinfo
		{author} {\bibfnamefont {A.}~\bibnamefont {Krause}}, \bibinfo {author}
		{\bibfnamefont {S.}~\bibnamefont {Groblacher}}, \bibinfo {author}
		{\bibfnamefont {M.}~\bibnamefont {Aspelmeyer}}, \ and\ \bibinfo {author}
		{\bibfnamefont {O.}~\bibnamefont {Painter}},\ }\href {\doibase
		10.1038/nature10461} {\bibfield  {journal} {\bibinfo  {journal} {Nature}\
		}\textbf {\bibinfo {volume} {478}},\ \bibinfo {pages} {89} (\bibinfo {year}
		{2011})}\BibitemShut {NoStop}%
	\bibitem [{\citenamefont {Peterson}\ \emph {et~al.}(2016)\citenamefont
		{Peterson}, \citenamefont {Purdy}, \citenamefont {Kampel}, \citenamefont
		{Andrews}, \citenamefont {Yu}, \citenamefont {Lehnert},\ and\ \citenamefont
		{Regal}}]{AmCooling1}%
	\BibitemOpen
	\bibfield  {author} {\bibinfo {author} {\bibfnamefont {R.~W.}\ \bibnamefont
			{Peterson}}, \bibinfo {author} {\bibfnamefont {T.~P.}\ \bibnamefont {Purdy}},
		\bibinfo {author} {\bibfnamefont {N.~S.}\ \bibnamefont {Kampel}}, \bibinfo
		{author} {\bibfnamefont {R.~W.}\ \bibnamefont {Andrews}}, \bibinfo {author}
		{\bibfnamefont {P.-L.}\ \bibnamefont {Yu}}, \bibinfo {author} {\bibfnamefont
			{K.~W.}\ \bibnamefont {Lehnert}}, \ and\ \bibinfo {author} {\bibfnamefont
			{C.~A.}\ \bibnamefont {Regal}},\ }\href {\doibase
		10.1103/PhysRevLett.116.063601} {\bibfield  {journal} {\bibinfo  {journal}
			{Phys. Rev. Lett.}\ }\textbf {\bibinfo {volume} {116}},\ \bibinfo {pages}
		{063601} (\bibinfo {year} {2016})}\BibitemShut {NoStop}%
	\bibitem [{\citenamefont {Schliesser}\ \emph {et~al.}(2006)\citenamefont
		{Schliesser}, \citenamefont {Del'Haye}, \citenamefont {Nooshi}, \citenamefont
		{Vahala},\ and\ \citenamefont {Kippenberg}}]{Cooling6}%
	\BibitemOpen
	\bibfield  {author} {\bibinfo {author} {\bibfnamefont {A.}~\bibnamefont
			{Schliesser}}, \bibinfo {author} {\bibfnamefont {P.}~\bibnamefont
			{Del'Haye}}, \bibinfo {author} {\bibfnamefont {N.}~\bibnamefont {Nooshi}},
		\bibinfo {author} {\bibfnamefont {K.~J.}\ \bibnamefont {Vahala}}, \ and\
		\bibinfo {author} {\bibfnamefont {T.~J.}\ \bibnamefont {Kippenberg}},\ }\href
	{\doibase 10.1103/PhysRevLett.97.243905} {\bibfield  {journal} {\bibinfo
			{journal} {Phys. Rev. Lett.}\ }\textbf {\bibinfo {volume} {97}},\ \bibinfo
		{pages} {243905} (\bibinfo {year} {2006})}\BibitemShut {NoStop}%
	\bibitem [{\citenamefont {Teufel}\ \emph {et~al.}(2009)\citenamefont {Teufel},
		\citenamefont {Donner}, \citenamefont {Castellanos-Beltran}, \citenamefont
		{Harlow},\ and\ \citenamefont {Lehnert}}]{Teufel2009}%
	\BibitemOpen
	\bibfield  {author} {\bibinfo {author} {\bibfnamefont {J.}~\bibnamefont
			{Teufel}}, \bibinfo {author} {\bibfnamefont {T.}~\bibnamefont {Donner}},
		\bibinfo {author} {\bibfnamefont {M.}~\bibnamefont {Castellanos-Beltran}},
		\bibinfo {author} {\bibfnamefont {J.}~\bibnamefont {Harlow}}, \ and\ \bibinfo
		{author} {\bibfnamefont {K.}~\bibnamefont {Lehnert}},\ }\href@noop {}
	{\bibfield  {journal} {\bibinfo  {journal} {Nature nanotechnology}\ }\textbf
		{\bibinfo {volume} {4}},\ \bibinfo {pages} {820} (\bibinfo {year}
		{2009})}\BibitemShut {NoStop}%
	\bibitem [{\citenamefont {Gil-Santos}\ \emph {et~al.}(2010)\citenamefont
		{Gil-Santos}, \citenamefont {Ramos}, \citenamefont {Martinez}, \citenamefont
		{Fernandez-Regulez}, \citenamefont {Garcia}, \citenamefont {San~Paulo},
		\citenamefont {Calleja},\ and\ \citenamefont {Tamayo}}]{Sensing1}%
	\BibitemOpen
	\bibfield  {author} {\bibinfo {author} {\bibfnamefont {E.}~\bibnamefont
			{Gil-Santos}}, \bibinfo {author} {\bibfnamefont {D.}~\bibnamefont {Ramos}},
		\bibinfo {author} {\bibfnamefont {J.}~\bibnamefont {Martinez}}, \bibinfo
		{author} {\bibfnamefont {M.}~\bibnamefont {Fernandez-Regulez}}, \bibinfo
		{author} {\bibfnamefont {R.}~\bibnamefont {Garcia}}, \bibinfo {author}
		{\bibfnamefont {A.}~\bibnamefont {San~Paulo}}, \bibinfo {author}
		{\bibfnamefont {M.}~\bibnamefont {Calleja}}, \ and\ \bibinfo {author}
		{\bibfnamefont {J.}~\bibnamefont {Tamayo}},\ }\href {\doibase
		10.1038/nnano.2010.151} {\bibfield  {journal} {\bibinfo  {journal} {Nat
				Nano}\ }\textbf {\bibinfo {volume} {5}},\ \bibinfo {pages} {641} (\bibinfo
		{year} {2010})}\BibitemShut {NoStop}%
	\bibitem [{\citenamefont {Hanay}\ \emph {et~al.}(2012)\citenamefont {Hanay},
		\citenamefont {Kelber}, \citenamefont {Naik}, \citenamefont {Chi},
		\citenamefont {Hentz}, \citenamefont {Bullard}, \citenamefont {Colinet},
		\citenamefont {Duraffourg},\ and\ \citenamefont {Roukes}}]{Sensing2}%
	\BibitemOpen
	\bibfield  {author} {\bibinfo {author} {\bibfnamefont {M.}~\bibnamefont
			{Hanay}}, \bibinfo {author} {\bibfnamefont {S.}~\bibnamefont {Kelber}},
		\bibinfo {author} {\bibfnamefont {A.}~\bibnamefont {Naik}}, \bibinfo {author}
		{\bibfnamefont {D.}~\bibnamefont {Chi}}, \bibinfo {author} {\bibfnamefont
			{S.}~\bibnamefont {Hentz}}, \bibinfo {author} {\bibfnamefont
			{E.}~\bibnamefont {Bullard}}, \bibinfo {author} {\bibfnamefont
			{E.}~\bibnamefont {Colinet}}, \bibinfo {author} {\bibfnamefont
			{L.}~\bibnamefont {Duraffourg}}, \ and\ \bibinfo {author} {\bibfnamefont
			{M.}~\bibnamefont {Roukes}},\ }\href@noop {} {\bibfield  {journal} {\bibinfo
			{journal} {Nature nanotechnology}\ }\textbf {\bibinfo {volume} {7}},\
		\bibinfo {pages} {602} (\bibinfo {year} {2012})}\BibitemShut {NoStop}%
	\bibitem [{\citenamefont {Weber}\ \emph {et~al.}(2016)\citenamefont {Weber},
		\citenamefont {G\"{u}ttinger}, \citenamefont {Noury}, \citenamefont
		{Vergara-Cruz},\ and\ \citenamefont {Bachtold}}]{ExpNumb6}%
	\BibitemOpen
	\bibfield  {author} {\bibinfo {author} {\bibfnamefont {P.}~\bibnamefont
			{Weber}}, \bibinfo {author} {\bibfnamefont {J.}~\bibnamefont
			{G\"{u}ttinger}}, \bibinfo {author} {\bibfnamefont {A.}~\bibnamefont
			{Noury}}, \bibinfo {author} {\bibfnamefont {J.}~\bibnamefont {Vergara-Cruz}},
		\ and\ \bibinfo {author} {\bibfnamefont {A.}~\bibnamefont {Bachtold}},\
	}\href {\doibase 10.1038/ncomms12496} {\bibfield  {journal} {\bibinfo
			{journal} {Nature Communications}\ }\textbf {\bibinfo {volume} {7}},\
		\bibinfo {pages} {12496} (\bibinfo {year} {2016})}\BibitemShut {NoStop}%
	\bibitem [{\citenamefont {Brooks}\ \emph {et~al.}(2012)\citenamefont {Brooks},
		\citenamefont {Botter}, \citenamefont {Schreppler}, \citenamefont {Purdy},
		\citenamefont {Brahms},\ and\ \citenamefont {Stamper-Kurn}}]{SqLight1}%
	\BibitemOpen
	\bibfield  {author} {\bibinfo {author} {\bibfnamefont {D.~W.}\ \bibnamefont
			{Brooks}}, \bibinfo {author} {\bibfnamefont {T.}~\bibnamefont {Botter}},
		\bibinfo {author} {\bibfnamefont {S.}~\bibnamefont {Schreppler}}, \bibinfo
		{author} {\bibfnamefont {T.~P.}\ \bibnamefont {Purdy}}, \bibinfo {author}
		{\bibfnamefont {N.}~\bibnamefont {Brahms}}, \ and\ \bibinfo {author}
		{\bibfnamefont {D.~M.}\ \bibnamefont {Stamper-Kurn}},\ }\href@noop {}
	{\bibfield  {journal} {\bibinfo  {journal} {Nature}\ }\textbf {\bibinfo
			{volume} {488}},\ \bibinfo {pages} {476} (\bibinfo {year}
		{2012})}\BibitemShut {NoStop}%
	\bibitem [{\citenamefont {Safavi-Naeini}\ \emph {et~al.}(2013)\citenamefont
		{Safavi-Naeini}, \citenamefont {Gr{\"o}blacher}, \citenamefont {Hill},
		\citenamefont {Chan}, \citenamefont {Aspelmeyer},\ and\ \citenamefont
		{Painter}}]{SqLight2}%
	\BibitemOpen
	\bibfield  {author} {\bibinfo {author} {\bibfnamefont {A.~H.}\ \bibnamefont
			{Safavi-Naeini}}, \bibinfo {author} {\bibfnamefont {S.}~\bibnamefont
			{Gr{\"o}blacher}}, \bibinfo {author} {\bibfnamefont {J.~T.}\ \bibnamefont
			{Hill}}, \bibinfo {author} {\bibfnamefont {J.}~\bibnamefont {Chan}}, \bibinfo
		{author} {\bibfnamefont {M.}~\bibnamefont {Aspelmeyer}}, \ and\ \bibinfo
		{author} {\bibfnamefont {O.}~\bibnamefont {Painter}},\ }\href@noop {}
	{\bibfield  {journal} {\bibinfo  {journal} {Nature}\ }\textbf {\bibinfo
			{volume} {500}},\ \bibinfo {pages} {185} (\bibinfo {year}
		{2013})}\BibitemShut {NoStop}%
	\bibitem [{\citenamefont {Purdy}\ \emph {et~al.}(2013)\citenamefont {Purdy},
		\citenamefont {Yu}, \citenamefont {Peterson}, \citenamefont {Kampel},\ and\
		\citenamefont {Regal}}]{SqLight3}%
	\BibitemOpen
	\bibfield  {author} {\bibinfo {author} {\bibfnamefont {T.~P.}\ \bibnamefont
			{Purdy}}, \bibinfo {author} {\bibfnamefont {P.-L.}\ \bibnamefont {Yu}},
		\bibinfo {author} {\bibfnamefont {R.~W.}\ \bibnamefont {Peterson}}, \bibinfo
		{author} {\bibfnamefont {N.~S.}\ \bibnamefont {Kampel}}, \ and\ \bibinfo
		{author} {\bibfnamefont {C.~A.}\ \bibnamefont {Regal}},\ }\href {\doibase
		10.1103/PhysRevX.3.031012} {\bibfield  {journal} {\bibinfo  {journal} {Phys.
				Rev. X}\ }\textbf {\bibinfo {volume} {3}},\ \bibinfo {pages} {031012}
		(\bibinfo {year} {2013})}\BibitemShut {NoStop}%
	\bibitem [{\citenamefont {M{\o}ller}\ \emph {et~al.}(2017)\citenamefont
		{M{\o}ller}, \citenamefont {Thomas}, \citenamefont {Vasilakis}, \citenamefont
		{Zeuthen}, \citenamefont {Tsaturyan}, \citenamefont {Jensen}, \citenamefont
		{Schliesser}, \citenamefont {Hammerer},\ and\ \citenamefont
		{Polzik}}]{QB-EvasionNBI}%
	\BibitemOpen
	\bibfield  {author} {\bibinfo {author} {\bibfnamefont {C.~B.}\ \bibnamefont
			{M{\o}ller}}, \bibinfo {author} {\bibfnamefont {R.~A.}\ \bibnamefont
			{Thomas}}, \bibinfo {author} {\bibfnamefont {G.}~\bibnamefont {Vasilakis}},
		\bibinfo {author} {\bibfnamefont {E.}~\bibnamefont {Zeuthen}}, \bibinfo
		{author} {\bibfnamefont {Y.}~\bibnamefont {Tsaturyan}}, \bibinfo {author}
		{\bibfnamefont {K.}~\bibnamefont {Jensen}}, \bibinfo {author} {\bibfnamefont
			{A.}~\bibnamefont {Schliesser}}, \bibinfo {author} {\bibfnamefont
			{K.}~\bibnamefont {Hammerer}}, \ and\ \bibinfo {author} {\bibfnamefont
			{E.~S.}\ \bibnamefont {Polzik}},\ }\href {\doibase
		http://dx.doi.org/10.1038/nature22980} {\bibfield  {journal} {\bibinfo
			{journal} {Nature}\ }\textbf {\bibinfo {volume} {547}},\ \bibinfo {pages}
		{191} (\bibinfo {year} {2017})}\BibitemShut {NoStop}%
	\bibitem [{\citenamefont {Ockeloen-Korppi}\ \emph {et~al.}(2016)\citenamefont
		{Ockeloen-Korppi}, \citenamefont {Damsk\"agg}, \citenamefont {Pirkkalainen},
		\citenamefont {Clerk}, \citenamefont {Woolley},\ and\ \citenamefont
		{Sillanp\"a\"a}}]{QB-Evasion2}%
	\BibitemOpen
	\bibfield  {author} {\bibinfo {author} {\bibfnamefont {C.~F.}\ \bibnamefont
			{Ockeloen-Korppi}}, \bibinfo {author} {\bibfnamefont {E.}~\bibnamefont
			{Damsk\"agg}}, \bibinfo {author} {\bibfnamefont {J.-M.}\ \bibnamefont
			{Pirkkalainen}}, \bibinfo {author} {\bibfnamefont {A.~A.}\ \bibnamefont
			{Clerk}}, \bibinfo {author} {\bibfnamefont {M.~J.}\ \bibnamefont {Woolley}},
		\ and\ \bibinfo {author} {\bibfnamefont {M.~A.}\ \bibnamefont
			{Sillanp\"a\"a}},\ }\href {\doibase 10.1103/PhysRevLett.117.140401}
	{\bibfield  {journal} {\bibinfo  {journal} {Phys. Rev. Lett.}\ }\textbf
		{\bibinfo {volume} {117}},\ \bibinfo {pages} {140401} (\bibinfo {year}
		{2016})}\BibitemShut {NoStop}%
	\bibitem [{\citenamefont {\text{Abbott et al.}}(2016)}]{Abbott2016}%
	\BibitemOpen
	\bibfield  {author} {\bibinfo {author} {\bibnamefont {\text{Abbott et al.}}}
		(\bibinfo {collaboration} {LIGO Scientific Collaboration and Virgo
			Collaboration}),\ }\href {\doibase 10.1103/PhysRevLett.116.061102} {\bibfield
		{journal} {\bibinfo  {journal} {Phys. Rev. Lett.}\ }\textbf {\bibinfo
			{volume} {116}},\ \bibinfo {pages} {061102} (\bibinfo {year}
		{2016})}\BibitemShut {NoStop}%
	\bibitem [{\citenamefont {Gr{\"o}blacher}\ \emph {et~al.}(2009)\citenamefont
		{Gr{\"o}blacher}, \citenamefont {Hammerer}, \citenamefont {Vanner},\ and\
		\citenamefont {Aspelmeyer}}]{Groblacher2009}%
	\BibitemOpen
	\bibfield  {author} {\bibinfo {author} {\bibfnamefont {S.}~\bibnamefont
			{Gr{\"o}blacher}}, \bibinfo {author} {\bibfnamefont {K.}~\bibnamefont
			{Hammerer}}, \bibinfo {author} {\bibfnamefont {M.~R.}\ \bibnamefont
			{Vanner}}, \ and\ \bibinfo {author} {\bibfnamefont {M.}~\bibnamefont
			{Aspelmeyer}},\ }\href@noop {} {\bibfield  {journal} {\bibinfo  {journal}
			{Nature}\ }\textbf {\bibinfo {volume} {460}},\ \bibinfo {pages} {724}
		(\bibinfo {year} {2009})}\BibitemShut {NoStop}%
	\bibitem [{\citenamefont {Lloyd}\ and\ \citenamefont
		{Braunstein}(1999)}]{LloydQC}%
	\BibitemOpen
	\bibfield  {author} {\bibinfo {author} {\bibfnamefont {S.}~\bibnamefont
			{Lloyd}}\ and\ \bibinfo {author} {\bibfnamefont {S.~L.}\ \bibnamefont
			{Braunstein}},\ }\href {\doibase 10.1103/PhysRevLett.82.1784} {\bibfield
		{journal} {\bibinfo  {journal} {Phys. Rev. Lett.}\ }\textbf {\bibinfo
			{volume} {82}},\ \bibinfo {pages} {1784} (\bibinfo {year}
		{1999})}\BibitemShut {NoStop}%
	\bibitem [{\citenamefont {Braginsky}\ and\ \citenamefont
		{Khalili}(1996)}]{BraginskyRev}%
	\BibitemOpen
	\bibfield  {author} {\bibinfo {author} {\bibfnamefont {V.~B.}\ \bibnamefont
			{Braginsky}}\ and\ \bibinfo {author} {\bibfnamefont {F.~Y.}\ \bibnamefont
			{Khalili}},\ }\href@noop {} {\bibfield  {journal} {\bibinfo  {journal}
			{Reviews of Modern Physics}\ }\textbf {\bibinfo {volume} {68}},\ \bibinfo
		{pages} {1} (\bibinfo {year} {1996})}\BibitemShut {NoStop}%
	\bibitem [{\citenamefont {Clerk}\ \emph {et~al.}(2010)\citenamefont {Clerk},
		\citenamefont {Devoret}, \citenamefont {Girvin}, \citenamefont {Marquardt},\
		and\ \citenamefont {Schoelkopf}}]{ClerkRev}%
	\BibitemOpen
	\bibfield  {author} {\bibinfo {author} {\bibfnamefont {A.~A.}\ \bibnamefont
			{Clerk}}, \bibinfo {author} {\bibfnamefont {M.~H.}\ \bibnamefont {Devoret}},
		\bibinfo {author} {\bibfnamefont {S.~M.}\ \bibnamefont {Girvin}}, \bibinfo
		{author} {\bibfnamefont {F.}~\bibnamefont {Marquardt}}, \ and\ \bibinfo
		{author} {\bibfnamefont {R.~J.}\ \bibnamefont {Schoelkopf}},\ }\href@noop {}
	{\bibfield  {journal} {\bibinfo  {journal} {Reviews of Modern Physics}\
		}\textbf {\bibinfo {volume} {82}},\ \bibinfo {pages} {1155} (\bibinfo {year}
		{2010})}\BibitemShut {NoStop}%
	\bibitem [{\citenamefont {Thompson}\ \emph {et~al.}(2008)\citenamefont
		{Thompson}, \citenamefont {Zwickl}, \citenamefont {Jayich}, \citenamefont
		{Marquardt}, \citenamefont {Girvin},\ and\ \citenamefont
		{Harris}}]{Marquardt}%
	\BibitemOpen
	\bibfield  {author} {\bibinfo {author} {\bibfnamefont {J.~D.}\ \bibnamefont
			{Thompson}}, \bibinfo {author} {\bibfnamefont {B.~M.}\ \bibnamefont
			{Zwickl}}, \bibinfo {author} {\bibfnamefont {A.~M.}\ \bibnamefont {Jayich}},
		\bibinfo {author} {\bibfnamefont {F.}~\bibnamefont {Marquardt}}, \bibinfo
		{author} {\bibfnamefont {S.~M.}\ \bibnamefont {Girvin}}, \ and\ \bibinfo
		{author} {\bibfnamefont {J.~G.~E.}\ \bibnamefont {Harris}},\ }\href {\doibase
		10.1038/nature06715} {\bibfield  {journal} {\bibinfo  {journal} {Nature}\
		}\textbf {\bibinfo {volume} {452}} (\bibinfo {year} {2008}),\
		10.1038/nature06715}\BibitemShut {NoStop}%
	\bibitem [{\citenamefont {Jayich}\ \emph {et~al.}(2008)\citenamefont {Jayich},
		\citenamefont {Sankey}, \citenamefont {Zwickl}, \citenamefont {Yang},
		\citenamefont {Thompson}, \citenamefont {Girvin}, \citenamefont {Clerk},
		\citenamefont {Marquardt},\ and\ \citenamefont {Harris}}]{MarquardtFollow}%
	\BibitemOpen
	\bibfield  {author} {\bibinfo {author} {\bibfnamefont {A.~M.}\ \bibnamefont
			{Jayich}}, \bibinfo {author} {\bibfnamefont {J.~C.}\ \bibnamefont {Sankey}},
		\bibinfo {author} {\bibfnamefont {B.~M.}\ \bibnamefont {Zwickl}}, \bibinfo
		{author} {\bibfnamefont {C.}~\bibnamefont {Yang}}, \bibinfo {author}
		{\bibfnamefont {J.~D.}\ \bibnamefont {Thompson}}, \bibinfo {author}
		{\bibfnamefont {S.~M.}\ \bibnamefont {Girvin}}, \bibinfo {author}
		{\bibfnamefont {A.~A.}\ \bibnamefont {Clerk}}, \bibinfo {author}
		{\bibfnamefont {F.}~\bibnamefont {Marquardt}}, \ and\ \bibinfo {author}
		{\bibfnamefont {J.~G.~E.}\ \bibnamefont {Harris}},\ }\href
	{http://stacks.iop.org/1367-2630/10/i=9/a=095008} {\bibfield  {journal}
		{\bibinfo  {journal} {New Journal of Physics}\ }\textbf {\bibinfo {volume}
			{10}},\ \bibinfo {pages} {095008} (\bibinfo {year} {2008})}\BibitemShut
	{NoStop}%
	\bibitem [{\citenamefont {Miao}\ \emph {et~al.}(2009)\citenamefont {Miao},
		\citenamefont {Danilishin}, \citenamefont {Corbitt},\ and\ \citenamefont
		{Chen}}]{YanbeiChen}%
	\BibitemOpen
	\bibfield  {author} {\bibinfo {author} {\bibfnamefont {H.}~\bibnamefont
			{Miao}}, \bibinfo {author} {\bibfnamefont {S.}~\bibnamefont {Danilishin}},
		\bibinfo {author} {\bibfnamefont {T.}~\bibnamefont {Corbitt}}, \ and\
		\bibinfo {author} {\bibfnamefont {Y.}~\bibnamefont {Chen}},\ }\href {\doibase
		10.1103/PhysRevLett.103.100402} {\bibfield  {journal} {\bibinfo  {journal}
			{Phys. Rev. Lett.}\ }\textbf {\bibinfo {volume} {103}},\ \bibinfo {pages}
		{100402} (\bibinfo {year} {2009})}\BibitemShut {NoStop}%
	\bibitem [{\citenamefont {Yanay}\ \emph {et~al.}(2016)\citenamefont {Yanay},
		\citenamefont {Sankey},\ and\ \citenamefont {Clerk}}]{Clerk2016}%
	\BibitemOpen
	\bibfield  {author} {\bibinfo {author} {\bibfnamefont {Y.}~\bibnamefont
			{Yanay}}, \bibinfo {author} {\bibfnamefont {J.~C.}\ \bibnamefont {Sankey}}, \
		and\ \bibinfo {author} {\bibfnamefont {A.~A.}\ \bibnamefont {Clerk}},\ }\href
	{\doibase 10.1103/PhysRevA.93.063809} {\bibfield  {journal} {\bibinfo
			{journal} {Phys. Rev. A}\ }\textbf {\bibinfo {volume} {93}},\ \bibinfo
		{pages} {063809} (\bibinfo {year} {2016})}\BibitemShut {NoStop}%
	\bibitem [{\citenamefont {Ludwig}\ \emph {et~al.}(2012)\citenamefont {Ludwig},
		\citenamefont {Safavi-Naeini}, \citenamefont {Painter},\ and\ \citenamefont
		{Marquardt}}]{MarquardtPRL}%
	\BibitemOpen
	\bibfield  {author} {\bibinfo {author} {\bibfnamefont {M.}~\bibnamefont
			{Ludwig}}, \bibinfo {author} {\bibfnamefont {A.~H.}\ \bibnamefont
			{Safavi-Naeini}}, \bibinfo {author} {\bibfnamefont {O.}~\bibnamefont
			{Painter}}, \ and\ \bibinfo {author} {\bibfnamefont {F.}~\bibnamefont
			{Marquardt}},\ }\href {\doibase 10.1103/PhysRevLett.109.063601} {\bibfield
		{journal} {\bibinfo  {journal} {Phys. Rev. Lett.}\ }\textbf {\bibinfo
			{volume} {109}},\ \bibinfo {pages} {063601} (\bibinfo {year}
		{2012})}\BibitemShut {NoStop}%
	\bibitem [{\citenamefont {Yanay}\ and\ \citenamefont {Clerk}(2017)}]{Yanay}%
	\BibitemOpen
	\bibfield  {author} {\bibinfo {author} {\bibfnamefont {Y.}~\bibnamefont
			{Yanay}}\ and\ \bibinfo {author} {\bibfnamefont {A.~A.}\ \bibnamefont
			{Clerk}},\ }\href@noop {} {\bibfield  {journal} {\bibinfo  {journal} {New
				Journal of Physics}\ }\textbf {\bibinfo {volume} {19}},\ \bibinfo {pages}
		{033014} (\bibinfo {year} {2017})}\BibitemShut {NoStop}%
	\bibitem [{\citenamefont {Ding}\ \emph {et~al.}(2017)\citenamefont {Ding},
		\citenamefont {Maslennikov}, \citenamefont {Habl\"utzel},\ and\ \citenamefont
		{Matsukevich}}]{AtPhon}%
	\BibitemOpen
	\bibfield  {author} {\bibinfo {author} {\bibfnamefont {S.}~\bibnamefont
			{Ding}}, \bibinfo {author} {\bibfnamefont {G.}~\bibnamefont {Maslennikov}},
		\bibinfo {author} {\bibfnamefont {R.}~\bibnamefont {Habl\"utzel}}, \ and\
		\bibinfo {author} {\bibfnamefont {D.}~\bibnamefont {Matsukevich}},\ }\href
	{\doibase 10.1103/PhysRevLett.119.193602} {\bibfield  {journal} {\bibinfo
			{journal} {Phys. Rev. Lett.}\ }\textbf {\bibinfo {volume} {119}},\ \bibinfo
		{pages} {193602} (\bibinfo {year} {2017})}\BibitemShut {NoStop}%
	\bibitem [{SM()}]{SM}%
	\BibitemOpen
	\href@noop {} {}\bibinfo {note} {For details, see Supplemental Material
		available online.}\BibitemShut {Stop}%
	\bibitem [{\citenamefont {Martin}\ and\ \citenamefont
		{Zurek}(2007)}]{QNDphonon2}%
	\BibitemOpen
	\bibfield  {author} {\bibinfo {author} {\bibfnamefont {I.}~\bibnamefont
			{Martin}}\ and\ \bibinfo {author} {\bibfnamefont {W.~H.}\ \bibnamefont
			{Zurek}},\ }\href {\doibase 10.1103/PhysRevLett.98.120401} {\bibfield
		{journal} {\bibinfo  {journal} {Phys. Rev. Lett.}\ }\textbf {\bibinfo
			{volume} {98}},\ \bibinfo {pages} {120401} (\bibinfo {year}
		{2007})}\BibitemShut {NoStop}%
	\bibitem [{\citenamefont {Devoret}\ \emph {et~al.}(1995)\citenamefont {Devoret}
		\emph {et~al.}}]{devoret1995quantum}%
	\BibitemOpen
	\bibfield  {author} {\bibinfo {author} {\bibfnamefont {M.~H.}\ \bibnamefont
			{Devoret}} \emph {et~al.},\ }\href@noop {} {\bibfield  {journal} {\bibinfo
			{journal} {Les Houches, Session LXIII}\ }\textbf {\bibinfo {volume} {7}}
		(\bibinfo {year} {1995})}\BibitemShut {NoStop}%
	\bibitem [{Ses()}]{Sesta}%
	\BibitemOpen
	\href@noop {} {}\bibinfo {note} {The linear term also leads to mechanically
		induced damping of the electrical circuit, but this is typically negligible
		compared to $\gamma_{t}$.}\BibitemShut {Stop}%
	\bibitem [{Pri()}]{Prima}%
	\BibitemOpen
	\href@noop {} {}\bibinfo {note} {The simple analytical expressions for
		$\lambda_{p}$ given in the text is valid for the parasitic elements of the
		circuit being larger than the non-parasitic ones.}\BibitemShut {Stop}%
	\bibitem [{\citenamefont {Song}\ \emph {et~al.}(2012)\citenamefont {Song},
		\citenamefont {Oksanen}, \citenamefont {Sillanpää}, \citenamefont
		{Craighead}, \citenamefont {Parpia},\ and\ \citenamefont
		{Hakonen}}]{ExpNumb1}%
	\BibitemOpen
	\bibfield  {author} {\bibinfo {author} {\bibfnamefont {X.}~\bibnamefont
			{Song}}, \bibinfo {author} {\bibfnamefont {M.}~\bibnamefont {Oksanen}},
		\bibinfo {author} {\bibfnamefont {M.~A.}\ \bibnamefont {Sillanpää}},
		\bibinfo {author} {\bibfnamefont {H.~G.}\ \bibnamefont {Craighead}}, \bibinfo
		{author} {\bibfnamefont {J.~M.}\ \bibnamefont {Parpia}}, \ and\ \bibinfo
		{author} {\bibfnamefont {P.~J.}\ \bibnamefont {Hakonen}},\ }\href {\doibase
		10.1021/nl203305q} {\bibfield  {journal} {\bibinfo  {journal} {Nano Letters}\
		}\textbf {\bibinfo {volume} {12}},\ \bibinfo {pages} {198} (\bibinfo {year}
		{2012})},\ \bibinfo {note} {pMID: 22141577},\ \Eprint
	{http://arxiv.org/abs/http://dx.doi.org/10.1021/nl203305q}
	{http://dx.doi.org/10.1021/nl203305q} \BibitemShut {NoStop}%
	\bibitem [{\citenamefont {Bunch}\ \emph {et~al.}(2007)\citenamefont {Bunch},
		\citenamefont {van~der Zande}, \citenamefont {Verbridge}, \citenamefont
		{Frank}, \citenamefont {Tanenbaum}, \citenamefont {Parpia}, \citenamefont
		{Craighead},\ and\ \citenamefont {McEuen}}]{ExpNumb3}%
	\BibitemOpen
	\bibfield  {author} {\bibinfo {author} {\bibfnamefont {J.~S.}\ \bibnamefont
			{Bunch}}, \bibinfo {author} {\bibfnamefont {A.~M.}\ \bibnamefont {van~der
				Zande}}, \bibinfo {author} {\bibfnamefont {S.~S.}\ \bibnamefont {Verbridge}},
		\bibinfo {author} {\bibfnamefont {I.~W.}\ \bibnamefont {Frank}}, \bibinfo
		{author} {\bibfnamefont {D.~M.}\ \bibnamefont {Tanenbaum}}, \bibinfo {author}
		{\bibfnamefont {J.~M.}\ \bibnamefont {Parpia}}, \bibinfo {author}
		{\bibfnamefont {H.~G.}\ \bibnamefont {Craighead}}, \ and\ \bibinfo {author}
		{\bibfnamefont {P.~L.}\ \bibnamefont {McEuen}},\ }\href {\doibase
		10.1126/science.1136836} {\bibfield  {journal} {\bibinfo  {journal}
			{Science}\ }\textbf {\bibinfo {volume} {315}},\ \bibinfo {pages} {490}
		(\bibinfo {year} {2007})}\BibitemShut {NoStop}%
	\bibitem [{\citenamefont {Singh}\ \emph {et~al.}(2014)\citenamefont {Singh},
		\citenamefont {Bosman}, \citenamefont {Schneider}, \citenamefont {Blanter},\
		and\ \citenamefont {Castellanos-Gomez}}]{ExpNumb4}%
	\BibitemOpen
	\bibfield  {author} {\bibinfo {author} {\bibfnamefont {V.}~\bibnamefont
			{Singh}}, \bibinfo {author} {\bibfnamefont {J.}~\bibnamefont {Bosman},
			\bibfnamefont {S.}}, \bibinfo {author} {\bibfnamefont {H.}~\bibnamefont
			{Schneider}, \bibfnamefont {B.}}, \bibinfo {author} {\bibfnamefont
			{M.}~\bibnamefont {Blanter}, \bibfnamefont {Y.}}, \ and\ \bibinfo {author}
		{\bibfnamefont {G.~A.}\ \bibnamefont {Castellanos-Gomez}, \bibfnamefont
			{A.~Steele}},\ }\href {\doibase 10.1038/nnano.2014.168} {\bibfield  {journal}
		{\bibinfo  {journal} {Nat Nano}\ }\textbf {\bibinfo {volume} {9}},\ \bibinfo
		{pages} {820} (\bibinfo {year} {2014})}\BibitemShut {NoStop}%
	\bibitem [{\citenamefont {Weber}\ \emph {et~al.}(2014)\citenamefont {Weber},
		\citenamefont {G\"{u}ttinger}, \citenamefont {Tsioutsios}, \citenamefont
		{Chang},\ and\ \citenamefont {Bachtold}}]{ExpNumb5}%
	\BibitemOpen
	\bibfield  {author} {\bibinfo {author} {\bibfnamefont {P.}~\bibnamefont
			{Weber}}, \bibinfo {author} {\bibfnamefont {J.}~\bibnamefont
			{G\"{u}ttinger}}, \bibinfo {author} {\bibfnamefont {I.}~\bibnamefont
			{Tsioutsios}}, \bibinfo {author} {\bibfnamefont {D.~E.}\ \bibnamefont
			{Chang}}, \ and\ \bibinfo {author} {\bibfnamefont {A.}~\bibnamefont
			{Bachtold}},\ }\href {\doibase 10.1021/nl500879k} {\bibfield  {journal}
		{\bibinfo  {journal} {Nano Letters}\ }\textbf {\bibinfo {volume} {14}},\
		\bibinfo {pages} {2854} (\bibinfo {year} {2014})},\ \bibinfo {note} {pMID:
		24745803},\ \Eprint
	{http://arxiv.org/abs/http://dx.doi.org/10.1021/nl500879k}
	{http://dx.doi.org/10.1021/nl500879k} \BibitemShut {NoStop}%
	\bibitem [{\citenamefont {Zeuthen}(2015)}]{EmilThesis}%
	\BibitemOpen
	\bibfield  {author} {\bibinfo {author} {\bibfnamefont {E.}~\bibnamefont
			{Zeuthen}},\ }\emph {\bibinfo {title} {Electro-Optomechanical Transduction \&
			Quantum Hard-Sphere Model for Dissipative Rydberg-EIT Media}},\ \href@noop {}
	{Ph.D. thesis},\ \bibinfo  {school} {The Niels Bohr Institute, Faculty of
		Science, University of Copenhagen} (\bibinfo {year} {2015})\BibitemShut
	{NoStop}%
	\bibitem [{\citenamefont {De~Alba}\ \emph {et~al.}(2016)\citenamefont
		{De~Alba}, \citenamefont {Massel}, \citenamefont {Storch}, \citenamefont
		{Abhilash}, \citenamefont {Hui}, \citenamefont {McEuen}, \citenamefont
		{Craighead},\ and\ \citenamefont {Parpia}}]{ExpNumb2}%
	\BibitemOpen
	\bibfield  {author} {\bibinfo {author} {\bibfnamefont {R.}~\bibnamefont
			{De~Alba}}, \bibinfo {author} {\bibfnamefont {F.}~\bibnamefont {Massel}},
		\bibinfo {author} {\bibfnamefont {R.}~\bibnamefont {Storch}, \bibfnamefont
			{I.}}, \bibinfo {author} {\bibfnamefont {S.}~\bibnamefont {Abhilash},
			\bibfnamefont {T.}}, \bibinfo {author} {\bibfnamefont {A.}~\bibnamefont
			{Hui}}, \bibinfo {author} {\bibfnamefont {L.}~\bibnamefont {McEuen},
			\bibfnamefont {P.}}, \bibinfo {author} {\bibfnamefont {G.}~\bibnamefont
			{Craighead}, \bibfnamefont {H.}}, \ and\ \bibinfo {author} {\bibfnamefont
			{M.}~\bibnamefont {Parpia}, \bibfnamefont {J.}},\ }\href {\doibase
		10.1038/nnano.2016.86} {\bibfield  {journal} {\bibinfo  {journal} {Nat Nano}\
		}\textbf {\bibinfo {volume} {11}},\ \bibinfo {pages} {741} (\bibinfo {year}
		{2016})}\BibitemShut {NoStop}%
	\bibitem [{\citenamefont {Mathew}\ \emph {et~al.}(2016)\citenamefont {Mathew},
		\citenamefont {Patel}, \citenamefont {Borah}, \citenamefont {VijayR.},\ and\
		\citenamefont {Deshmukh}}]{GrapheneTuning1}%
	\BibitemOpen
	\bibfield  {author} {\bibinfo {author} {\bibfnamefont {J.~P.}\ \bibnamefont
			{Mathew}}, \bibinfo {author} {\bibfnamefont {R.~N.}\ \bibnamefont {Patel}},
		\bibinfo {author} {\bibfnamefont {A.}~\bibnamefont {Borah}}, \bibinfo
		{author} {\bibnamefont {VijayR.}}, \ and\ \bibinfo {author} {\bibfnamefont
			{M.~M.}\ \bibnamefont {Deshmukh}},\ }\href {\doibase 10.1038/nnano.2016.94}
	{\bibfield  {journal} {\bibinfo  {journal} {Nat Nano}\ }\textbf {\bibinfo
			{volume} {11}},\ \bibinfo {pages} {747} (\bibinfo {year} {2016})}\BibitemShut
	{NoStop}%
	\bibitem [{\citenamefont {Chen}\ \emph {et~al.}(2009)\citenamefont {Chen},
		\citenamefont {Rosenblatt}, \citenamefont {Bolotin}, \citenamefont {Kalb},
		\citenamefont {Kim}, \citenamefont {Kymissis}, \citenamefont {Stormer},
		\citenamefont {Heinz},\ and\ \citenamefont {Hone}}]{GrapheneTuning2}%
	\BibitemOpen
	\bibfield  {author} {\bibinfo {author} {\bibfnamefont {C.}~\bibnamefont
			{Chen}}, \bibinfo {author} {\bibfnamefont {S.}~\bibnamefont {Rosenblatt}},
		\bibinfo {author} {\bibfnamefont {K.~I.}\ \bibnamefont {Bolotin}}, \bibinfo
		{author} {\bibfnamefont {W.}~\bibnamefont {Kalb}}, \bibinfo {author}
		{\bibfnamefont {P.}~\bibnamefont {Kim}}, \bibinfo {author} {\bibfnamefont
			{I.}~\bibnamefont {Kymissis}}, \bibinfo {author} {\bibfnamefont {H.~L.}\
			\bibnamefont {Stormer}}, \bibinfo {author} {\bibfnamefont {T.~F.}\
			\bibnamefont {Heinz}}, \ and\ \bibinfo {author} {\bibfnamefont
			{J.}~\bibnamefont {Hone}},\ }\href {\doibase 10.1038/nnano.2009.267}
	{\bibfield  {journal} {\bibinfo  {journal} {Nat Nano}\ }\textbf {\bibinfo
			{volume} {4}},\ \bibinfo {pages} {861} (\bibinfo {year} {2009})}\BibitemShut
	{NoStop}%
	\bibitem [{\citenamefont {Rocheleau}\ \emph {et~al.}(2010)\citenamefont
		{Rocheleau}, \citenamefont {Ndukum}, \citenamefont {Macklin}, \citenamefont
		{Hertzberg}, \citenamefont {Clerk},\ and\ \citenamefont
		{Schwab}}]{TemperatureRising}%
	\BibitemOpen
	\bibfield  {author} {\bibinfo {author} {\bibfnamefont {T.}~\bibnamefont
			{Rocheleau}}, \bibinfo {author} {\bibfnamefont {T.}~\bibnamefont {Ndukum}},
		\bibinfo {author} {\bibfnamefont {C.}~\bibnamefont {Macklin}}, \bibinfo
		{author} {\bibfnamefont {J.}~\bibnamefont {Hertzberg}}, \bibinfo {author}
		{\bibfnamefont {A.}~\bibnamefont {Clerk}}, \ and\ \bibinfo {author}
		{\bibfnamefont {K.}~\bibnamefont {Schwab}},\ }\href@noop {} {\bibfield
		{journal} {\bibinfo  {journal} {Nature}\ }\textbf {\bibinfo {volume} {463}},\
		\bibinfo {pages} {72} (\bibinfo {year} {2010})}\BibitemShut {NoStop}%
	\bibitem [{\citenamefont {Braunstein}\ and\ \citenamefont {van
			Loock}(2005)}]{Braunstein2005}%
	\BibitemOpen
	\bibfield  {author} {\bibinfo {author} {\bibfnamefont {S.~L.}\ \bibnamefont
			{Braunstein}}\ and\ \bibinfo {author} {\bibfnamefont {P.}~\bibnamefont {van
				Loock}},\ }\href {\doibase 10.1103/RevModPhys.77.513} {\bibfield  {journal}
		{\bibinfo  {journal} {Rev. Mod. Phys.}\ }\textbf {\bibinfo {volume} {77}},\
		\bibinfo {pages} {513} (\bibinfo {year} {2005})}\BibitemShut {NoStop}%
	\bibitem [{\citenamefont {He}\ \emph {et~al.}(2017)\citenamefont {He},
		\citenamefont {Yang}, \citenamefont {Lin},\ and\ \citenamefont
		{Xiao}}]{Linearization}%
	\BibitemOpen
	\bibfield  {author} {\bibinfo {author} {\bibfnamefont {B.}~\bibnamefont
			{He}}, \bibinfo {author} {\bibfnamefont {L.}~\bibnamefont {Yang}}, \bibinfo
		{author} {\bibfnamefont {Q.}~\bibnamefont {Lin}}, \ and\ \bibinfo {author}
		{\bibfnamefont {M.}~\bibnamefont {Xiao}},\ }\href {\doibase
		10.1103/PhysRevLett.118.233604} {\bibfield  {journal} {\bibinfo  {journal}
			{Phys. Rev. Lett.}\ }\textbf {\bibinfo {volume} {118}},\ \bibinfo {pages}
		{233604} (\bibinfo {year} {2017})}\BibitemShut {NoStop}%
	\bibitem [{\citenamefont {Dalibard}\ \emph {et~al.}(1992)\citenamefont
		{Dalibard}, \citenamefont {Castin},\ and\ \citenamefont
		{M{\o}lmer}}]{Dalibard1992}%
	\BibitemOpen
	\bibfield  {author} {\bibinfo {author} {\bibfnamefont {J.}~\bibnamefont
			{Dalibard}}, \bibinfo {author} {\bibfnamefont {Y.}~\bibnamefont {Castin}}, \
		and\ \bibinfo {author} {\bibfnamefont {K.}~\bibnamefont {M{\o}lmer}},\
	}\href@noop {} {\bibfield  {journal} {\bibinfo  {journal} {Physical review
				letters}\ }\textbf {\bibinfo {volume} {68}},\ \bibinfo {pages} {580}
		(\bibinfo {year} {1992})}\BibitemShut {NoStop}%
\end{thebibliography}
\end{document}

\renewcommand{\theequation}{S\arabic{equation}}
	\renewcommand{\thesection}{S\arabic{section}}
	\renewcommand{\thefigure}{S\arabic{figure}}
	
\title{Supplementary material: Quantum nondemolition measurement of mechanical motion quanta}% Force line breaks with \\
\author{Luca Dellantonio}
\affiliation{The Niels Bohr Institute, University of Copenhagen, Blegdamsvej 17, DK-2100 Copenhagen \O, Denmark}
\affiliation{Center for Hybrid Quantum Networks (Hy-Q), Niels Bohr Institute, University of Copenhagen, Blegdamsvej 17, DK-2100 Copenhagen \O, Denmark}
\author{Oleksandr Kyriienko}%
\affiliation{The Niels Bohr Institute, University of Copenhagen, Blegdamsvej 17, DK-2100 Copenhagen \O, Denmark}
\affiliation{NORDITA, KTH Royal Institute of Technology and Stockholm University, Roslagstullsbacken 23, SE-106 91 Stockholm, Sweden}
\author{Florian Marquardt}
\affiliation{Institute for Theoretical Physics, University Erlangen-N\"{u}rnberg, Staudstra\ss e 7, 91058 Erlangen, Germany}
\affiliation{Max Planck Institute for the Science of Light, G\"{u}nther-Scharowsky-Stra\ss e 1, 91058 Erlangen, Germany}
\author{Anders S. S\o rensen}
\affiliation{The Niels Bohr Institute, University of Copenhagen, Blegdamsvej 17, DK-2100 Copenhagen \O, Denmark}
\affiliation{Center for Hybrid Quantum Networks (Hy-Q), Niels Bohr Institute, University of Copenhagen, Blegdamsvej 17, DK-2100 Copenhagen \O, Denmark}

\date{\today}% It is always \today, today,

\maketitle

Here we provide supplemental material for the article ``Quantum nondemolition measurement of mechanical motion quanta'', which is structured as follows. In the first section (\ref{sec:RLCcircuit}) we carefully derive the results presented in the main text for the $RLC$ circuit. In subsection \ref{sec:QNDsimple} we describe how the QND interaction allows for reading out the mechanical state, while in subsection \ref{sec:HeatingSimple} we study the dynamics induced by the linear coupling. The second section (\ref{sec:CompleteCirc}) generalizes the results by deriving the figure of merit $\lambda$ for the circuit in Fig. 1(c) of the main text. Considering the symmetric case, analytical results are derived in subsections \ref{sec:QNDcomplete} and \ref{sec:HeatComplBalanced}, while the most general case of the antisymmetric system is treated in subsection \ref{sec:HeatingSim}. In the third section (\ref{sec:Measure}) we look at a possible measurement scheme, both analytically and numerically. Finally, in section \ref{sec:ExptEst} we present an investigation of the required experimental parameters for a concrete realization of our setup. Throughout the supplemental material we present several numerical simulations to support our analytical results (subsections \ref{sec:HeatingSim}, \ref{sec:HeatComplBalanced} and section \ref{sec:Measure}).

\section{RLC circuit}\label{sec:RLCcircuit}

\begin{figure}[htbp]
	\centering
	\includegraphics[width=8 cm]{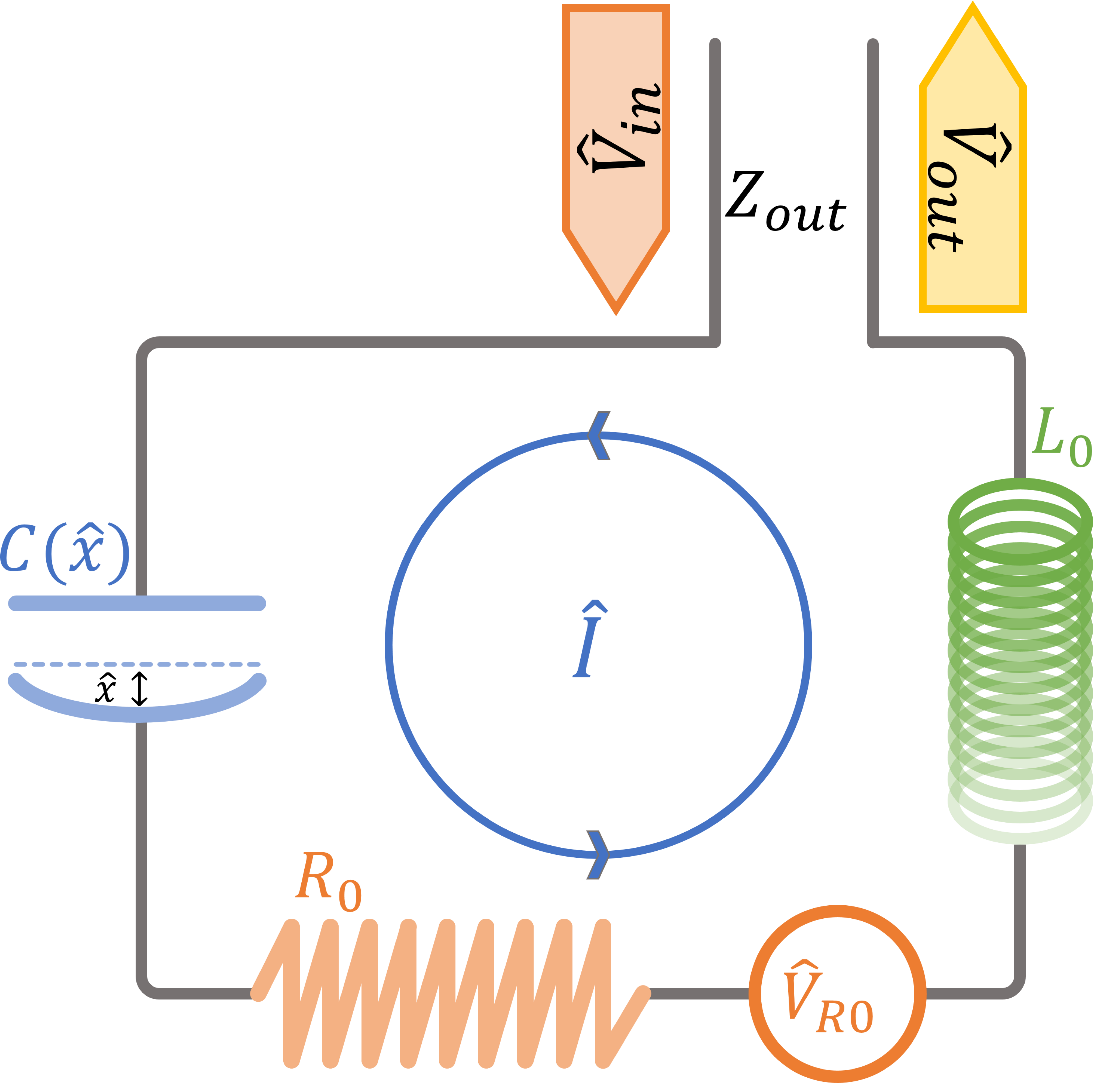}
	\caption[FigS1]{Circuit diagram for an RLC circuit coupled capacitively to a mechanical oscillator, with the position operator denoted as $\hat{x}$. $R_{0}$ and $L_{0}$ are the inductance and resistance of the electrical circuit, while $C(\hat{x})$ is the position--dependent capacitance. The setup is driven by the microwave (MW) field $\hat{V}_{in}$ through the semi--infinite transmission line of impedance $Z_{out}$. $\hat{V}_{out}$ is the reflected signal, $\hat{V}_{R0}$ the Johnson--Nyquist noise associated with $R_{0}$, and $\hat{I}$ is the current flowing in the circuit.}
	\label{Fig:FigS1}
\end{figure}

In the following we consider an $RLC$ circuit where the capacitor contains an oscillating element. Following the standard procedure for quantizing an electrical circuit \cite{devoret1995quantum}, we can write the Hamiltonian of the setup presented in Fig. \ref{Fig:FigS1} as
\begin{equation}
\hat{\mathcal{H}} = \hbar \omega_{m} \hat{b}^{\dagger}\hat{b} +\frac{\hat{\Phi}^{2}}{2L_{0}}+\frac{\hat{Q}^{2}}{2C_{0}}+\frac{g_{1} \omega_{s}L_{0}}{2}\hat{Q}^{2}(\hat{b}+\hat{b}^{\dagger})+\frac{g_{2} \omega_{s}L_{0}}{2}\hat{Q}^{2}\left(\hat{b}^{\dagger}\hat{b}+\frac{\hat{b}\hat{b}+\hat{b}^{\dagger}\hat{b}^{\dagger}}{2}\right) - 2 \hat{Q} (\hat{V}_{in}+\hat{V}_{R0})- x_{0}\hat{F}_{b}(\hat{b}+\hat{b}^{\dagger}), \label{Eq:HamFinalSimple}
\end{equation}
where the conjugate position $\hat{Q}$ and momentum $\hat{\Phi}=L_{0}\hat{I}$ are the electrical charge and flux, respectively. $x_{0}=\sqrt{\hbar/(2 m \omega_{m})}$ is the zero--point motion amplitude for a membrane of mass $m$, and $\hat{b}$ ($\hat{b}^{\dagger}$) denotes the mechanical annihilation (creation) operator. $\omega_{m}$ and $\omega_{s}=(C_{0} L_{0})^{-\frac{1}{2}}$ are the mechanical and electrical resonance frequencies. $C_{0}$, $g_{1}$ and $g_{2}$ are derived from the expansion of the inverse capacitance,
\begin{equation}
C^{-1}(\hat{b}+\hat{b}^{\dagger}) \simeq C^{-1}_{0} + g_{1} L_{0} \omega_{s} ( \hat{b}+\hat{b}^{\dagger} ) + g_{2} L_{0} \omega_{s} \left(\hat{b}+\hat{b}^{\dagger}\right)^{2},
\end{equation}
where the coupling required for the QND interaction comes from the rotating wave approximation: $(\hat{b}+\hat{b}^{\dagger})^{2}\simeq 2\hat{n}_{b} + 1$. Finally, $\hat{V}_{in}$ is the input field, $\hat{V}_{R0}$ is the Johnson--Nyquist noise associated with the resistor $R_{0}$, and $\hat{F}_{b}$ is the random force related to the mechanical reservoir that slowly thermalizes the membrane.
%%%
%%%
From Eq. \eqref{Eq:HamFinalSimple} we can derive the Heisenberg equations of motion for the electromechanical system operators $\hat{Q}$, $\hat{\Phi}$ and $\hat{b}$. Adding decays and noises to the equations of motion we have
\begin{subequations}\label{Eq:EOMsimple}
	\begin{align}
	\dot{\hat{Q}} = & \frac{\hat{\Phi}}{L_{0}}, \label{Eq:EOMqS} \\
	\dot{\hat{\Phi}} = & -\frac{\hat{Q}}{C_{0}}-g_{1} \omega_{s}L_{0} \hat{Q}\left(\hat{b}+\hat{b}^{\dagger}\right)-g_{2}\omega_{s}L_{0}\hat{Q}\left(\hat{n}_{b}+\frac{\hat{b}\hat{b}+\hat{b}^{\dagger}\hat{b}^{\dagger}}{2}\right) -(\gamma_{t}+\gamma_{r})\hat{\Phi}+2\left(\hat{V}_{in}+\hat{V}_{R0}\right), \label{Eq:EOMphiS}\\ 
	\dot{\hat{b}} = & -i \omega_{m}\hat{b}-g_{1}\frac{i\omega_{s}L_{0}\hat{Q}^{2}}{2\hbar}-g_{2}\frac{i\omega_{s}L_{0}\hat{Q}^{2}}{2\hbar}\left(\hat{b}+\hat{b}^{\dagger} \right) - \frac{\gamma_{b}}{2}\hat{b}+i \frac{x_{0}}{\hbar}\hat{F}_{b}. \label{Eq:EOMbS}
	\end{align}
\end{subequations}
Importantly, these expressions are just the familiar Kirchoff's laws, which provide the form of the electrical decay rates $\gamma_{r}=R_{0}/L_{0}$ and $\gamma_{t}=Z_{out}/L_{0}$. The mechanical decay rate $\gamma_{b}$ is an intrinsic property of the membrane, and we have assumed that the mechanical reservoir can be treated using the Markov approximation. Finally, the reflected signal is determined by the input/output relations
\begin{equation}
\hat{V}_{out} = \hat{V}_{in}-\gamma_{t} \hat{\Phi}, \label{Eq:In/Out}
\end{equation}
that read the same in the time and frequency domains.

\subsection{QND measurement of the phonon number}\label{sec:QNDsimple}

As explained in the main text, there are three mechanisms with which the mechanical system influences the electrical one. These are the two terms $g_{1}(\hat{b}+\hat{b}^{\dagger} )$ and $g_{2}(\hat{b}\hat{b}+\hat{b}^{\dagger}\hat{b}^{\dagger} )$, which generate sidebands at frequencies $\omega_{s}\pm \omega_{m}$ and $\omega_{s}\pm 2\omega_{m}$, and a phonon--dependent frequency shift proportional to $g_{2}\hat{n}_{b}$ [see Eq. \eqref{Eq:EOMphiS}]. When we perform homodyne measurement at the resonant frequency of the electrical circuit $\omega_{s}$, the sidebands contribution to the measurement outcome averages out, and we are left with the phonon--dependent frequency shift. The main detrimental effect of $g_{1}$ will be to cause heating. We consider this in subsection \ref{sec:HeatingSimple}, and in the following discussion of the readout we thus assume that the phonon number is conserved, $\hat{n}_{b}(t)\rightarrow\hat{n}_{b}=\text{constant}$. Under these assumptions, the electrical readout is independent of the mechanical dynamics, and can be described by
\begin{subequations}\label{Eq:qndEOMsimple}
	\begin{align}
	\dot{\hat{Q}} = & \frac{\hat{\Phi}}{L_{0}}, \label{Eq:qndEOMqS} \\
	\dot{\hat{\Phi}} = & -\frac{\hat{Q}}{C_{0}}-g_{2}\omega_{s}L_{0}\hat{Q}\hat{n}_{b} -(\gamma_{t}+\gamma_{r})\hat{\Phi}+2\left(\hat{V}_{in}+\hat{V}_{R0}\right). \label{Eq:qndEOMphiS}
	\end{align}
\end{subequations}

As it is instructive to look at these equations in the frequency domain, we introduce the Fourier series of any operator $\hat{O}(t)$ by
\begin{equation}
\hat{O}(t)=\sum_{k=-\infty}^{\infty}\hat{O}[\Omega_{k}]\frac{e^{-i\Omega_{k}t}}{\sqrt{\tau}},\label{Eq:FourierDef}
\end{equation}
with $\Omega_{k} = 2\pi k/\tau$ ($k\in \mathbb{Z}$) being the allowed frequencies, and $\tau$ is the time period used to define the Fourier series. For now we will let $\tau$ be equal to the measurement time $T$, but later we shall consider a larger value to describe heating on longer time scales. The Fourier coefficients $\hat{O}[\Omega_{k}]$ are then defined by
\begin{equation}
\hat{O}[\Omega_{k}]=\int_{0}^{\tau}\hat{O}(t)\frac{e^{i \Omega_{k} t}}{\sqrt{\tau}}dt. \label{Eq:FourierCoeffDefSimple}
\end{equation}
It is possible to rewrite equations \eqref{Eq:qndEOMsimple} in the frequency domain:
\begin{subequations}\label{Eq:EOMSimpleQuad}
	\begin{align}
	-i\Omega_{k}\hat{\Phi}[\Omega_{k}] & = -\frac{\hat{Q}[\Omega_{k}]}{C_{0}} - g_{2}\omega_{s}L_{0}\hat{Q}[\Omega_{k}] \hat{n}_{b}-(\gamma_{r}+\gamma_{t})\hat{\Phi}[\Omega_{k}]+2 (\hat{V}_{in}[\Omega_{k}]+\hat{V}_{R0}[\Omega_{k}]), \label{Eq:EOMPhiSimpleQuad} \\
	-i\Omega_{k}\hat{Q}[\Omega_{k}] & = \frac{\hat{\Phi}[\Omega_{k}]}{L_{0}}. \label{Eq:EOMQSimpleQuad}
	\end{align}
\end{subequations}
From these two relations we derive $\hat{\Phi}[\Omega_{k}]$, which can then be used in Eq. \eqref{Eq:In/Out} for determining $\hat{V}_{out}[\Omega_{k}]$:
\begin{subequations}\label{Eq:VoutSimple}
	\begin{align}
	\hat{V}_{out}[\Omega_{k}] & = \left[1-\zeta(\Omega_{k},\hat{n}_{b})\right]\hat{V}_{in}[\Omega_{k}] - \zeta(\Omega_{k},\hat{n}_{b})\hat{V}_{R0}[\Omega_{k}], \label{Eq:VoutSimpleForm} \\
	\zeta(\Omega_{k},\hat{n}_{b}) & = \frac{2 \gamma_{t} \Omega_{k} }{-i(\Omega_{k}^{2}-\omega_{s}^{2}-g_{2}\omega_{s}\hat{n}_{b})+\Omega_{k} (\gamma_{r}+\gamma_{t})}. \label{Eq:VoutSimpleCoeff}
	\end{align}
\end{subequations}
These expressions describe the principle of the QND measurement: the quadratic interaction shifts the resonance frequency of the circuit by an amount proportional to $g_{2}n_{b}$ (denominator of Eq. \eqref{Eq:VoutSimpleCoeff}). By sending a signal resonant with the electrical circuit, it is possible to detect this frequency change as a phase shift of the reflected signal.

From the quantized form of $\hat{V}_{in}[\Omega_{k}]$ and $\hat{V}_{R0}[\Omega_{k}]$ \cite{ClerkRev}, we have the Fourier components
\begin{subequations}
	\begin{align}
	\hat{V}_{in}[\Omega_{k}] & = \sqrt{\frac{\hbar \Omega_{k} Z_{out}}{2}} \hat{a}_{in,k}, \label{Eq:VinSimple} \\
	\hat{V}_{R0}[\Omega_{k}] & = \sqrt{\frac{\hbar \Omega_{k} R_{0}}{2}}\hat{a}_{R0,k}, \label{Eq:VROSimlpe}
	\end{align}
\end{subequations}
where $\hat{a}_{in,k}$ and $\hat{a}_{R0,k}$ are the annihilation operators for the input field and the electrical reservoir, and satisfy the standard commutation relations. The measurement outcome $\hat{V}_{M}$ is obtained by homodyne detection, meaning that we have access to
\begin{equation}
\hat{V}_{M}[\Omega_{k}]=\frac{e^{i\frac{\theta}{2}}\hat{V}_{out}[\Omega_{k}]+e^{-i\frac{\theta}{2}}\hat{V}_{out}^{\dagger}[\Omega_{k}]}{2}, \label{Eq:VmSimple}
\end{equation}
where $\theta$ is an arbitrary phase that allows choosing the quadrature of the reflected signal to be measured. 

\begin{figure}[htbp]
	\centering
	\includegraphics[width=17 cm]{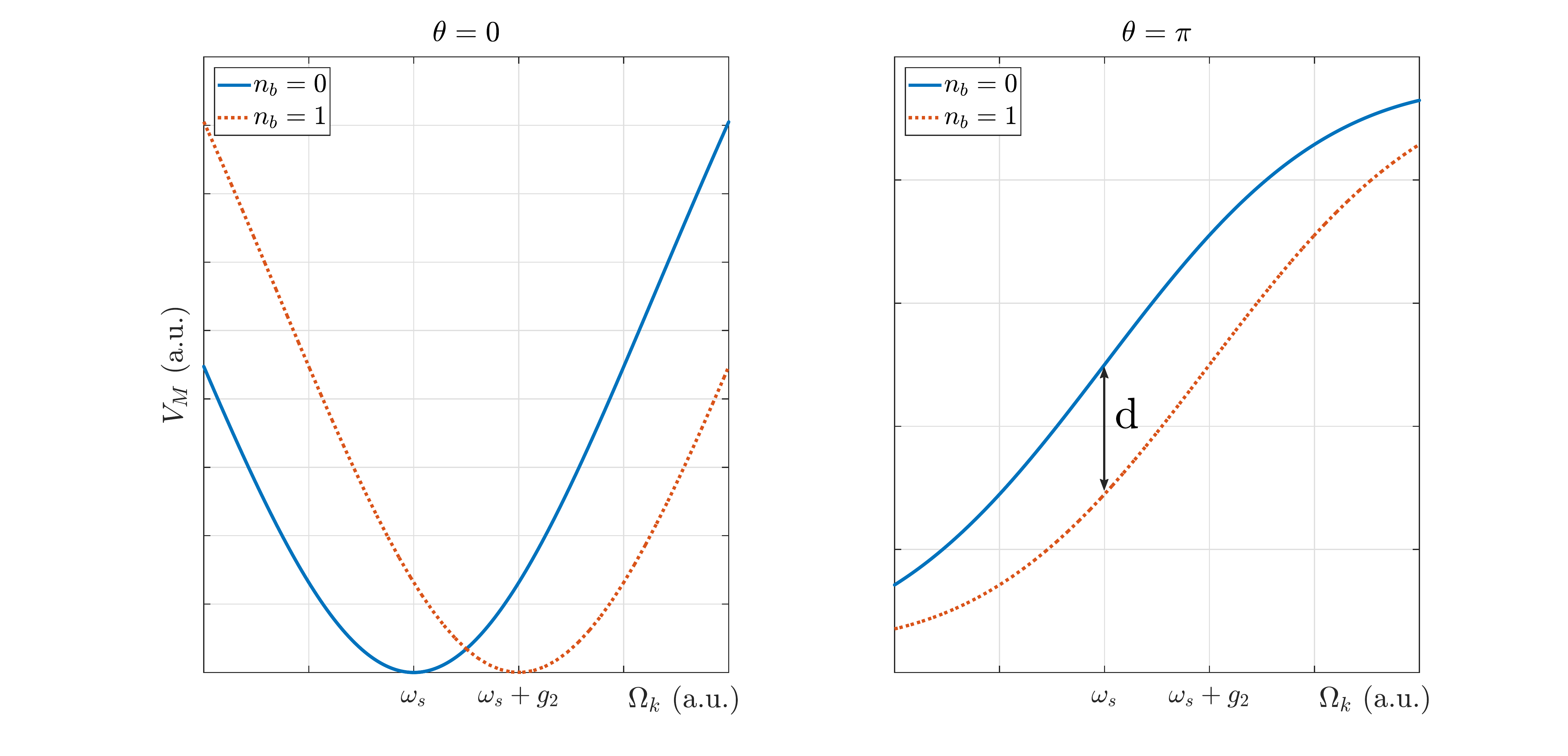}
	\caption[FigS2]{Example of amplitude \textbf{(left)} and phase \textbf{(right)} quadratures of the measured operator $\hat{V}_{M}$. The plain and dotted curves refer, respectively, to the cases in which the membrane is in its ground ($n_b=0$) or in its first excited ($n_b=1$) state. As one can see, the quadratic coupling $g_{2}$ induces a frequency shift, proportional to the phonon number. The quantity $d$, derived analytically in Eq. \eqref{Eq:dSimple}, is indicated. We assumed optimally loaded setups with $\gamma_{r}=\gamma_{t}$. Compared to the realistic scenario $g_{2}\ll\gamma_{t}$, the frequency shift is here exaggerated. Whenever $g_{2}\ll\gamma_{t}$, the frequency shift is more easily detected by looking at the phase quadrature $\theta = \pi$ of the reflected signal.}
	\label{Fig:FigS2}
\end{figure}

Having now all the operator equations describing the system in the Heisenberg formalism, we need to define the input state $\hat{\rho}_{in}$. Assuming that the semi--infinite transmission line of impedance $Z_{out}$ and the resistor $R_{0}$ are connected to reservoirs at the same temperature $T_{e}$, and that we drive the circuit with a coherent field at the resonant frequency $\omega_{s}$, we have:
\begin{equation}
\hat{\rho}_{in}(\tilde{\alpha},\beta_{R}) = \frac{1}{\mathbf{Z}}\hat{D}(\tilde{\alpha})e^{-\beta_{R}\hat{\mathcal{H}}_{R}}\hat{D}^{\dagger}(\tilde{\alpha}). \label{Eq:DensityMatrixSimple}
\end{equation}
Here, $\mathbf{Z}=\text{Tr}\left\lbrace e^{-\beta_{R}\hat{\mathcal{H}}_{R}} \right\rbrace$ is the partition function, $\hat{D}(\tilde{\alpha})=\exp\left(\int_{0}^{T} \tilde{\alpha}(t)\hat{a}^{\dagger}_{in,s}(t) - \tilde{\alpha}^{*}(t)\hat{a}_{in,s} (t) dt \right)$ is the displacement operator, $\lvert \tilde{\alpha}\rvert^{2}$ is the photon flux, $\hat{\mathcal{H}}_{R} = \sum\limits_{k}\big\lbrace \hbar \Omega_{k}\hat{a}^{\dagger}_{in,k}\hat{a}_{in,k}+ \hbar \Omega_{k}\hat{a}^{\dagger}_{R0,k}\hat{a}_{R0,k} \big\rbrace$ is the reservoir Hamiltonian, and $\beta_{R}^{-1}=k_{b} T_{e}$ with $k_{b}$ being the Boltzmann constant. Using $\hat{\rho}_{in}(\tilde{\alpha},\beta_{R})$ it is possible to calculate the average measured signal $V_{M}=\langle \hat{V}_{M} \rangle$ and the variance $(\Delta \hat{V}_{M} )^{2}=\langle \hat{V}_{M}^{2}  \rangle-\langle \hat{V}_{M} \rangle^{2}$ for an incident field at the resonance frequency $\omega_{s}$:
\begin{subequations}
	\begin{align}
	V_{M} (n_{b}) & = -\alpha\sqrt{\frac{\hbar \omega_{s} Z_{out}}{2}}\text{Im}\lbrace \zeta(\omega_{s},n_{b})\rbrace, \label{Eq:AvgVMSimple} \\
	(\Delta \hat{V}_{M} )^{2} & = \frac{\hbar \omega_{s} Z_{out}}{2}\left[ 1+2 \bar{n}_{e}(\omega_{s},T_{e}) \right], \label{Eq:VarVMSimple}
	\end{align}
\end{subequations}
where $\lvert\alpha\rvert^{2}=\int_{0}^{T}\lvert\tilde{\alpha}(t)\rvert^{2}dt$ is the number of photons sent into the circuit during the measurement. The imaginary part is denoted with $\text{Im}\{ \cdot \}$, and $\bar{n}_{e}(\omega_{s},T_{e})=\left[ \exp\left( \hbar \beta_{R} \omega_{s} \right)-1 \right]^{-1} $ is the average number of photons at frequency $\omega_{s}$ in the electrical reservoir.
Note that we have taken $\theta=\pi$ in Eq. \eqref{Eq:VmSimple}, and $\tilde{\alpha}(t)=\alpha/\sqrt{T}$ to be purely real. With this choice, we measure the phase quadrature, which optimizes the signal for small $g_{2}$ (see Fig. \ref{Fig:FigS2}).
%%%
%%%

It is now possible to derive the parameter $D^{2}=d^{2}/\sigma^{2}$ introduced in the main text. We defined $d$ to be the difference between two outcomes with one and zero phonons respectively,
\begin{equation}
d=V_{M} (n_{b}=1)-V_{M} (n_{b}=0)=\alpha\sqrt{\frac{\hbar \omega_{s} Z_{out}}{2}}\text{Im}\lbrace\zeta(\omega_{s},n_{b}=0)-\zeta(\omega_{s},n_{b}=1)\rbrace. \label{Eq:dSimple}
\end{equation}
Considering that $\sigma^{2}$ is the variance $(\Delta \hat{V}_{M} )^{2}$, we find
\begin{equation}
D^{2}=\frac{d^{2}}{\sigma^{2}}=\frac{4}{1+2 \bar{n}_{e}}\frac{g_{2}^{2}\lvert\alpha\rvert^{2}\gamma_{t}^{2}}{\left[ g_{2}^{2} + \left( \gamma_{t}+\gamma_{r} \right)^{2} \right]^{2}}. \label{Eq:ParDsqSimple}
\end{equation}
This last equation reduces to the form given in the main text by setting $\gamma_{r}=\gamma_{t}$ and noticing that typically $g_{2} \ll \gamma_{r}+\gamma_{t}$.

\subsection{Membrane heating}\label{sec:HeatingSimple}

In this subsection we study the time evolution of the average phonon number $n_{b}(t)=\langle \hat{n}_{b}(t)\rangle$, and determine the parameter $\Delta n_{b}$ that characterizes the probability for the mechanical state to jump during the measurement time $T$. It is convenient to introduce creation $\hat{a}^{\dagger}$ and annihilation $\hat{a}$ operators for the electrical charge $\hat{Q}$ and flux $\hat{\Phi}$ operators,
\begin{subequations}
	\begin{align}
	\hat{Q} & = \sqrt{\frac{\hbar C_{0} \omega_{s}}{2}}(\hat{a}+\hat{a}^{\dagger}), \label{Eq:QaSimple}\\
	\hat{\Phi} & = i\sqrt{\frac{\hbar}{2 C_{0}\omega_{s} }}(\hat{a}^{\dagger}-\hat{a}).\label{Eq:PhiaSimple}
	\end{align}
\end{subequations}
Using these we can rewrite the Hamiltonian in Eq. \eqref{Eq:HamFinalSimple} as
\begin{equation}
\hat{\mathcal{H}} = \hbar \omega_{s} \hat{a}^{\dagger}\hat{a} + \hbar \omega_{m} \hat{b}^{\dagger}\hat{b} + \frac{\hbar g_{1} }{2}\hat{a}^{\dagger}\hat{a} (\hat{b}+\hat{b}^{\dagger})+\frac{\hbar g_{2} }{2}\hat{a}^{\dagger}\hat{a}\left( \hat{b}^{\dagger}\hat{b} + \frac{\hat{b}\hat{b}+\hat{b}^{\dagger}\hat{b}^{\dagger}}{2} \right) -\hbar \sqrt{\frac{2 C_{0} \omega_{s}}{\hbar}} (\hat{a}+\hat{a}^{\dagger}) (\hat{V}_{in}+\hat{V}_{R0})- x_{0}\hat{F}_{b}(\hat{b}+\hat{b}^{\dagger}), \label{Eq:HamLinearSimple}
\end{equation}
where we have used the rotating wave approximation to neglect terms which are off--resonant with the electrical frequency $\omega_{s}$. Given that the coupling coefficients $g_{1}$ and $g_{2}$ are small compared to all other parameters, we linearize $\hat{a}$ and $\hat{a}^{\dagger}$ such that only deviations from their steady states are considered: $\hat{a}^{(\dagger)}=\langle \hat{a}^{(\dagger)}\rangle + \hat{\delta a}^{(\dagger)}$, where
\begin{equation}
\left\langle \hat{a}(t) \right\rangle \xrightarrow{\text{\tiny steady}} \frac{2 i \alpha \sqrt{\gamma_{t}} e^{-i \omega_{s} t }}{\sqrt{T} (\gamma_{t}+\gamma_{r})}. \label{Eq:AvgASimple}
\end{equation}
If the experiment is performed in a pulsed fashion with incoming pulses varying on a timescale comparable to the circuit's lifetime $(\gamma_{r}+\gamma_{t})^{-1}$, Eq. \eqref{Eq:AvgASimple} should be replaced by a suitable expression that takes into account the transient dynamics \cite{Linearization}. Here we restrict ourselves to the cases in which either fields are applied continuously, or pulses are slowly varying on the timescale identified by $(\gamma_{r}+\gamma_{t})^{-1}$. In the latter case, the simple replacement $\alpha\rightarrow \alpha(t)$ in Eq. \eqref{Eq:AvgASimple} is sufficient.

Considering that $g_{2} \ll g_{1}$, we can neglect the quadratic coupling in the Hamiltonian Eq. \eqref{Eq:HamLinearSimple} for calculating the heating. The Hamiltonian can then be rewritten in the form
\begin{equation}
\hat{\mathcal{H}} = \hbar \omega_{m} \hat{b}^{\dagger}\hat{b} + i \frac{\hbar g_{1} \alpha }{\gamma_{t}+\gamma_{r}}\sqrt{\frac{\gamma_{t}}{T}}(\hat{\delta a}-\hat{\delta a}^{\dagger}) (\hat{b}+\hat{b}^{\dagger}) - x_{0}\hat{F}_{b}(\hat{b}+\hat{b}^{\dagger}) -\hbar \sqrt{\frac{2 C_{0} \omega_{s}}{\hbar}} \left[ \hat{\delta a}^{\dagger}\left( \hat{\delta V}_{in}+\hat{V}_{R0} \right) + \hat{\delta a} \left( \hat{\delta V}_{in}^{\dagger}+\hat{V}_{R0}^{\dagger} \right) \right], \label{Eq:HamLinFinalSimple}
\end{equation}
where we have switched to the rotating frame using the unitary transformation $\hat{U}=e^{it\omega_{s}\hat{\delta a}^{\dagger}\hat{\delta a}}$. Notice that we linearized the operator $\hat{V}_{in}=\langle \hat{V}_{in} \rangle + \hat{\delta V}_{in}$, with its average given by the coherent field $\alpha$ at the frequency $\omega_{s}$, and neglected a term $\propto \lvert \alpha \rvert^{2}(\hat{b} + \hat{b}^{\dagger})$, that can be removed by changing the rest position of the mechanical oscillator. 

The equations of motion for the operators $\hat{b}$ and $\hat{\delta a}$ are then given by
\begin{subequations}
	\begin{align}
	\dot{\hat{\delta a}} & =  \frac{g_{1}\alpha}{\gamma_{t}+\gamma_{r}}\sqrt{\frac{\gamma_{t}}{T}}(\hat{b}+\hat{b}^{\dagger}) - \frac{\gamma_{t}+\gamma_{r}}{2}\hat{\delta a}+i \sqrt{\frac{2 C_{0} \omega_{s}}{\hbar}} (\hat{\delta V}_{in}+\hat{V}_{R_{0}}), \label{Eq:EOMASimple}\\
	\dot{\hat{b}} & =  -i \omega_{m} \hat{b} - \frac{g_{1}\alpha}{\gamma_{t}+\gamma_{r}}\sqrt{\frac{\gamma_{t}}{T}}(\hat{\delta a} - \hat{\delta a}^{\dagger})  - \frac{\gamma_{b}}{2}\hat{b}+i\frac{x_{0}}{\hbar}\hat{F}_{b}, \label{Eq:EOMbSimple}
	\end{align}
\end{subequations}
where we have linearised in the operators $\hat{\delta a}$ and $\hat{\delta a}^{\dagger}$.
A formal solution in the time domain for these two differential operator equations then reads
\begin{subequations}
	\begin{align}
	\begin{split}
	\hat{\delta a}(t) = & \hat{\delta a}(0)e^{-\frac{\gamma_{t}+\gamma_{r}}{2}t}+  \frac{g_{1}\alpha}{\gamma_{t}+\gamma_{r}}\sqrt{\frac{\gamma_{t}}{T}}\int_{0}^{t}e^{-\frac{\gamma_{t}+\gamma_{r}}{2}(t-\tau)} \left[ \hat{b}(\tau)+\hat{b}^{\dagger}(\tau)\right] d\tau  \\ &+ i\sqrt{\frac{2 C_{0} \omega_{s}}{\hbar}} \int_{0}^{t}e^{-\frac{\gamma_{t}+\gamma_{r}}{2}(t-\tau)} \left[\hat{\delta V}_{in}(\tau)+\hat{V}_{R_{0}}(\tau)\right] d\tau, \label{Eq:SolASimple} 
	\end{split}\\
	\begin{split}
	\hat{b}(t)  = & \hat{b}(0)e^{-\left( i \omega_{m} + \frac{\gamma_{b}}{2} \right)t} - \frac{g_{1}\alpha}{\gamma_{t}+\gamma_{r}}\sqrt{\frac{\gamma_{t}}{T}}\int_{0}^{t}e^{-\left( i \omega_{m} + \frac{\gamma_{b}}{2} \right)(t-\tau)} \left[ \hat{\delta a}(\tau) - \hat{\delta a}^{\dagger}(\tau)\right] d\tau \\& + i\frac{x_{0}}{\hbar}\int_{0}^{t}e^{-\left( i \omega_{m} + \frac{\gamma_{b}}{2} \right)(t-\tau)}\hat{F}_{b}(\tau)d\tau.
	\end{split} \label{Eq:SolbSimple}
	\end{align}
\end{subequations}
Since we are interested in the average phonon number $\langle \hat{b}^{\dagger}\hat{b} (t) \rangle$, we can substitute Eq. \eqref{Eq:SolASimple} into Eq. \eqref{Eq:SolbSimple}, to derive the time evolution of the operator $\hat{b}$ as a function of the noises $\hat{\delta V}_{in}$, $\hat{V}_{R0}$ and $\hat{F}_{b}$:
\begin{equation}
\begin{split}
\hat{b}(t) & = \hat{b}(0)e^{-\left( i \omega_{m} + \frac{\gamma_{b}}{2} \right)t}+ i\frac{x_{0}}{\hbar}\int_{0}^{t}e^{-\left( i \omega_{m} + \frac{\gamma_{b}}{2} \right)(t-\tau)}\hat{F}_{b}(\tau)d\tau \\ & - \frac{2 g_{1} \alpha\left[ \hat{\delta a}(0) - \hat{\delta a}^{\dagger}(0) \right]}{(\gamma_{t}+\gamma_{r})(\gamma_{t}+\gamma_{r}-\gamma_{b}-2 i \omega_{m})}\sqrt{\frac{\gamma_{t}}{T}}\left( e^{-\frac{\gamma_{t}+\gamma_{r}}{2}t} - e^{-\left( i \omega_{m}+\frac{\gamma_{b}}{2} \right)t} \right) \\ & + \frac{i g_{1}\alpha}{\gamma_{t}+\gamma_{r}}\sqrt{\frac{\gamma_{t}}{T}} \sqrt{\frac{2 C_{0} \omega_{s}}{\hbar}}\int_{0}^{t}\int_{0}^{\tau_{1}} \Bigg\lbrace e^{-\left( i \omega_{m}+ \frac{\gamma_{b}}{2} \right)(t - \tau_{1})} e^{-\left( \frac{\gamma_{t}+\gamma_{r}}{2} \right)(\tau_{1}-\tau_{2})} \\ & \times \left[ \hat{\delta V}_{in}(\tau_{2})+\hat{\delta V}_{in}^{\dagger}(\tau_{2}) + \hat{V}_{R0}(\tau_{2})+\hat{V}_{R0}^{\dagger}(\tau_{2})  \right]\Bigg\rbrace d\tau_{1}d\tau_{2}. \label{Eq:SolbLinFinSimple}
\end{split}
\end{equation}
Importantly, because we choose to probe the system at the electrical resonance frequency $\omega_{s}$, the oscillating terms proportional to $\hat{b}(\tau)$ and $\hat{b}^{\dagger}(\tau)$ in Eq. \eqref{Eq:SolASimple} cancel each other, such that the result simplifies to Eq. \eqref{Eq:SolbLinFinSimple}. The above expression is an integral operator equation, the solution of which fully describes the mechanical annihilation operator $\hat{b}(t)$. An analogous relation can be obtained for the creation operator $\hat{b}^{\dagger}(t)$, by taking the adjoint of Eq. \eqref{Eq:SolbLinFinSimple}. In general, this integral is difficult to evaluate. However, assuming the Markov approximation, we can compute the second moments of all the noise operators involved. Thus, even if we cannot solve Eq. \eqref{Eq:SolbLinFinSimple} for $\hat{b}(t)$ or $\hat{b}^{\dagger}(t)$, we can use \eqref{Eq:SolbLinFinSimple} for determining $n_{b}(t) = \langle \hat{b}^{\dagger}(t) \hat{b}(t) \rangle$. In particular, we use:
\begin{subequations}\label{Eq:SecMomSimple}
	\begin{align}
	\left\langle \hat{\delta V}_{in}^{\dagger}(\tau_{1})\hat{\delta V}_{in}(\tau_{2}) \right\rangle & = \frac{\hbar \gamma_{t}}{2 C_{0} \omega_{s}} \bar{n}_{e}(\omega_{s},T_{e}) \delta(\tau_{2}-\tau_{1}), \label{Eq:SecMomVinSimple1}\\
	\left\langle \hat{\delta V}_{in}(\tau_{1})\hat{\delta V}_{in}^{\dagger}(\tau_{2}) \right\rangle & = \frac{\hbar \gamma_{t}}{2 C_{0} \omega_{s}} \left[\bar{n}_{e}(\omega_{s},T_{e}) + 1\right] \delta(\tau_{2}-\tau_{1}), \label{Eq:SecMomVinSimple2}\\
	\left\langle \hat{V}_{R0}^{\dagger}(\tau_{1})\hat{V}_{R0}(\tau_{2}) \right\rangle & = \frac{\hbar \gamma_{r}}{2 C_{0} \omega_{s}} \bar{n}_{e}(\omega_{s},T_{e}) \delta(\tau_{2}-\tau_{1}), \label{Eq:SecMomVR0Simple1}\\
	\left\langle \hat{V}_{R0}(\tau_{1})\hat{V}_{R0}^{\dagger}(\tau_{2}) \right\rangle & = \frac{\hbar \gamma_{r}}{2 C_{0} \omega_{s}} \left[\bar{n}_{e}(\omega_{s},T_{e}) + 1\right] \delta(\tau_{2}-\tau_{1}), \label{Eq:SecMomVR0Simple2}\\
	\left\langle \hat{F}_{b}^{\dagger}(\tau_{1})\hat{F}_{b}(\tau_{2}) \right\rangle & = 2\hbar m \omega_{m} \bar{n}_{m}(\omega_{m},T_{m}) \delta(\tau_{2}-\tau_{1}), \label{Eq:SecMomFbSimple1}\\
	\left\langle \hat{F}_{b}(\tau_{1})\hat{F}_{b}^{\dagger}(\tau_{2}) \right\rangle & = 2\hbar m \omega_{m} \left[\bar{n}_{m}(\omega_{m},T_{m}) + 1\right] \delta(\tau_{2}-\tau_{1}),   \label{Eq:SecMomFbSimple2}
	\end{align}
\end{subequations}
where $\bar{n}_{e}(\omega_{s},T_{e})$ and $\bar{n}_{m}(\omega_{m},T_{m})$ are the average thermal occupation numbers of the electrical and mechanical reservoirs at temperatures $T_{e}$ and $T_{m}$, respectively. In the following, we omit their argument, assuming that they refer to the frequencies $\omega_{s}$ (electrical) and $\omega_{m}$ (mechanical), and temperatures $T_{e}$ (electrical) and $T_{m}$ (mechanical) of the respective systems. The requirements on the reservoirs necessary to write the second moments of the noises as in Eqs. \eqref{Eq:SecMomSimple} are the following. First, the mechanical force $\hat{F}_{b}$ needs to vary on a timescale much faster than the membrane's decay time, which is fulfilled for $\gamma_{b}\ll\omega_{m}$. Second, we have neglected the difference between the electrical excitation number $\bar{n}_{e}$ at the central frequency and the mechanical sidebands. This means that the electrical resonance frequency has to be much bigger than the mechanical one ($\omega_{s}\gg\omega_{m}$). Both these requirements are satisfied for the considered experimental parameters.

Knowing the second moments of all noise operators, and neglecting the term proportional to $\langle\hat{\delta a}^{\dagger}(0)\hat{\delta a}(0)\rangle$, using Eq. \eqref{Eq:SolbLinFinSimple} we can find $n_{b}(t)=\left\langle \hat{n}_{b} (t) \right\rangle$ to be
\begin{equation}
\begin{split}
n_{b}(t) & = n_{b}(0)e^{-\gamma_{b}t} + \bar{n}_{m}\left( 1-e^{-\gamma_{b}t} \right) - \frac{4 g_{1}^{2}\lvert\alpha\rvert^{2} \gamma_{t}(1+2 \bar{n}_{e})e^{-(\gamma_{t}+\gamma_{r})t}}{T(\gamma_{t}+\gamma_{r})^{2}\left[ (-\gamma_{b}+\gamma_{t}+\gamma_{r})^{2}+4 \omega_{m}^{2} \right]} \\ & + \frac{8 g_{1}^{2}\lvert\alpha\rvert^{2}\gamma_{t}(1+2\bar{n}_{e})e^{-\frac{\gamma_{b}+\gamma_{t}+\gamma_{r}}{2}t}}{T(\gamma_{t}+\gamma_{r})\left[ (-\gamma_{b}+\gamma_{t}+\gamma_{r})^{2}+4 \omega_{m}^{2} \right]} \Bigg( \frac{e^{-i \omega_{m} t}}{\gamma_{b}+\gamma_{t}+\gamma_{r}+2 i \omega_{m}} \\ &+\frac{e^{i \omega_{m} t}}{\gamma_{b}+\gamma_{t}+\gamma_{r}-2 i \omega_{m}} \Bigg) + \frac{4 g_{1}^{2}\lvert\alpha\rvert^{2} \gamma_{t}(\gamma_{b}+\gamma_{t}+\gamma_{r})(1+2 \bar{n}_{e})}{T\gamma_{b}(\gamma_{t}+\gamma_{r})^{2}\left[(-\gamma_{b}+\gamma_{t}+\gamma_{r})^{2}+4 \omega_{m}^{2} \right]} \\ & - \frac{4 g_{1}^{2}\lvert\alpha\rvert^{2} \gamma_{t}(1+2 \bar{n}_{e})e^{-\gamma_{b}t}}{T\gamma_{b}(\gamma_{t}+\gamma_{r})\left[ (-\gamma_{b}+\gamma_{t}+\gamma_{r})^{2}+4 \omega_{m}^{2} \right]}, \label{Eq:PhononNumbEvComplSimple}
\end{split}
\end{equation}
where the first two terms $n_{b}(0)e^{-\gamma_{b}t} + \bar{n}_{m}\left( 1-e^{-\gamma_{b}t} \right)$ describe the usual time evolution of a free membrane influenced by its own thermal bath, and everything else is the dynamics induced from the electrical system. 

\begin{figure}[htbp]
	\centering
	\includegraphics[width=18 cm]{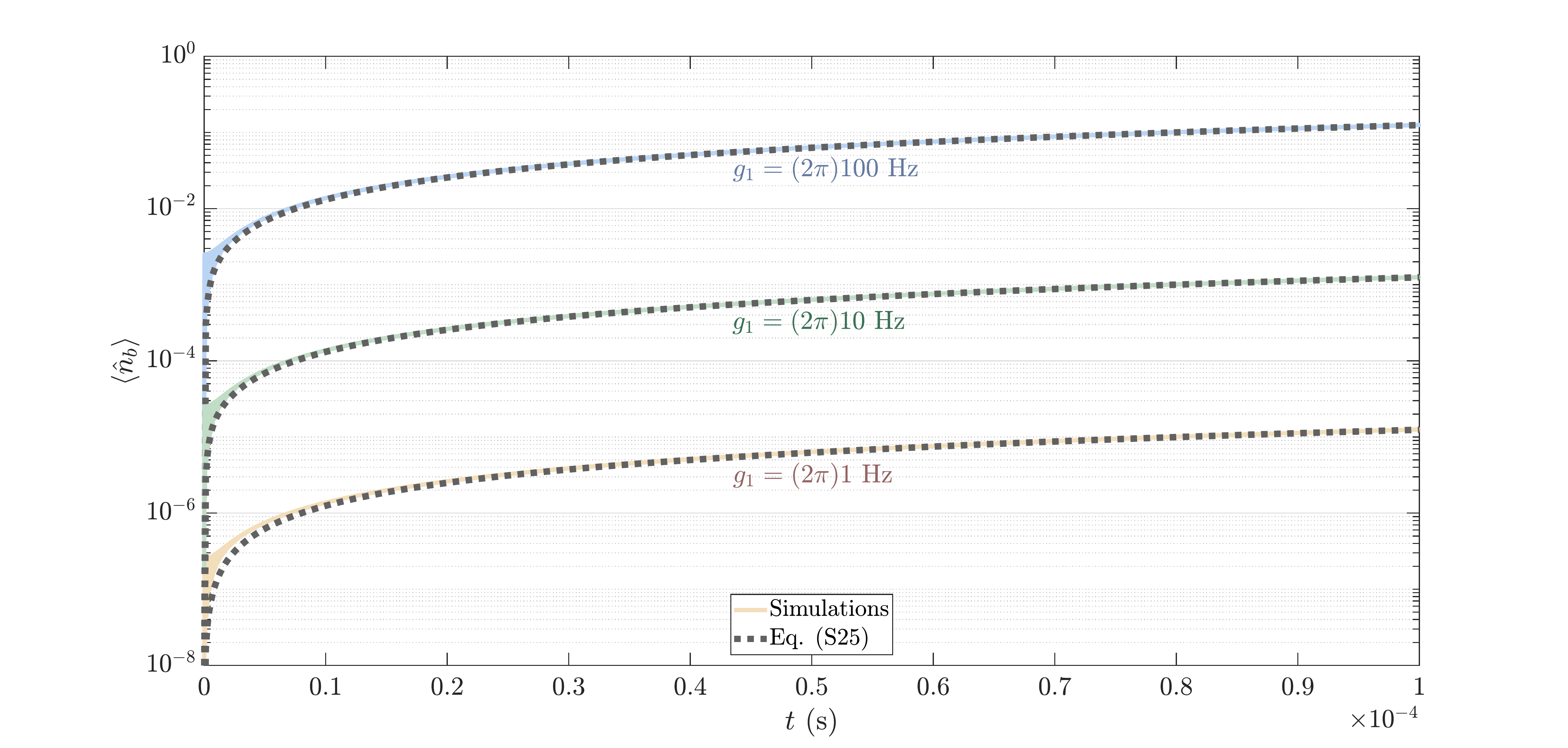}
	\caption[FigS3]{Examples for the dynamics of the average mechanical phonon number $n_{b}(t)$. The dotted curves are the analytical results given in Eq. \eqref{Eq:PhononNumbEvAppSimple}. The solid curves comes from numerically solving the master equation with the Hamiltonian of Eq. \eqref{Eq:HamLinFinalSimple} and the electrical and mechanical reservoirs (using the Markov approximation). The three sets differ from the parameter $g_{1}$, that has been taken to be equal to $(2\pi)100$ Hz (top set), $(2\pi)10$ Hz (middle set), and $(2\pi)1$ Hz (bottom set). We assume that $\lvert\alpha\rvert^{2}=10^{12}$ photons are sent into the system for the duration $T$ of the measurement, and we used the parameters: $\omega_{m}=(2\pi)100$ MHz, $\omega_{s}=(2\pi)5$ GHz, $\gamma_{r}=\gamma_{t}=(2\pi)1$ MHz, $\gamma_{b}=(2\pi)100$ Hz, $T=100$ $\mu$s, $\bar{n}_{e}=0$ and $\bar{n}_{m}=0$.}
	\label{Fig:FigS3}
\end{figure}

Eq. \eqref{Eq:PhononNumbEvComplSimple} represents an exact result in the limit in which the photonic ($\bar{n}_{e}$) and phononic ($\bar{n}_{m}$) reservoirs are Markovian. However, it can be better understood if we assume the mechanical damping $\gamma_{b}$ to be much smaller than the electrical one, $\gamma_{t}+\gamma_{r}$. In particular, if we are probing the system on a time scale that is much longer than the electrical lifetime, i.e. $t\gg (\gamma_{t}+\gamma_{r})^{-1}$, we can rewrite Eq. \eqref{Eq:PhononNumbEvComplSimple} as
\begin{equation}
n_{b}(t) = n_{b}(0)e^{-\gamma_{b}t} + \left[\bar{n}_{m} + \frac{\Gamma_{b}}{\gamma_{b}}(1+2 \bar{n}_{e}) \right]\left( 1-e^{-\gamma_{b}t} \right),  \label{Eq:PhononNumbEvAppSimple}
\end{equation}
where we have defined the induced heating $\Gamma_{b}$ to be
\begin{equation}
\Gamma_{b} = \frac{4 g_{1}^{2} \lvert\alpha\rvert^{2} \gamma_{t}}{T(\gamma_{t}+\gamma_{r})\left[ (\gamma_{t}+\gamma_{r})^{2} + 4 \omega_{m}^{2} \right]}.  \label{Eq:GammaBSimple}
\end{equation}
Equation \eqref{Eq:PhononNumbEvAppSimple} has been compared with numerical simulations of the master equation of the Hamiltonian in Eq. \eqref{Eq:HamLinFinalSimple}, taking into account the electrical and mechanical reservoirs. Examples are given in Fig. \ref{Fig:FigS3}.
%%%
%%%

From the above expressions, we can determine the parameter $\Delta n_{b}$ by finding the first order expansion of the function $n_{b}(t)$ in Eq. \eqref{Eq:PhononNumbEvAppSimple} with $n_{b}(0)=0$,
\begin{equation}
\Delta n_{b} = \gamma_{b}\bar{n}_{m}T + \Gamma_{b}(1+2 \bar{n}_{e})T.
\end{equation}
The form of $\Delta n_{b}$ given in the main text assumes $\gamma_{b}  \ll \frac{\Gamma_{b}}{\bar{n}_{m}}$. Having determined both $D^{2}$ and $\Delta n_{b}$, it is possible to derive the general form of the parameter $\lambda$, including the contribution from the mechanical reservoir.

\section{``Double arm'' circuit}\label{sec:CompleteCirc}

\begin{figure}[htbp]
	\centering
	\includegraphics[width=8 cm]{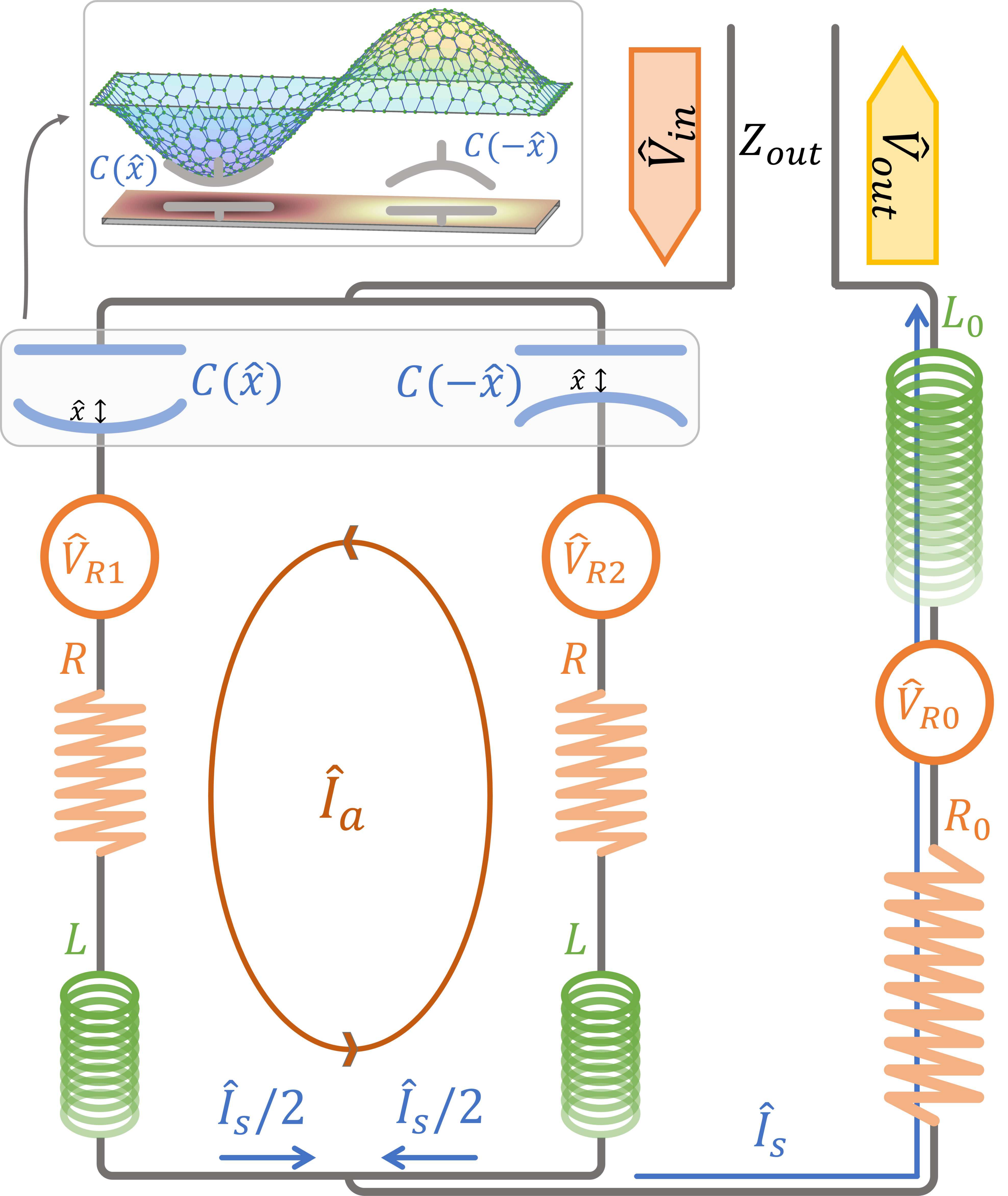}
	\caption[FigS4]{``Double arm'' model of the considered electromechanical setup, with parasitic resistances ($R$) and inductances ($L$). We take into account the Johnson--Nyquist noises associated with the main resistor $R_{0}$ and the two parasitic ones: $\hat{V}_{R0}$, $\hat{V}_{R1}$ and $\hat{V}_{R2}$, respectively. $\hat{I}_{s}$ and $\hat{I}_{a}$ are the two electrical currents considered in our analysis, and $C(\pm \hat{x})$ represent the two halves of the capacitor. In the inset we sketch how $C(\pm \hat{x})$ may arise.}
	\label{Fig:FigS4}
\end{figure}

In the following we derive the parameter $\lambda$ for the ``double arm'' circuit, introduced to take into account the coupling to the antisymmetric mode associated with the redistribution of charge on the membrane. This is depicted in Fig. \ref{Fig:FigS4}. We include possible asymmetries in the fabrication process, the effect of which can be described by a residual linear coupling $g_{r}$. The contribution to $\lambda$ from other kinds of asymmetries -- different parasitic resistances and inductances -- will only be investigated numerically, since the analytical results are too long and complicated in this case.
%%%
%%%

Using the currents $\hat{I}_{s}$ and $\hat{I}_{a}$ and their associated charges $\hat{Q}_{s}$ and $\hat{Q}_{a}$, we can determine the Hamiltonian of the circuit in Fig. \ref{Fig:FigS4} to be
\begin{equation}
\begin{split}
\hat{\mathcal{H}}  =& \hbar \omega_{m} \hat{b}^{\dagger} \hat{b} - x_{0} \left( \hat{b}^{\dagger} + \hat{b}\right)\hat{F}_{b} + \frac{\hat{\Phi}_{a}^{2}}{4L}+\frac{\hat{Q}_{a}^{2}}{C_{0}}+\frac{\hat{\Phi}_{s}^{2}}{L+2L_{0}}+\frac{\hat{Q}_{s}^{2}}{4C_{0}}  - \hat{Q}_{s}(2\hat{V}_{in}+2\hat{V}_{R0}+\hat{V}_{R1}+\hat{V}_{R2})\\ & -2\hat{Q}_{a}(\hat{V}_{R2}-\hat{V}_{R1}) + \frac{g_{1}}{C_{0} \omega_{s}}\hat{Q}_{a}\hat{Q}_{s}\left( \hat{b}+\hat{b}^{\dagger} \right) + \frac{g_{2}}{C_{0} \omega_{s}}\hat{Q}_{a}^{2}\hat{b}^{\dagger}\hat{b} + \frac{g_{2}}{4 C_{0} \omega_{s}}\hat{Q}_{s}^{2}\hat{b}^{\dagger}\hat{b},
\end{split}
\label{Eq:HamQuantCompleteBal}
\end{equation}
where the magnetic fluxes $\hat{\Phi}_{a}$ and $\hat{\Phi}_{s}$ are defined below. In principle, there are other terms proportional to $g_{2}\hat{b}\hat{b}$ and $g_{2}\hat{b}^{\dagger}\hat{b}^{\dagger}$ in the Hamiltonian. These are responsible for sidebands at frequencies $\omega_{s}\pm 2\omega_{m}$, that we have already encountered in section \ref{sec:RLCcircuit}. For the same reasons explained there, these terms do not contribute neither to the electrical readout (homodyne measurement at frequency $\omega_{s}$) nor to the heating ($g_{2} \ll g_{1}$), and we shall therefore ignore them in the following. A quantitative reason for neglecting these two--phonon processes, is the following. With Fermi Golden rule, we can determine the rate at which these processes happen to be
\begin{equation}\label{Eq:RateTwoPhon}
\frac{g_{2}^2 \lvert\alpha\rvert^{2} \gamma_{t}}{\left( \frac{\gamma_{r}+\gamma_{t}}{2} \right)^{2} + \left( 2 \omega_{m} \right)^{2}},
\end{equation}
that is orders of magnitude lower than $\Delta n_{b}/T$ (the one--phonon processes induced by the linear coupling), as can be estimated later in section \ref{sec:ExptEst}, where parameters for a proposed implementation are given. Notice that the rate in Eq. \eqref{Eq:RateTwoPhon} describes the two phonon processes relative to the term $\propto g_{2} \hat{Q}_{s}^{2}$ in the Hamiltonian in Eq. \eqref{Eq:HamQuantCompleteBal}. The other two phonon process, $\propto g_{2} \hat{Q}_{a}^{2}$, is even more suppressed, as the asymmetric field is not directly driven by the input $\hat{V}_{in}$.

From Eq. \eqref{Eq:HamQuantCompleteBal}, it is possible to derive the equations of motion for the fields $\hat{Q}_{a}$, $\hat{Q}_{s}$, $\hat{\Phi}_{a}$, $\hat{\Phi}_{s}$ and $\hat{b}$. The electrical decays are included using Kirchoff laws, resulting in:
\begin{subequations}\label{Eq:EOMQBComp}
	\begin{align}
	\dot{\hat{\Phi}}_{a}  = & -2\frac{\hat{Q}_{a}}{C_{0}}-\frac{g_{1}}{C_{0} \omega_{s}}\hat{Q}_{s}\left(\hat{b}+\hat{b}^{\dagger}\right)-\frac{2 g_{2}}{C_{0} \omega_{s}}\hat{Q}_{a}\hat{n}_{b}-\gamma_{l}\hat{\Phi}_{a}+2\left( \hat{V}_{R2}-\hat{V}_{R1} \right), \label{Eq:EOMQBPhiAComplete}\\
	\dot{\hat{Q}}_{a}  = & \frac{\hat{\Phi}_{a}}{2L}, \label{Eq:EOMQBQAComplete}  \\
	\dot{\hat{\Phi}}_{s}  = & -\frac{\hat{Q}_{s}}{2C_{0}}-\frac{g_{1}}{C_{0} \omega_{s}}\hat{Q}_{a}\left(\hat{b}+\hat{b}^{\dagger}\right)-\frac{g_{2}}{2C_{0} \omega_{s}}\hat{Q}_{s}\hat{n}_{b} -(\gamma_{t}+\gamma_{r})\hat{\Phi}_{s}+2\hat{V}_{in}+2\hat{V}_{R0} + \hat{V}_{R1}+\hat{V}_{R2}, \label{Eq:EOMQBPhisComplete}  \\
	\dot{\hat{Q}}_{s}  = & \frac{2\hat{\Phi}_{s}}{L+2L_{0}},  \label{Eq:EOMQBQsComplete}  \\
	\dot{\hat{b}}  = & -i \omega_{m} \hat{b} -\frac{i}{\hbar}\frac{g_{1}}{ C_{0} \omega_{s}}\hat{Q}_{a}\hat{Q}_{s}-\frac{i}{\hbar}\frac{g_{2}}{ C_{0} \omega_{s}}\left( \hat{Q}_{a}^{2}+\frac{\hat{Q}_{s}^{2}}{4} \right)\hat{b} -\frac{\gamma_{b}}{2} \hat{b}+\frac{x_{0}}{\hbar}\hat{F}_{b}. \label{Eq:EOMQBbComplete}
	\end{align}
\end{subequations}
To enable a direct comparison with the $RLC$ circuit presented in section \ref{sec:RLCcircuit} we have here used the same notation. This requires small differences in the definitions, to take into account the parasitic resistances and inductances: 
\begin{subequations}\label{Eq:CnstCompl}
	\begin{align}
	\omega_{s}^{2} & = \frac{1}{C_{0}(L+2L_{0})},\\
	\omega_{a}^{2} & = \frac{1}{C_{0}L},\\
	\gamma_{t} & = \frac{2Z_{out}}{L+2L_{0}},\\
	\gamma_{r} & = \frac{R+2R_{0}}{L+2L_{0}},\\
	\gamma_{l} & = \frac{R}{L}.
	\end{align}
\end{subequations}
Moreover, the linear and quadratic couplings $g_{1}$ and $g_{2}$ come from the expansions of \textit{each} one of the two capacitors $C(\pm \hat{x})$ in the mechanical position $\hat{x}$, as suggested by the inset in Fig. \ref{Fig:FigS4}. Following the same procedure used for the $RLC$ circuit, we first study the measurement of $\hat{n}_{b}$, and later identify the conditions under which the measurement is effectively QND.

Notice that, depending on the specific experimental setup, other circuit scheme may be better suited for describing the system. For instance, in the experiment of Ref. \cite{Bagci2014} the mechanical oscillator is not directly connected to the circuit, and different equivalent circuit would be required. Here, we restrict ourselves to the setup in Fig. \ref{Fig:FigS4} and defer other setup for later investigation \cite{FollowUp}.

\subsection{QND measurement of the phonon number, and comparison with the ``membrane in the middle'' setup \cite{YanbeiChen}}\label{sec:QNDcomplete}

Assuming that the mechanical state is unchanged during the whole measurement time $T$, we can neglect all sources of heating in the system of equations \eqref{Eq:EOMQBComp}, and take $\hat{n}_{b}$ to be constant in time. Importantly, for the setup in Fig. \ref{Fig:FigS4} there are two mechanisms that shift the resonant frequency $\omega_{s}$ and thus allow for the QND measurement. The first one we encountered before for the $RLC$ circuit, and relates to the quadratic electromechanical coupling: $\omega_{s}\rightarrow \omega_{s}+g_{2}n_{b}$. The second is more involved, and is the same one considered in the optomechanical setup of Ref. \cite{Marquardt}. It relies on an effective quadratic coupling proportional to $g_{1}^{2}$, that arises once we substitute Eqs. \eqref{Eq:EOMQBPhiAComplete} and \eqref{Eq:EOMQBQAComplete} into Eqs. \eqref{Eq:EOMQBPhisComplete} and \eqref{Eq:EOMQBQsComplete}. To better understand this process, it is instructive to look at the final equation for $\hat{\Phi}_{s}$ in the frequency domain:
\begin{equation}
\begin{split}
\Bigg( \Omega_{k}^{2}-\omega_{s}^{2}+ & \overbrace{\frac{(\hat{n}_{b}+1)g_{1}^{2}\omega_{a}^{2}}{\omega_{a}^{2}+(\Omega_{k}-\omega_{m})(-i \gamma_{l} - \Omega_{k}+\omega_{m})} + \frac{\hat{n}_{b}g_{1}^{2}\omega_{a}^{2}}{\omega_{a}^{2}-(\Omega_{k}+\omega_{m})(i \gamma_{l} + \Omega_{k}+\omega_{m})}}^{g_{1}^{2}\text{ shift}} -\overbrace{g_{2} \omega_{s} \hat{n}_{b}}^{g_{2}\text{ shift}} \Bigg)\hat{\Phi}_{s}[\Omega_{k}] = \\ & i \Omega_{k} (\gamma_{r}+\gamma_{t})\hat{\Phi}_{s}[\Omega_{k}] + 2 i \Omega_{k} \hat{V}_{in}[\Omega_{k}]+2 i \Omega_{k} \hat{V}_{R0}[\Omega_{k}]+ i \Omega_{k} \hat{V}_{R1}[\Omega_{k}]+ i \Omega_{k} \hat{V}_{R2}[\Omega_{k}],
\end{split}\label{Eq:PhiCompleteEx}
\end{equation}
where we neglected off--resonant terms and assumed $g_{2} \ll \omega_{s}$. As indicated explicitly in Eq. \eqref{Eq:PhiCompleteEx}, it is possible to see the frequency shifts induced by the linear $g_{1}$ and quadratic $g_{2}$ couplings. Whether the QND interaction is dominated by the quadratic coupling $g_{2}$ or the effective quadratic interaction $\propto g_{1}^{2}$, depends on the resonance condition of the symmetric $\hat{I}_{s}$ and antisymmetric $\hat{I}_{a}$ modes in Fig. \ref{Fig:FigS4}. To have a sizeable effect of the $g_{1}$ term, we need to be near resonance with the antisymmetric mode $\Omega_{k}\simeq \omega_{a} \gg \omega_{m},\gamma_{l}$. At the same time, for the QND detection we probe the system at the resonance frequency $\Omega_{k}\simeq\omega_{s}$. Hence, the $g_{1}$ term is dominant when we allow for a strong hybridization of the two electrical modes: $\omega_{s}\simeq \omega_{a}$. In such situation, by determining the heating rate for the electromechanical setup (see Sec. \ref{sec:HeatComplBalanced}), we can derive the condition for a feasible QND detection:
\begin{equation}
\lambda = \frac{2}{(1+\bar{n}_{e})(1+2 \bar{n}_{e})}\frac{ g_{1}^{2}}{(\gamma_{r}+\gamma_{t})^{2}}\frac{\gamma_{t}}{\gamma_{l}}\gg 1. \label{Eq:LambdaEffLin}
\end{equation}
This condition is similar to the one found in Ref. \cite{YanbeiChen} for the experiment in Ref. \cite{Marquardt}, where strong heating was proven to forbid the QND detection. Compared to that work, however, we gain the factor $\gamma_{t}/\gamma_{l}$, that comes from the asymmetry between the damping of the two electrical modes. Hence, our setup does have some gain compared to the $RLC$ circuit, if the damping of the antisymmetric mode is small: $\gamma_{l} \ll \gamma_{t}$. However, in this case the QND measurement still faces the challenge of a strong heating of the mechanical motion due to the large coupling between the electrical modes, and requires $g_{1}^{2} \gg \gamma_{l}\gamma_{t}$.

In the following, we will investigate the opposite limit, where the antisymmetric mode $\hat{I}_{a}$ is far--off resonant from the symmetric one $\hat{I}_{s}$. In this case, the quadratic coupling $g_{2}$ is dominant \cite{Terza}, and the heating induced to the mechanical mode will be strongly suppressed. If we neglect the effective quadratic coupling $\propto g_{1}^{2}$, we only need to consider Eqs. \eqref{Eq:EOMQBPhisComplete} and \eqref{Eq:EOMQBQsComplete} for determining the measurement signal. In the absence of heating, these are uncoupled from both the mechanics and the other electrical modes. We can then derive relations for $\hat{\Phi}_{s}[\Omega_{k}]$ and $\hat{Q}_{s}[\Omega_{k}]$ similar to Eqs. \eqref{Eq:EOMSimpleQuad}:
\begin{subequations}\label{Eq:EOMCompleteQuad}
	\begin{align}
	-i\Omega_{k}\hat{\Phi}_{s}[\Omega_{k}] & = -\frac{\hat{Q}_{s}[\Omega_{k}]}{2 C_{0}} - \frac{g_{2}\omega_{s}L_{0}}{2}\hat{Q}_{s}[\Omega_{k}] \hat{n}_{b}-(\gamma_{r}+\gamma_{t})\hat{\Phi}_{s}[\Omega_{k}]+2 (\hat{V}_{in}[\Omega_{k}]+\hat{V}_{R0}[\Omega_{k}])+\hat{V}_{R1}[\Omega_{k}]+\hat{V}_{R2}[\Omega_{k}], \label{Eq:EOMPhiCompleteQuad} \\
	-i\Omega_{k}\hat{Q}_{s}[\Omega_{k}] & = \frac{2\hat{\Phi}_{s}[\Omega_{k}]}{L+2L_{0}}. \label{Eq:EOMQCompleteQuad}
	\end{align}
\end{subequations}
The reflected field $\hat{V}_{out}[\Omega_{k}]$ is
\begin{equation}
\hat{V}_{out}[\Omega_{k}] = \left[1-\zeta(\Omega_{k},\hat{n}_{b})\right]\hat{V}_{in}[\Omega_{k}] - \zeta(\Omega_{k},\hat{n}_{b})\left(\hat{V}_{R0}[\Omega_{k}]+\frac{\hat{V}_{R1}[\Omega_{k}]+\hat{V}_{R2}[\Omega_{k}]}{2}\right), \label{Eq:VoutCompleteForm}
\end{equation}
where we have used the input/output relation $\hat{V}_{out} = \hat{V}_{in}-\gamma_{t}\hat{\Phi}_{s}$. The coefficient $\zeta(\Omega_{k},\hat{n}_{b})$ is the same as in Eq. \eqref{Eq:VoutSimpleCoeff}, except for a factor $2$ coming from the fact that, here, we are considering the two halves of our capacitor. With the outcome of the homodyne detection being described by the same operator $\hat{V}_{M}$ defined in Eq. \eqref{Eq:VmSimple}, we can find the parameter $D^{2}$ for the setup in Fig. \ref{Fig:FigS4} to be
\begin{equation}
D^{2}=\frac{d^{2}}{\sigma^{2}}=\frac{16}{1+2 \bar{n}_{e}}\frac{g_{2}^{2}\lvert\alpha\rvert^{2}\gamma_{t}^{2}}{\left[ g_{2}^{2} + \left( \gamma_{t}+\gamma_{r} \right)^{2} \right]^{2}}. \label{Eq:ParDsqCompl}
\end{equation}
Notice that for deriving Eq. \eqref{Eq:ParDsqCompl} from Eq. \eqref{Eq:VoutCompleteForm} we assumed that the electrical reservoirs are in a thermal state with average photon number $\bar{n}_{e}$, and that the drive is a coherent state $\alpha$  at the frequency $\omega_{s}$. Consistent with previous sections, $\alpha$ is chosen to be real and the phase of the homodyne measurement is fixed such that $\theta=\pi$ in Eq. \eqref{Eq:VmSimple}.

\subsection{Membrane heating}\label{sec:HeatComplBalanced}

Given that $g_{2} \ll g_{1}$, we set the quadratic coupling to zero in this subsection. We can then rewrite the Hamiltonian in Eq. \eqref{Eq:HamQuantCompleteBal} in the form
\begin{equation}
\begin{split}
\hat{\mathcal{H}} & = \hbar \omega_{m} \hat{b}^{\dagger} \hat{b} - x_{0} \left( \hat{b}^{\dagger} + \hat{b}\right)\hat{F}_{b} + \frac{\hat{\Phi}_{a}^{2}}{4L}+\frac{\hat{Q}_{a}^{2}}{C_{0}}+\frac{\hat{\Phi}_{s}^{2}}{L+2L_{0}}+\frac{\hat{Q}_{s}^{2}}{4C_{0}}  - \hat{Q}_{s}(2\hat{V}_{in}+2\hat{V}_{R0}+\hat{V}_{R1}+\hat{V}_{R2})\\ & -2\hat{Q}_{a}(\hat{V}_{R2}-\hat{V}_{R1}) + \frac{g_{1}}{C_{0} \omega_{s}}\hat{Q}_{a}\hat{Q}_{s}\left( \hat{b}+\hat{b}^{\dagger} \right).
\end{split}
\label{Eq:HamQuantCompleteBalLinWithS}
\end{equation}
Moreover, by looking at Eqs. \eqref{Eq:EOMQBComp}, it is possible to see that $\hat{Q}_{s}$ and $\hat{\Phi}_{s}$ are the only driven fields in the system (with $\hat{V}_{R0}$, $\hat{V}_{R1}$ and $\hat{V}_{R2}$ being in a thermal state). Therefore, we can neglect perturbations induced by the mechanical motion, and substitute them with their average values $\langle \hat{Q}_{s}\rangle$ and $\langle\hat{\Phi}_{s}\rangle$. This assumption will be verified in subsection \ref{sec:HeatingSim}, where we simulate the dynamics of the electromechanical system in the general case in which the parasitic elements may differ from each other. From the Hamiltonian Eq. \eqref{Eq:HamQuantCompleteBalLinWithS} we can determine the equations of motion for $\langle \hat{Q}_{s}\rangle$ and $\langle\hat{\Phi}_{s}\rangle$:
\begin{subequations}\label{Eq:EOMQBsModeLinAvg}
	\begin{align}
	\langle \dot{\hat{\Phi}}_{s}(t) \rangle  = & -\frac{\langle\hat{Q}_{s}(t)\rangle}{2C_{0}} -(\gamma_{t}+\gamma_{r})\langle\hat{\Phi}_{s}(t)\rangle + 2\langle\hat{V}_{in}(t)\rangle, \label{Eq:EOMQBPhisCompLinAvg}  \\
	\langle\dot{\hat{Q}}_{s}(t)\rangle  = & \frac{2\langle\hat{\Phi}_{s}(t)\rangle}{L+2L_{0}}, \label{Eq:EOMQBQsComptLinAvg} 
	\end{align}
\end{subequations}
where $\langle\hat{V}_{in}(t)\rangle = \alpha\sqrt{2\hbar \omega_{s}Z_{out}/T}\cos(\omega_{s} t)$. Similar to above, we are mainly interested in the case of constant incident fields or long pulses, such that we ignore transient behaviours. The steady state solution for the charge $\langle\hat{Q}_{s}(t)\rangle$ is then given by
\begin{equation}
\langle\hat{Q}_{s}(t)\rangle = -\frac{2 i \alpha}{\gamma_{t}+\gamma_{r}}\sqrt{\frac{\hbar C_{0}\omega_{s}\gamma_{t}}{T}}\left(e^{i\omega_{s}t}- e^{-i\omega_{s}t} \right). \label{Eq:AvgQsCompl}
\end{equation}
Assuming that the strongly driven symmetric fields $\hat{Q}_{s}$ and $\hat{\Phi}_{s}$ are not perturbed by the mechanical dynamics, we can substitute Eq. \eqref{Eq:AvgQsCompl} into the Hamiltonian of Eq. \eqref{Eq:HamQuantCompleteBalLinWithS}, and obtain
\begin{equation}
\hat{\mathcal{H}} =  \hbar \omega_{m} \hat{b}^{\dagger} \hat{b} - x_{0} \left( \hat{b}^{\dagger} + \hat{b}\right)\hat{F}_{b} + \frac{\hat{\Phi}_{a}^{2}}{4L}+\frac{\hat{Q}_{a}^{2}}{C_{0}} - 2\hat{Q}_{a}(\hat{V}_{R1}-\hat{V}_{R1}) - \frac{2 i\hbar g_{1}\alpha}{\gamma_{t}+\gamma_{r}}\sqrt{\frac{\gamma_{t}}{\hbar T C_{0} \omega_{s}}}\left(e^{i\omega_{s}t}- e^{-i\omega_{s}t} \right)\hat{Q}_{a} \left( \hat{b}+\hat{b}^{\dagger} \right).
\label{Eq:HamQuantCompleteBalLin}
\end{equation}
Differential equations for $\hat{Q}_{a}$, $\hat{\Phi}_{a}$, and the mechanical annihilation operator $\hat{b}$ can finally be derived:
\begin{subequations}\label{Eq:EOMQBCompLin}
	\begin{align}
	\dot{\hat{\Phi}}_{a} = & -2\frac{\hat{Q}_{a}}{C_{0}} + \frac{2 i g_{1}\alpha}{\gamma_{t}+\gamma_{r}}\sqrt{\frac{\hbar\gamma_{t}}{ T C_{0} \omega_{s}}}\left( e^{-i\omega_{s}t}-e^{i\omega_{s}t} \right)\left(\hat{b}+\hat{b}^{\dagger}\right) -\gamma_{l}\hat{\Phi}_{a}+2\left( \hat{V}_{R2}-\hat{V}_{R1} \right), \label{Eq:EOMQBPhiACompLin}\\
	\dot{\hat{Q}}_{a} = & \frac{\hat{\Phi}_{a}}{2L}, \label{Eq:EOMQBQACompLin}  \\
	\dot{\hat{b}}  = & -i \omega_{m} \hat{b} - \frac{2 g_{1}\alpha}{\gamma_{t}+\gamma_{r}}\sqrt{\frac{\gamma_{t}}{\hbar T C_{0} \omega_{s}}}\left( e^{-i\omega_{s}t}-e^{i\omega_{s}t} \right)\hat{Q}_{a}-\frac{\gamma_{b}}{2} \hat{b}+i\frac{x_{0}}{\hbar}\hat{F}_{b}. \label{Eq:EOMQBbCompLin}
	\end{align}
\end{subequations}
These equations describe the coupling of the mechanical mode to the antisymmetric electrical one. This coupling is enhanced by driving the symmetric mode with the coherent state $\alpha$, and is responsible for heating up the membrane, similarly to Ref. \cite{YanbeiChen}. Therefore, we need to assess to which degree this is deleterious for the QND measurement of $\hat{n}_{b}$. Equations \eqref{Eq:EOMQBCompLin} are nontrivial, and exact solutions are not accessible. However, under reasonable assumptions, we are able to find an approximate analytical solution that will be subsequently confirmed by our numerical approach.

Our first step for dealing with the system \eqref{Eq:EOMQBCompLin} is to switch to the Fourier domain:
\begin{subequations}\label{Eq:EOMQBCompLinFou}
	\begin{align}
	\begin{split}
	-i\Omega_{k}\hat{\Phi}_{a}\left[\Omega_{k}\right]  = & -2\frac{\hat{Q}_{a}\left[\Omega_{k}\right]}{C_{0}} + \frac{2 i g_{1}\alpha}{\gamma_{t}+\gamma_{r}}\sqrt{\frac{\hbar\gamma_{t}}{ T C_{0} \omega_{s}}}\Big(\hat{b}\left[\Omega_{k}-\omega_{s}\right]+\hat{b}^{\dagger}\left[\Omega_{k}-\omega_{s}\right] \\ & - \hat{b}\left[\Omega_{k}+\omega_{s}\right]-\hat{b}^{\dagger}\left[\Omega_{k}+\omega_{s}\right]\Big) -\gamma_{l}\hat{\Phi}_{a}\left[\Omega_{k}\right]+2\left( \hat{V}_{R2}\left[\Omega_{k}\right]-\hat{V}_{R1}\left[\Omega_{k}\right] \right),
	\end{split} \label{Eq:EOMQBPhiACompLinFou}\\
	-i\Omega_{k}\hat{Q}_{a}\left[\Omega_{k}\right]  = & \frac{\hat{\Phi}_{a}\left[\Omega_{k}\right]}{2L}, \label{Eq:EOMQBQACompLinFou}  \\
	\begin{split}
	-i\Omega_{k}\hat{b}\left[\Omega_{k}\right]  = & -i \omega_{m} \hat{b}\left[\Omega_{k}\right] - \frac{2 g_{1}\alpha}{\gamma_{t}+\gamma_{r}}\sqrt{\frac{\gamma_{t}}{\hbar T C_{0} \omega_{s}}}\left( \hat{Q}_{a}\left[\Omega_{k}-\omega_{s}\right] -\hat{Q}_{a}\left[\Omega_{k}+\omega_{s}\right] \right)\\ &-\frac{\gamma_{b}}{2} \hat{b}\left[\Omega_{k}\right]+i\frac{x_{0}}{\hbar}\hat{F}_{b}\left[\Omega_{k}\right].
	\end{split}\label{Eq:EOMQBbCompLinFou}
	\end{align}
\end{subequations}
We can employ equations \eqref{Eq:EOMQBPhiACompLinFou} and \eqref{Eq:EOMQBQACompLinFou} in order to obtain an expression for $\hat{b}\left[\Omega_{k}\right]$ that depends on the noises only
\begin{equation}
	\begin{split}
	\hat{b}\left[\Omega_{k}\right]&\left( \frac{\gamma_{b}+\Gamma_{b}(\Omega_{k})}{2} - i\left[\Omega_{k}-\omega_{m}-\omega_{b}(\Omega_{k})\right] \right)  =  i\frac{x_{0}}{\hbar}\hat{F}_{b}\left[\Omega_{k}\right]+\frac{2 g_{1} \alpha}{\gamma_{t}+\gamma_{r}} \omega_{a}^{2}\sqrt{\frac{\gamma_{t}C_{0}}{\hbar T \omega_{s}}}\times \\& \times\Bigg( \frac{\hat{V}_{R2}\left[\Omega_{k}-\omega_{s}\right]-\hat{V}_{R1}\left[\Omega_{k}-\omega_{s}\right]}{-\omega_{a}^{2}+\left( \Omega_{k}-\omega_{s} \right)\left( i\gamma_{l} + \Omega_{k}-\omega_{s} \right)} - \frac{\hat{V}_{R2}\left[\Omega_{k}+\omega_{s}\right]-\hat{V}_{R1}\left[\Omega_{k}+\omega_{s}\right]}{-\omega_{a}^{2}+\left( \Omega_{k}+\omega_{s} \right)\left( i\gamma_{l} + \Omega_{k}+\omega_{s} \right)} \Bigg),
	\end{split} \label{Eq:EOMQBMechLinFou}
\end{equation}
where we have defined the effective decay $\Gamma_{b}(\Omega_{k})$ and frequency shift $\omega_{b}(\Omega_{k})$, resulting from the electrical influence on the mechanical motion:
\begin{subequations}
	\begin{align}
	\begin{split}
	\Gamma_{b}(\Omega_{k}) &= \text{Re}\left\lbrace \frac{-4i \gamma_{t}}{T \omega_{s}}\left(\frac{\alpha g_{1}\omega_{a} }{\gamma_{r}+\gamma_{t}}\right)^{2}\left( \frac{1}{-\omega_{a}^{2}+(\Omega_{k}-\omega_{s})(i \gamma_{l}+\Omega_{k}-\omega_{s})}+\frac{1}{-\omega_{a}^{2}+(\Omega_{k}+\omega_{s})(i \gamma_{l}+\Omega_{k}+\omega_{s})} \right) \right\rbrace,
	\end{split}\label{Eq:GammaB}\\
	\begin{split}
	\omega_{b}(\Omega_{k}) &= \text{Im}\left\lbrace \frac{-2i \gamma_{t}}{T \omega_{s}}\left(\frac{\alpha g_{1}\omega_{a} }{\gamma_{r}+\gamma_{t}}\right)^{2}\left( \frac{1}{-\omega_{a}^{2}+(\Omega_{k}-\omega_{s})(i \gamma_{l}+\Omega_{k}-\omega_{s})}+\frac{1}{-\omega_{a}^{2}+(\Omega_{k}+\omega_{s})(i \gamma_{l}+\Omega_{k}+\omega_{s})} \right) \right\rbrace.	\end{split}\label{Eq:OmegaShift}
	\end{align}
\end{subequations}
Note that, for deriving Eq. \eqref{Eq:EOMQBMechLinFou}, we neglected the off resonant terms $\hat{b}[\Omega_{k}\pm 2 \omega_{s}]$, $\hat{b}^{\dagger}[\Omega_{k}\pm 2 \omega_{s}]$, and $\hat{b}^{\dagger}[\Omega_{k}]$ (we verify this approximation numerically below in Fig. \ref{Fig:FigS6}). Moreover, the Fourier components of the noises are written in their most general form, meaning that $\Omega_{k}$ is now allowed to be negative, with the additional constraints $\hat{V}_{R1}[\Omega_{k}]=\hat{V}^{\dagger}_{R1}[-\Omega_{k}]$, and $\hat{V}_{R2}[\Omega_{k}]=\hat{V}^{\dagger}_{R2}[-\Omega_{k}]$. 

In order to have an analytical expression for $n_{b}(t)\equiv\langle \hat{n}_{b}(t) \rangle$, we assume that the damping $\gamma_{b}$ of the mechanical motion is much smaller than its own natural frequency $\omega_{m}$. If $\gamma_{b}\ll\omega_{m}$, the set of frequencies contributing to $\hat{b}[\Omega_{k}]$ and $\hat{b}^{\dagger}[\Omega_{k}]$ will be located around $\omega_{m}$ and $-\omega_{m}$ respectively. This, in turn, implies that we can substitute the effective damping $\Gamma_{b}(\Omega_{k})$ and shift $\omega_{b}(\Omega_{k}) $ with the values taken at the relevant frequencies $\pm \omega_{m}$. Defining the constant effective decay $\Gamma_{b}=\Gamma_{b}(\omega_{m})=\Gamma_{b}(-\omega_{m})$ and frequency shift $\omega_{b}=\omega_{b}(\omega_{m})=-\omega_{b}(-\omega_{m})$, we can rewrite Eq. \eqref{Eq:EOMQBMechLinFou} in the following way \cite{Quarta}:
\begin{equation}\label{Eq:EOMQBMechLinFouPar}
	\begin{split}
	\hat{b}&\left[\Omega_{k}\right]\left( \frac{\gamma_{b}+\Gamma_{b}}{2} - i\left[\Omega_{k}-\omega_{m}-\omega_{b}\right] \right)  =  i\frac{x_{0}}{\hbar}\hat{F}_{b}\left[\Omega_{k}\right]+\frac{2 g_{1} \alpha}{\gamma_{t}+\gamma_{r}}\omega_{a}^{2}\sqrt{\frac{\gamma_{t}C_{0}}{\hbar T \omega_{s}}}\times \\& \times\Bigg( \frac{\hat{V}_{R2}\left[\Omega_{k}-\omega_{s}\right]-\hat{V}_{R1}\left[\Omega_{k}-\omega_{s}\right]}{-\omega_{a}^{2}+\left( \Omega_{k}-\omega_{s} \right)\left( i\gamma_{l} + \Omega_{k}-\omega_{s} \right)} - \frac{\hat{V}_{R2}\left[\Omega_{k}+\omega_{s}\right]-\hat{V}_{R1}\left[\Omega_{k}+\omega_{s}\right]}{-\omega_{a}^{2}+\left( \Omega_{k}+\omega_{s} \right)\left( i\gamma_{l} + \Omega_{k}+\omega_{s} \right)} \Bigg),
	\end{split}
\end{equation}
from which we can see that $\Gamma_{b}$ affects the dynamics as an effective decay and $\omega_{b}$ serves as a frequency shift.

We can now take the Fourier series of Eq. \eqref{Eq:EOMQBMechLinFouPar}, in order to go back to the time domain, and solve formally for $\hat{b}(t)$. %The frequency domain was essential in order to take the approximations described above, that are easy to understand in such context. 
Defining the noise operator $\hat{N}(t)=\hat{N}^{\dagger}(t)$ to be
\begin{subequations}\label{Eq:NoiseCompl}
	\begin{align}
	\begin{split}
	\hat{N}(t) & = \sum_{k=-\infty}^{\infty}\hat{N}\left[\Omega_{k}\right]\frac{e^{-i\Omega_{k}t}}{\sqrt{T}},
	\end{split}\label{Eq:NoiseOpCompl}\\
	\begin{split}
	\hat{N}\left[\Omega_{k}\right] & = \left( \frac{\hat{V}_{R2}\left[\Omega_{k}-\omega_{s}\right]-\hat{V}_{R1}\left[\Omega_{k}-\omega_{s}\right]}{-\omega_{a}^{2}+\left( \Omega_{k}-\omega_{s} \right)\left( i\gamma_{l} + \Omega_{k}-\omega_{s} \right)} - \frac{\hat{V}_{R2}\left[\Omega_{k}+\omega_{s}\right]-\hat{V}_{R1}\left[\Omega_{k}+\omega_{s}\right]}{-\omega_{a}^{2}+\left( \Omega_{k}+\omega_{s} \right)\left( i\gamma_{l} + \Omega_{k}+\omega_{s} \right)} \right),
	\end{split}\label{Eq:CoeffNoiseOpCompl}
	\end{align}
\end{subequations}
we obtain
\begin{equation}
	\dot{\hat{b}}(t) = \hat{b}(t)\left( - i\left( \omega_{m}+\omega_{b} \right) -\frac{\gamma_{b}+\Gamma_{b}}{2} \right) +i\frac{x_{0}}{\hbar}\hat{F}_{b}(t)+\frac{2 g_{1}\alpha}{\gamma_{t}+\gamma_{r}}\omega_{a}^{2}\sqrt{\frac{\gamma_{t}C_{0}}{T \omega_{s} \hbar}}\hat{N}(t). \label{Eq:DotBCompl}
\end{equation}
From Eq. \eqref{Eq:DotBCompl} it is now possible to obtain the exact solution for $\hat{b}(t)$ as a function of the quantum noises $\hat{F}_{b}(t)$ and $\hat{N}(t)$:
\begin{equation}
	\begin{split}
	\hat{b}(t) = \hat{b}(0) e^{-\left[\frac{\gamma_{b}+\Gamma_{b}}{2}+i\left( \omega_{m}+\omega_{b} \right)\right]t} + i \frac{x_{0}}{\hbar} \int_{0}^{t} e^{-\left[\frac{\gamma_{b}+\Gamma_{b}}{2}+i\left( \omega_{m}+\omega_{b} \right)\right](t-\tau_{1})} \hat{F}_{b}(\tau_{1})d\tau_{1} \\ +\frac{2 g_{1}\alpha}{\gamma_{t}+\gamma_{r}}\omega_{a}^{2}\sqrt{\frac{\gamma_{t}C_{0}}{T \omega_{s} \hbar}} \int_{0}^{t} e^{-\left[\frac{\gamma_{b}+\Gamma_{b}}{2}+i\left( \omega_{m}+\omega_{b} \right)\right](t-\tau_{1})}\hat{N}(\tau_{1}) d\tau_{1},
	\end{split}\label{Eq:BCompl}
\end{equation}
where $\hat{b}(0)$ is the annihilation operator at the initial time $t=0$. From here one can verify that the mechanical creation and annihilation operators satisfy the standard commutation relation, $\left[\hat{b}(t),\hat{b}^{\dagger}(t)\right]=1$. 

With the above results, we can finally determine $n_{b}(t)$ to be
\begin{equation}
\begin{split}
n_{b} & (t) = n_{b}(0)e^{-(\gamma_{t}+\gamma_{r})t} \\ & + \frac{1}{2\hbar m \omega_{m}} \int_{0}^{t}\int_{0}^{t} e^{-\frac{\gamma_{b}+\Gamma_{b}}{2}\left( 2t-\tau_{1}-\tau_{2} \right)}e^{-i\left( \omega_{m}+\omega_{b} \right)\left( \tau_{1}-\tau_{2} \right)} \left\langle \hat{F}^{\dagger}_{b}(\tau_{1})\hat{F}_{b}(\tau_{2}) \right\rangle d \tau_{1} d\tau_{2} \\ & +4 \frac{g_{1}^{2}\lvert\alpha\rvert^{2}\omega_{a}^{4}}{\left(\gamma_{t}+\gamma_{r}\right)^{2}}\frac{\gamma_{r} C_{0}}{T \omega_{s} \hbar }\int_{0}^{t}\int_{0}^{t} e^{-\frac{\gamma_{b}+\Gamma_{b}}{2}\left( 2t-\tau_{1}-\tau_{2} \right)}e^{-i\left( \omega_{m}+\omega_{b} \right)\left( \tau_{1}-\tau_{2} \right)} \left\langle \hat{N}(\tau_{1})\hat{N}(\tau_{2}) \right\rangle d \tau_{1} d\tau_{2}.
\end{split} \label{Eq:NbTimeEvCompl}
\end{equation}
At this stage, the whole problem is reduced to computing the variances of the mechanical noise $\left\langle \hat{F}^{\dagger}_{b}(\tau_{1})\hat{F}_{b}(\tau_{2}) \right\rangle$ and the electric fields $\left\langle \hat{N}(\tau_{1})\hat{N}(\tau_{2}) \right\rangle$. As a consequence of our assumption $\gamma_{b} \ll \omega_{m}$, we can neglect off--resonant contributions in $\left\langle \hat{F}^{\dagger}_{b}(\tau_{1})\hat{F}_{b}(\tau_{2}) \right\rangle$ and write it as
\begin{equation}
\left\langle \hat{F}^{\dagger}_{b}(\tau_{1})\hat{F}_{b}(\tau_{2}) \right\rangle \simeq 2 \hbar m \omega_{m} \gamma_{b}\bar{n}_{m}\delta(\tau_{2}-\tau_{1}). \label{Eq:VarMechNoiseCompl}
\end{equation}
The determination of the other variance, $\left\langle \hat{N}(\tau_{1})\hat{N}(\tau_{2}) \right\rangle$, is more involved. In general, taking the definition of $\hat{N}(t)$ in Eqs. \eqref{Eq:NoiseCompl} and substituting it into expression for the correlator, one encounters a number of products of the form
\begin{equation}
\frac{\left\langle \hat{V}_{Ri}\left[ \Omega_{h} \right]\hat{V}_{Rj}\left[ \Omega_{m} \right] \right\rangle e^{-i \Omega_{h}t}e^{-i\Omega_{m}t}}{\text{Den}\left(\Omega_{h}, \Omega_{m}\right)}, \label{Eq:ContrVarNoiseCompl}
\end{equation}
where $i,j=1,2$, and $\Omega_{h}$, $\Omega_{m}$ are generic Fourier variables. The denominators $\text{Den}\left(\Omega_{h}, \Omega_{m}\right)$ can be determined from Eq. \eqref{Eq:CoeffNoiseOpCompl}. Considering that
\begin{subequations}\label{Eq:VarianceElNoise}
	\begin{align}
	\left\langle \hat{V}_{Ri}\left[ \Omega_{h} \right]\hat{V}_{Rj}\left[ \Omega_{m} \right] \right\rangle = & 0, \;\;\; \forall i\neq j, \label{Eq:VarianceElNoiseDiff}\\
	\begin{split}
	\left\langle \hat{V}_{Ri}\left[ \Omega_{h} \right]\hat{V}_{Ri}\left[ \Omega_{m} \right] \right\rangle = & \frac{\hbar \Omega_{h} R}{2}\bar{n}_{e}(\Omega_{h},T_{e})\delta\left(\Omega_{h}+\Omega_{m}\right)\theta\left(-\Omega_{h}\right) \\ & +\frac{\hbar \Omega_{h} R}{2}\left[\bar{n}_{e}(\Omega_{h},T_{e})+1\right]\delta\left(\Omega_{h}+\Omega_{m}\right)\theta\left(\Omega_{h}\right),
	\end{split}\label{Eq:VarianceElNoiseSame}
	\end{align}
\end{subequations}
with $\bar{n}_{e}(\Omega_{h},T_{e})$ being the occupation number of the electrical reservoir at the considered frequency $\Omega_{h}$ and temperature $T_{e}$, we get
\begin{equation}
\begin{split}
\left\langle \hat{N}(\tau_{1})\hat{N}(\tau_{2}) \right\rangle = & \frac{(\omega_{s}-\omega_{m})\hbar R \bar{n}_{e}\left(\omega_{s}-\omega_{m},T_{e}\right)\delta\left(\tau_{2}-\tau_{1}\right)}{\omega_{a}^{4}-2\omega_{a}^{2}\left(\omega_{m}-\omega_{s}\right)^{2}+\left(\omega_{m}-\omega_{s}\right)^{2}\left(\gamma_{l}^{2}+\left(\omega_{m}-\omega_{s}\right)^{2}\right)} \\  & +\frac{(\omega_{s}+\omega_{m}) \hbar R \bar{n}_{e}\left(\omega_{s}+\omega_{m},T_{e}\right)\delta\left(\tau_{2}-\tau_{1}\right)}{\omega_{a}^{4}-2\omega_{a}^{2}\left(\omega_{m}+\omega_{s}\right)^{2}+\left(\omega_{m}+\omega_{s}\right)^{2}\left(\gamma_{l}^{2}+\left(\omega_{m}+\omega_{s}\right)^{2}\right)}.
\end{split} \label{Eq:VarianceNoisesComp}
\end{equation}
We have here assumed both resistances $R$ to be at the same temperature $T_{e}$, and neglected fast--oscillating contributions to $\left\langle \hat{N}(\tau_{1})\hat{N}(\tau_{2}) \right\rangle$.  In Eq. \eqref{Eq:VarianceElNoiseSame}, we also denoted the Heaviside step function with the letter $\theta$.

With this last result, we are finally able to determine the time evolution for the average phonon number operator $n_{b}(t)$. In fact, using equations \eqref{Eq:VarMechNoiseCompl} and \eqref{Eq:VarianceNoisesComp}, it is possible to calculate the integrals in Eq. \eqref{Eq:NbTimeEvCompl} and therefore obtain a clear analytical expression for $n_{b}(t)$. In its most general form, the expression is rather involved. For simplicity, we restrict ourselves to a reasonable approximation that allows us to better understand the final result. Considering the limit $\omega_{a}\gg\omega_{s}\gg\omega_{m}$, from Eqs. \eqref{Eq:GammaB} and \eqref{Eq:OmegaShift} with $\Omega_{k}\rightarrow\omega_{m}$, we can rewrite the effective decay $\Gamma_{b}$ and the frequency shift $\omega_{b}$ to be
\begin{subequations}
	\begin{align}
	\Gamma_{b} & = \frac{8 g_{1}^{2} \lvert\alpha\rvert^{2} \gamma_{l}\gamma_{t}\omega_{m}}{T\left(\gamma_{t}+\gamma_{r}\right)^{2}\omega_{s}\omega_{a}^{2}}, \label{Eq:GammaBApp}\\
	\omega_{b} & = -\frac{4 g_{1}^{2}\lvert\alpha\rvert^{2}\gamma_{t}}{T\left(\gamma_{t}+\gamma_{r}\right)^{2}\omega_{s}}.	\label{Eq:OmegaShiftApp}
	\end{align}
\end{subequations}
From now on, referring to $\Gamma_{b}$ and $\omega_{b}$, we implicitly consider the form given in these last two equations. Note, however, that Eq. \eqref{Eq:GammaB} describes the damping due to the difference in sideband strengths for a mode at frequency $\omega_{a}$ driven at frequency $\omega_{s}$. In the limit where $\omega_{a}$ is large, this induced damping is very small. The shift of Eq. \eqref{Eq:OmegaShift} may be sizeable, but merely leads to a new mechanical resonance frequency, which is not important for the present discussion. Furthermore, as a consequence of $\omega_{s} \gg \omega_{m}$, we assume that $\bar{n}_{e}(\omega_{s}-\omega_{m},T_{e})$ and $\bar{n}_{e}(\omega_{s}+\omega_{m},T_{e})$ are approximatively equal in Eq. \eqref{Eq:VarianceNoisesComp}, such that
\begin{equation}
\bar{n}_{e}(\omega_{s}\pm\omega_{m},T_{e})\simeq\bar{n}_{e}(\omega_{s},T_{e})\equiv\bar{n}_{e}.
\end{equation}
Putting all these results together we can finally derive $n_{b}(t)$ to be
\begin{equation}
\begin{split}
n_{b}(t) = n_{b}(0)e^{-(\gamma_{b}+\Gamma_{b})t} + \frac{\gamma_{b}}{\gamma_{b}+\Gamma_{b}}\bar{n}_{m}\left(1-e^{-(\gamma_{b}+\Gamma_{b})t}\right) + \frac{\Gamma_{b}}{\gamma_{b}+\Gamma_{b}}\frac{\omega_{s}}{\omega_{m}}\left(\bar{n}_{e}+\frac{1}{2}\right)\left(1-e^{-(\gamma_{b}+\Gamma_{b})t}\right).
\end{split}
\label{Eq:NbTimeComplete}
\end{equation}

As a conclusive part of this discussion, let us determine $\Delta n_{b}$. Recalling that $\Delta n_{b}$ is the change in the phonon number for the membrane initially cooled in its ground state, we get
\begin{equation}
\Delta n_{b}= \gamma_{b}\bar{n}_{m}T+\frac{\omega_{s}}{\omega_{m}}\Gamma_{b} \left(\bar{n}_{e}+\frac{1}{2}\right)T, \label{Eq:DeltaNbCompl}
\end{equation}
where we truncated the expansion at first order in $T$. This is the parameter used as denominator in $\lambda_{b}=D^{2}/\Delta n_{b}$, presented in the main text neglecting the contribution from the mechanical reservoir.

The parameter $\Delta n_{b}$ derived above takes into account \textit{only} the heating induced by the redistribution of charges on the capacitor. Asymmetries are well described analytically by considering a linear coupling $\delta g_{1}$ and capacitance $\delta C$ such that $C^{-1}(\pm\hat{x})\propto (C_{0}\pm \delta C)^{-1} \pm (g_{1} \pm \delta g_{1})\hat{x} +g_{2} \hat{n}_{b}$. These are the only asymmetries directly affecting the electromechanical coupling; other ones only enter as higher order perturbation, as we will see later in subsection \ref{sec:HeatingSim}. In this case we can find an \textit{overall} residual linear coupling $g_{r}=\frac{1}{2}C_{0}x_{0}\omega_{s}\partial_{x}\left[C(x)+C(-x)\right]^{-1}> 0$ of the membrane equal to
\begin{equation}
g_{r} = \delta g_{1} + 2 \frac{g_{1} \delta C}{C_{0}} + \frac{\delta g_{1} \delta C^{2}}{C_{0}^{2}}.\label{Eq:GeffPar}
\end{equation}
Under the assumption that the parasitic elements $R$ and $L$ of the circuit are smaller than the non-parasitic ones $R_{0}$ and $L_{0}$, we can neglect the coupling to the antisymmetric mode to lowest order (this mode is far off resonant, while the symmetric one is resonant). The heating is then dominated by the residual linear coupling $g_{r}$, similarly to the situation encountered above in subsection \ref{sec:HeatingSimple}. We can thus use the result in Eq. \eqref{Eq:PhononNumbEvAppSimple} obtained for the $RLC$ circuit to identify the contribution to the heating:
\begin{equation}
n_{b}(t)=\frac{\tilde{\Gamma}_{b}}{\gamma_{b}+\Gamma_{b}} (1+2\bar{n}_{e})\left(1-e^{-(\gamma_{b}+\Gamma_{b})t}\right). \label{Eq:HeatAsyCon}
\end{equation}
Here, $\tilde{\Gamma}_{b}$ has been defined from $\Gamma_{b}$ in Eq. \eqref{Eq:GammaBSimple} by substituting $g_{1}$ with $g_{r}$: 
\begin{equation}
\tilde{\Gamma}_{b} = \frac{4 g_{r}^{2} \lvert\alpha\rvert^{2} \gamma_{t}}{T(\gamma_{t}+\gamma_{r})\left[ (\gamma_{t}+\gamma_{r})^{2} + 4 \omega_{m}^{2} \right]}.
\end{equation}
From Eq. \eqref{Eq:HeatAsyCon} we can determine $\Delta n_{b}$,
\begin{equation}\label{Eq:HeatingPar}
\Delta n_{b} = \tilde{\Gamma}_{b}(1+2\bar{n}_{e})T,
\end{equation}
being the denominator of the parameter $\lambda_{p}=D^{2}/\Delta n_{b}$ presented in the main text for $\delta C\rightarrow 0 $.

Putting together Eqs. \eqref{Eq:NbTimeComplete} and \eqref{Eq:HeatAsyCon}, we can find the time evolution of the average phonon number $n_{b}(t)$, taking into account all sources of heating:
\begin{equation}
n_{b}(t) =n_{b}(0)e^{-(\gamma_{b}+\Gamma_{b})t} + \left[ \gamma_{b}\bar{n}_{m} + \left(\Gamma_{b}\frac{\omega_{s}}{2\omega_{m}} + \tilde{\Gamma}_{b} \right)\left(1+2\bar{n}_{e}\right)\right]\frac{1-e^{-(\gamma_{b}+\Gamma_{b})t}}{\gamma_{b}+\Gamma_{b}}. \label{Eq:PhononEvCompl}
\end{equation}
This result will be supported by our simulations, presented in subsection \ref{sec:HeatingSim}.

\subsection{Heating simulation} \label{sec:HeatingSim}

In this subsection we present two methods for simulating the membrane heating, described by $n_{b}(t)$. For the ``double arm'' circuit in Fig. \ref{Fig:FigS4} we cannot employ the wave function Monte--Carlo approach \cite{Dalibard1992} that was used for Fig. \ref{Fig:FigS3}. Quantum jump simulations apply to Markovian reservoirs, and thereby assume that there is the same number of thermal photons for all frequencies. In our case, there is a big difference in the resonance frequencies of the symmetric $\omega_{s}$ and antisymmetric $\omega_{a}$ electrical modes. Therefore, we need to address the frequency dependence of the reservoir occupation. At a given temperature $T_{e}$ of the electrical reservoirs, the antisymmetric mode will come to equilibrium with $\bar{n}_{e}(\omega_{a},T_{e}) = [\exp(\hbar \omega_{a}/k_{B} T_{e})-1]^{-1}$ photons at frequency $\omega_{a}$. On the other hand, the coupling between the mechanical oscillator and the antisymmetric field is mediated by the driving, which is resonant with the symmetric mode. Hence, the membrane couples to a reservoir containing $\bar{n}_{e}(\omega_{s},T_{e})\gg \bar{n}_{e}(\omega_{a},T_{e})$ photons, as indicated in Eq. \eqref{Eq:NbTimeComplete}. Therefore, more advanced methods are required for simulating the present situation. 

We will investigate the approximations taken for deriving Eq. \eqref{Eq:PhononEvCompl} on two different levels. First, we consider Eqs. \eqref{Eq:EOMQBCompLinFou} and solve them in the Fourier domain, keeping the off-resonant terms that have been ignored in the derivation of Eq. \eqref{Eq:NbTimeComplete} above. Second, we allow for any possible deviation from the balanced circuit in Fig. \ref{Fig:FigS4}, by introducing unbalanced parasitic resistances $R_{1}$ and $R_{2}$, inductances $L_{1}$ and $L_{2}$, rest capacitances $C_{1}$ and $C_{2}$, and linear couplings $g_{1}+\delta g_{1}$ and $g_{1}-\delta g_{1}$. In the following, since we are interested in the time evolution of the mechanical subsystem for times longer than $T$, we will explicitly consider a time variable $\tau > T$ in the Fourier expansion of Eq. \eqref{Eq:FourierDef}

Let us start our analysis by rewriting Eqs. \eqref{Eq:EOMQBCompLinFou} with the creation $\hat{a}^{\dagger}$ and annihilation $\hat{a}$ operators for the electrical charge $\hat{Q}_{a}$ and flux $\hat{\Phi}_{a}$:
\begin{subequations}\label{Eq:FourierComponentsSim}
	\begin{align}
	\begin{split}
	\hat{a}[\Omega_{k}]\left( -i \Omega_{k} - i \omega_{a} + \frac{\gamma_{l}}{2} \right) =&  \frac{\gamma_{l}}{2}\hat{a}^{\dagger}[\Omega_{k}] -i \frac{g_{1} \alpha }{\gamma_{r}+\gamma_{t}}\sqrt{\frac{\gamma_{t} \omega_{a}}{T \omega_{s}}}\left( \hat{b}[\Omega_{k}-\omega_{s}] + \hat{b}^{\dagger}[\Omega_{k}-\omega_{s}] -\hat{b}[\Omega_{k}+\omega_{s}]- \hat{b}^{\dagger}[\Omega_{k}+\omega_{s}] \right) \\ & +i \sqrt{\frac{C_{0} \omega_{a}}{\hbar}}\left( \hat{V}_{R2}[\Omega_{k}]-\hat{V}_{R1}[\Omega_{k}] \right) + \frac{\hat{a}(t=0)}{\sqrt{\tau}} -  \frac{\hat{a}(t=\tau)}{\sqrt{\tau}},
	\end{split}\label{Eq:AFouSimCom}\\
	\begin{split}
	\hat{b}[\Omega_{k}]\left( -i \Omega_{k} - i \omega_{m} + \frac{\gamma_{b}}{2} \right) = & -i\frac{g_{1} \alpha }{\gamma_{r}+\gamma_{t}}\sqrt{\frac{\gamma_{t} \omega_{a}}{T \omega_{s}}}\left( \hat{a}[\Omega_{k}-\omega_{s}] + \hat{a}^{\dagger}[\Omega_{k}-\omega_{s}] -\hat{a}[\Omega_{k}+\omega_{s}] - \hat{a}^{\dagger}[\Omega_{k}+\omega_{s}] \right) \\ & +i \frac{x_{0}}{\hbar}\hat{F}_{b}[\Omega_{k}]	+ \frac{\hat{b}(t=0)}{\sqrt{\tau}} -  \frac{\hat{b}(t=\tau)}{\sqrt{\tau}},
	\end{split}	\label{Eq:BFouSimCom}
	\end{align}
\end{subequations}
where $\hat{Q}_{a} = \sqrt{\hbar C_{0} \omega_{a}}(\hat{a}+\hat{a}^{\dagger})/2$ and $\hat{\Phi}_{a} = i\sqrt{\hbar/(C_{0} \omega_{a})}(\hat{a}^{\dagger}-\hat{a})$. Note that in Eqs. \eqref{Eq:FourierComponentsSim} we have ignored $g_{2}$ contributions ($g_{2} \ll g_{1}$), and added boundary terms, that are required to describe non--periodic dynamics in the Fourier series. Without those, we would obtain the steady state evolution, that cannot capture the thermalisation from the ground or any other state than the equilibrium. The additional terms can be adjusted to represent the desired initial state.

\begin{figure}[htbp]
	\centering
	\includegraphics[width=17 cm]{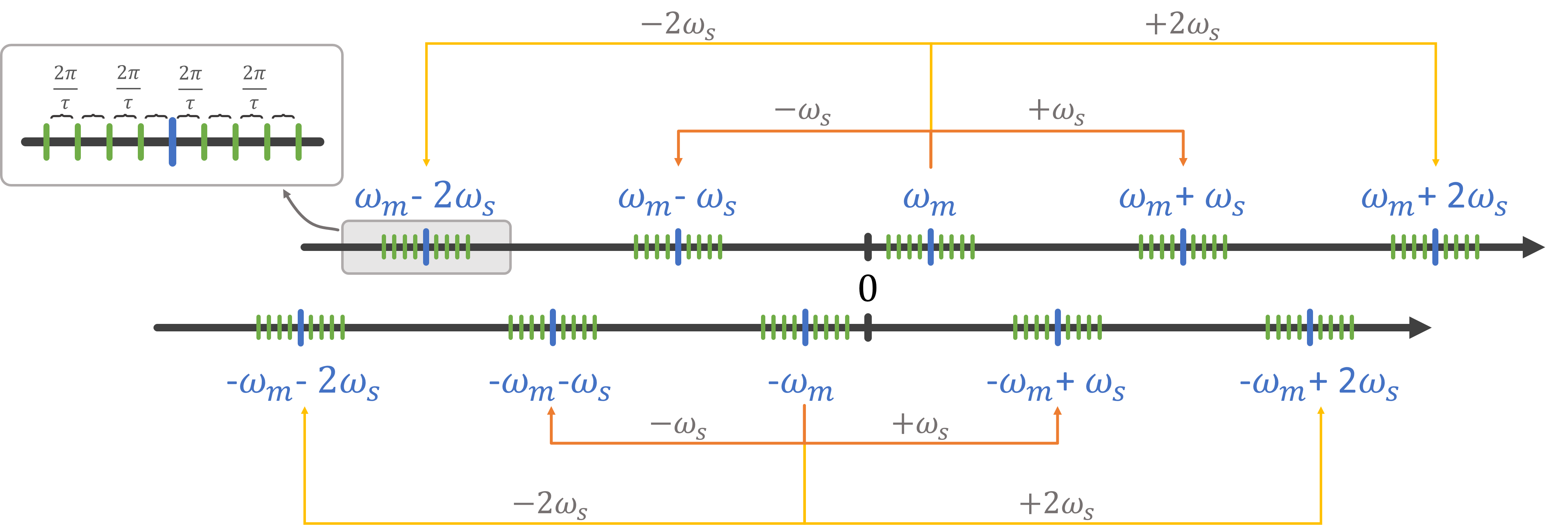}
	\caption[FigS5]{Example for a possible set of relevant frequencies in the truncated Fourier series of Eq. \eqref{Eq:TruncatedFourier}. In this particular case, we choose $N_{j}=2$ and $N_{f}=4$. In the \textbf{upper arrow} the frequencies obtained starting from $\omega_{m}$, in the \textbf{lower arrow} from $-\omega_{m}$.}
	\label{Fig:FigS5}
\end{figure}

\begin{figure}[htbp]
	\centering
	\includegraphics[width=18 cm]{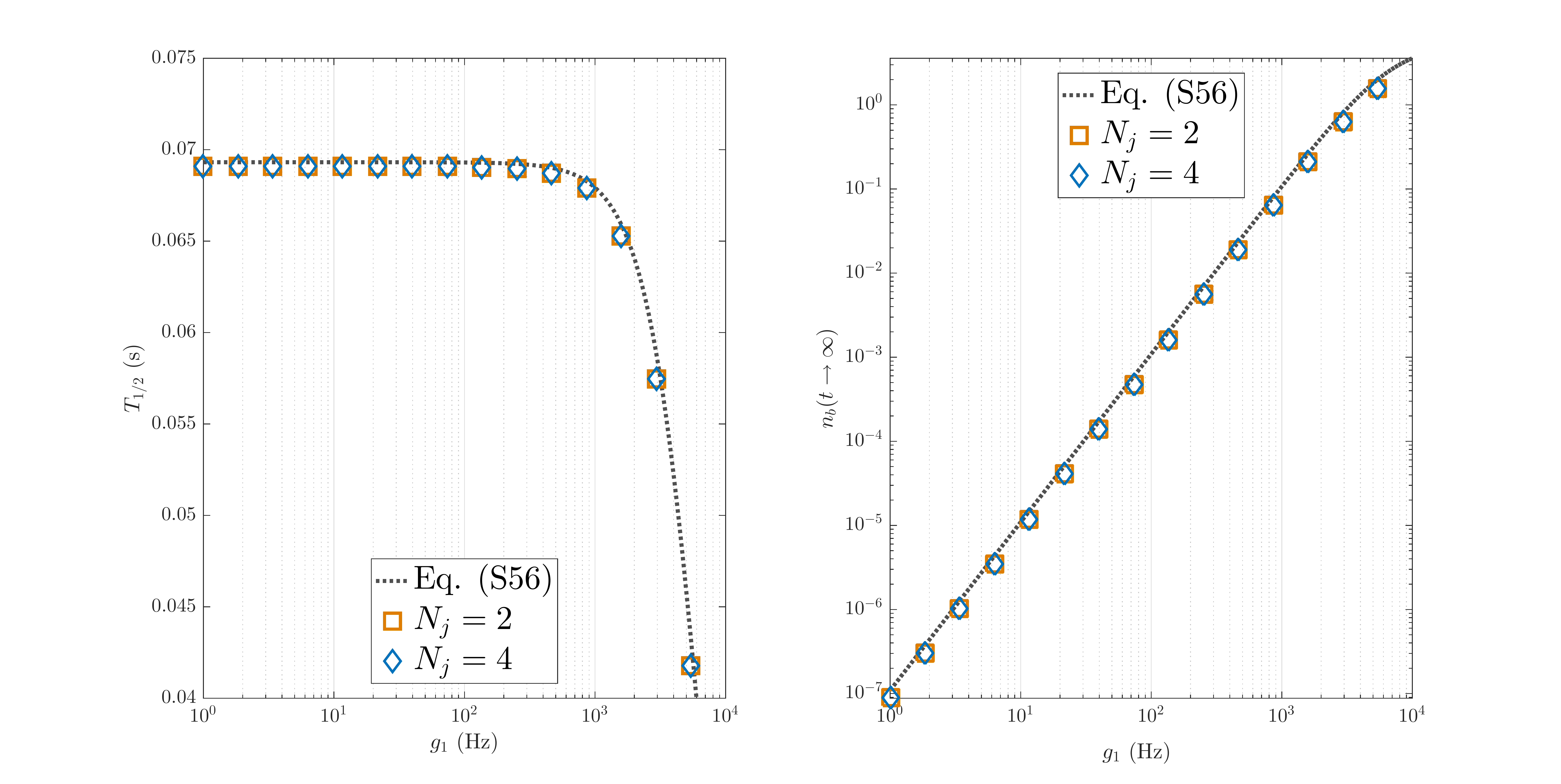}
	\caption[FigS6]{Comparison between Eq. \eqref{Eq:NbTimeComplete} and simulations. In the \textbf{left plot}, we show $T_{1/2}$ versus the linear coupling $g_{1}$, with $T_{1/2}$ defined by the relation $n_{b}(T_{1/2})=n_{b}(t\rightarrow\infty)/2$. In the \textbf{right plot}, the average mechanical steady state population $n_{b}(t\rightarrow\infty)$ is presented, as a function of $g_{1}$. Dotted lines are the analytical predictions obtained from Eq. \eqref{Eq:NbTimeComplete}. Red squares and blue diamonds come from simulations. The first one corresponds to the case where we allow for only one sideband ($N_{j}=2$), and the second for two ($N_{j}=4$). As it is possible to see, two--sidebands corrections are negligible. We assumed $L/L_{0}=10^{-4}$, $R/Z_{out}=0.1$, $\omega_{s} =(2\pi) 3$ GHz, $\omega_{m} = (2\pi)314$ MHz, $\gamma_{t}\simeq\gamma_{r}=(2\pi)100$ kHz, $\gamma_{b} = (2\pi)10$ Hz, $\lvert\alpha\rvert^{2}/\tau=10^{12}$ s$^{-1}$ and $\bar{n}_{e}=\bar{n}_{m}=0$ \cite{Quinta}. Here, $\tau$ is the arbitrary time of the Fourier series in Eq. \eqref{Eq:FourierDef}.}
	\label{Fig:FigS6}
\end{figure}

In subsection \ref{sec:HeatComplBalanced} we truncated the Fourier series by keeping only the terms resonant with the mechanical frequency $\omega_{m}$. Instead, we will now include additional terms at frequencies centred around $\pm \omega_{m}$,
\begin{equation}
\hat{O}(t) = \sum_{k=-N_{j}}^{N_{j}}\sum_{l=-N_{f}}^{N_{f}}\hat{O}\left[\omega_{m}+k \omega_{s}+l\frac{2 \pi}{\tau}\right]\frac{e^{-i\left( \omega_{m} + k \omega_{s} + l\frac{2 \pi}{\tau} \right)}}{\sqrt{\tau}} + \hat{O}\left[-\omega_{m}-k \omega_{s}-l\frac{2 \pi}{\tau}\right]\frac{e^{i\left( \omega_{m} + k \omega_{s} + l\frac{2 \pi}{\tau} \right)}}{\sqrt{\tau}},\label{Eq:TruncatedFourier}
\end{equation}
where $\hat{O}$ can be any operator $\hat{a}^{(\dagger)}$ or $\hat{b}^{(\dagger)}$, and $k$, $l$ are integers corresponding to the expansion. In Eq. \eqref{Eq:TruncatedFourier}, the parameter $N_{j}$ indicates how many sidebands we consider, while $N_{f}$ is related with the convergence of the Fourier series. By setting $N_{j}=0$, we are assuming that the mechanical and electrical subsystems are completely decoupled, while $N_{j}=2$ is the case studied above analytically. Corrections are given for $N_{j}>2$. An example of the considered frequencies is illustrated in Fig. \ref{Fig:FigS5} for $N_{j}=2$ and $N_{f}=4$.
%%%
%%%
The Fourier components $\hat{a}\left[\pm\left(\omega_{m}+k \omega_{s}+l\frac{2 \pi}{\tau}\right)\right]$ and $\hat{b}\left[\pm\left(\omega_{m}+k \omega_{s}+l\frac{2 \pi}{\tau}\right)\right]$ are then derived from the set of coupled linear equations \eqref{Eq:FourierComponentsSim}, that can be solved efficiently using the matrix notation. A comparison between the result of Eq. \eqref{Eq:NbTimeComplete} and the so--derived simulations is presented in Fig. \ref{Fig:FigS6}. In the left plot we present $T_{1/2}$ for different values of the linear coupling $g_{1}$. Here, $T_{1/2}$ is the time for which the average mechanical occupation is half the one of the steady state $n_{b}(t\rightarrow\infty)$. In the right plot, it is possible to see the $g_{1}$--dependence of the mechanical steady state population. As it is possible to see, the simulations converge to the analytical curve of Eq. \eqref{Eq:NbTimeComplete} independently from the parameter $N_{j}$. This means that off-resonant terms do not play a major role, and that the approximations taken for deriving Eq. \eqref{Eq:NbTimeComplete} are appropriate. The small discrepancy between analytical and simulated results is dominated by a poor convergence of the Fourier series. By taking bigger $N_{f}$, this discrepancy would be eventually removed.
%%%
%%%

\begin{figure}[htbp]
	\centering
	\includegraphics[width=16 cm]{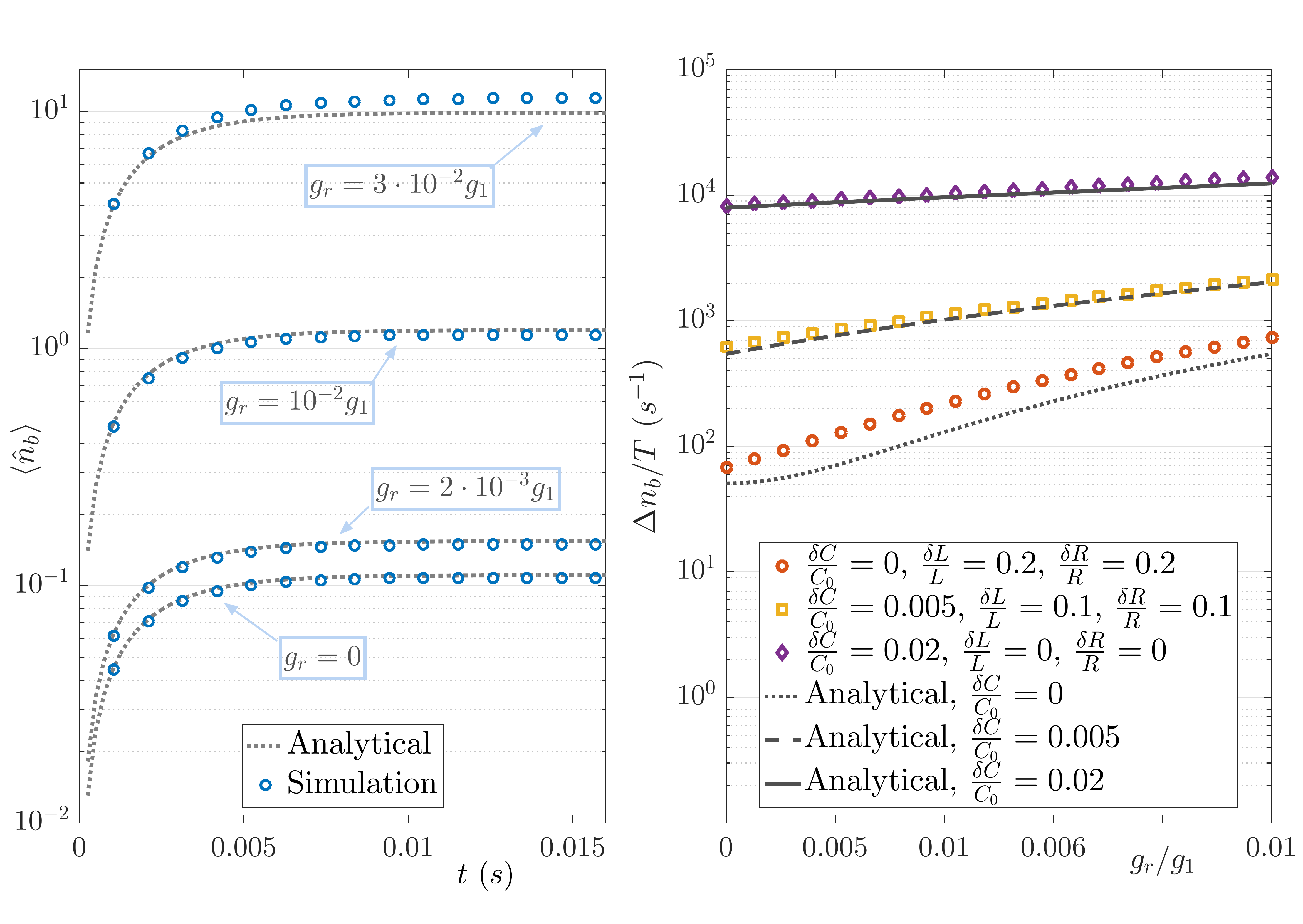}
	\caption[FigS7]{\textbf{Left} plot: Comparison between the analytical curves given by Eq. \eqref{Eq:PhononEvCompl} and the full simulations of the system \eqref{Eq:EOMunbalanced}. From the bottom to the top we have set $\delta g_{1}/g_{1}$ to be $0$, $2\cdot 10^{-3}$, $10^{-2}$ and $3\cdot 10^{-2}$. We use $\delta R = \delta L = \delta C = 0$. \textbf{Right} plot: Heating rate $h=\partial_{t}n_{b}(t)=\Delta n_{b}/T$ as a function of the normalised residual linear coupling $\delta g_{1}/g_{1}$. Here we test Eq. \eqref{Eq:PhononEvCompl} in presence of asymmetries in the parasitic elements of the circuit. The three dark grey lines are the analytical predictions for $\delta C/C_{0}$ being equal to $0$ (dotted), $0.05$ (dashed) and $0.2$ (plain). The circles, squares and diamonds are the simulated results for the values $\delta R/R$, $\delta L /L$, and $\delta C/C$ reported in the legend. We have assumed $L/L_{0}=10^{-2}$, $R/Z_{out} = 10^{-1}$, $\omega_{s}=(2\pi)7$ GHz, $\omega_{m} = (2\pi)80$ MHz, $\gamma_{r}\simeq\gamma_{t}=(2\pi)0.15$ MHz, $\gamma_{b} = (2\pi)80$ Hz, $g_{1}=(2\pi)7$ kHz, and $\bar{n}_{e} = \bar{n}_{m} = 0$. The constant photon flux $\lvert\tilde{\alpha}\rvert^{2}=1.15\cdot 10^{15}$ s$^{-1}$ has been chosen to have $n_{b}(t\rightarrow\infty)=1$ for $\delta g_{1}/g_{1}=10^{-2}$ in the left plot.}
	\label{Fig:FigS7}
\end{figure}

In the second part of this subsection we simulate the electromechanical system presented in Fig. \ref{Fig:FigS4}, allowing all the parasitic elements to differ between the two arms. In particular, we introduce deviations for resistances $R\rightarrow R\pm \delta R$, inductances $L\rightarrow L\pm \delta L$, bare capacitances $C_{0}\rightarrow C_{0}\pm \delta C$, and the linear couplings $g_{1}\rightarrow g_{1}\pm \delta g_{1}$ in the left and right arms. Once we have done this, following the standard approach presented above we can derive the equations of motion of our electromechanical setup to be:
\begin{subequations}\label{Eq:EOMunbalanced}
	\begin{align}
	\begin{split}
	\dot{\hat{Q}}_{a} = & \frac{\hat{\Phi}_{a}}{2 L}+\frac{\delta L^{2} \hat{\Phi}_{a}-2L \delta L \hat{\Phi}_{s}}{2L\left( L^{2}+2L L_{0}-\delta L^{2} \right)},
	\end{split}\label{Eq:EOMunbalancedQa}\\
	\begin{split}
	\dot{\hat{\Phi}}_{a} = & -\frac{2\hat{Q}_{a}}{C_{0}}-\frac{g_{1}}{C_{0}\omega_{s}}\hat{Q}_{s}\left( \hat{b} + \hat{b}^{\dagger} \right) - \frac{2 g_{2}}{C_{0}\omega_{s}}\hat{Q}_{a} \hat{b}^{\dagger} \hat{b} -\gamma_{l}\hat{\Phi}_{a} + 2 \left( \hat{V}_{R2}-\hat{V}_{R1} \right) - \frac{2 \delta C^{2} \hat{Q}_{a} }{C_{0} \left( C_{0}^{2}-\delta C^{2} \right)} + \frac{\delta C \hat{Q}_{s} }{ C_{0}^{2}-\delta C^{2}} \\ & - \frac{2 \delta g_{1}}{C_{0} \omega_{s}}\hat{Q}_{a}\left( \hat{b} + \hat{b}^{\dagger} \right) + \frac{\delta L ( -R \delta L + L \delta R )}{L\left( L^{2}+2L L_{0} - \delta L^{2} \right)} \hat{\Phi}_{a} + \frac{2R \delta L -2 L \delta R}{L^{2} + 2L L_{0} - \delta L^{2}} \hat{\Phi}_{s},
	\end{split}	\label{Eq:EOMunbalancedPhia} \\
		\begin{split}
	\dot{\hat{Q}}_{s} = & \frac{2\hat{\Phi}_{s}}{L + 2 L_{0}} - \frac{\delta L \hat{\Phi}_{a} }{\left( L^{2}+2L L_{0}-2\delta L^{2} \right)} + \frac{2 \delta L^{2} \hat{\Phi}_{s} }{(L + 2 L_{0})\left( L^{2}+2L L_{0}-2\delta L^{2} \right)},
	\end{split}\label{Eq:EOMunbalancedQs}\\
	\begin{split}
	\dot{\hat{\Phi}}_{s} = & -\frac{\hat{Q}_{s}}{2 C_{0}}-\frac{g_{1}}{C_{0}\omega_{s}}\hat{Q}_{a}\left( \hat{b} + \hat{b}^{\dagger} \right) - \frac{2 g_{2}}{C_{0}\omega_{s}}\hat{Q}_{s} \hat{b}^{\dagger} \hat{b} -(\gamma_{r}+\gamma_{t}) \hat{\Phi}_{s} +2 \left( \hat{V}_{in} + \hat{V}_{R0} \right) + \hat{V}_{R1} + \hat{V}_{R2} \\ & + \frac{\delta C \hat{Q}_{a} }{C_{0}^{2}-\delta C^{2}} - \frac{\delta C^{2} \hat{Q}_{s} }{2C_{0}\left( C_{0}^{2}-\delta C^{2} \right)} - \frac{ \delta g_{1}}{2 C_{0} \omega_{s}}\hat{Q}_{s}\left( \hat{b} + \hat{b}^{\dagger} \right) + \frac{R \delta L +2(R_{0}+Z_{out}) \delta L -(L+2L_{0})\delta R }{2\left( L^{2}+2L L_{0} - \delta L^{2} \right)} \hat{\Phi}_{a} \\& + \frac{\delta L \left[ \delta R (L+2L_{0}) -\left( R+2(R_{0}+Z_{out})  \right) \delta L \right]}{\left( L+2L_{0} \right)\left( L^{2} + 2L L_{0} - \delta L^{2}\right)} \hat{\Phi}_{s},
	\end{split}	\label{Eq:EOMunbalancedPhis} \\
	\begin{split}
	\dot{\hat{b}} = & -i \omega_{m} \hat{b}-i\frac{g_{1} }{\hbar C_{0} \omega_{s}}\hat{Q}_{a}\hat{Q}_{s} - i \frac{g_{2}}{ \hbar C_{0} \omega_{s}}\left( \hat{b} + \hat{b}^{\dagger} \right)\left( \hat{Q}_{a}^{2} + \frac{\hat{Q}_{s}^{2}}{4} \right) -\frac{\gamma_{b}}{2} \hat{b} + i \frac{x_{0}}{\hbar} \hat{F}_{b} + \frac{i \delta g_{1} }{\hbar C_{0} \omega_{s}} \left( \hat{Q}_{a}^{2} + \frac{\hat{Q}_{s}^{2}}{4} \right).
	\end{split}	\label{Eq:EOMunbalancedB}
	\end{align}
\end{subequations}
For simulating this set of equations we follow a similar approach to the one presented in subsection \ref{sec:HeatComplBalanced}, neglecting off--resonant terms that do not contribute substantially to the mechanical heating, and the quadratic coupling (since $g_{2} \ll g_{1}$). In order to get a solution, we have linearised the symmetric and antisymmetric electrical fields, and the mechanical operators. As opposed to the previous investigations, now our simulations start at a specific time. We thus consider transient dynamics for the average mechanical operators, that we can set to have any initial value (generally zero, as for the ground state). Our simulations are thus performed with the mechanical average values being time dependent \cite{Linearization}. As it is possible to see from Fig. \ref{Fig:FigS7}, the agreement between the analytical results and the simulations is excellent, as far as the asymmetries $\delta R$, $\delta L$ and $\delta C$ are small enough. The system is more susceptible to relative increase of $\delta C$ as compared to $\delta R$ and $\delta L$. This is because $\delta R$ and $\delta L$ are asymmetries of parasitic elements of the circuit, while $\delta C$ affects the main (and only) capacitor. A non--vanishing value of $\delta C$ directly affects the residual linear coupling $g_{r}$ [see Eq. \eqref{Eq:GeffPar}] and thus has a larger influence, that is included in our analytical prediction of Eq. \eqref{Eq:PhononEvCompl}. Roughly speaking, whenever $\delta C/C$, $\delta R/R$ and $\delta L /L$ are smaller than $25\%$, the agreement between Eq. \eqref{Eq:PhononEvCompl} and the simulations is very good.
%%%
%%%

In summary, we have tested the assumptions made in the derivation of our analytical results. We have proven with the Fourier analysis that off--resonant terms do not contribute significantly to the mechanical heating, while simulations of Eqs. \eqref{Eq:EOMunbalanced} ensured that asymmetries play a secondary role in the electromechanical dynamics.  %Remarkably, even if throughout this subsection we referred to the derived results as ``simulation'', everything presented here is completely analytical. However, the expressions are so long that cannot be understood or, sometimes, even displayed by our machines. In this sense, we can describe our approach as ``numeritical'', something in between analytical and numerical. 

As a final comment, we note that the derivation of the parameter $D^{2}$ including asymmetries can be done analytically with the procedure introduced in subsections \ref{sec:QNDsimple} and \ref{sec:QNDcomplete}. We do not report the result here since including $\delta R$, $\delta L$ and $\delta C$ makes the expressions long and complicated. In principle, however, the parameter $\lambda$ for the most general antisymmetric setup can be derived analytically, with the simple expressions for $\lambda_{b}$ and $\lambda_{p}$ (given in the main text) being valid for small asymmetries.

\section{Measurement}\label{sec:Measure}

\begin{figure}[htbp]
	\centering
	\includegraphics[width=15 cm]{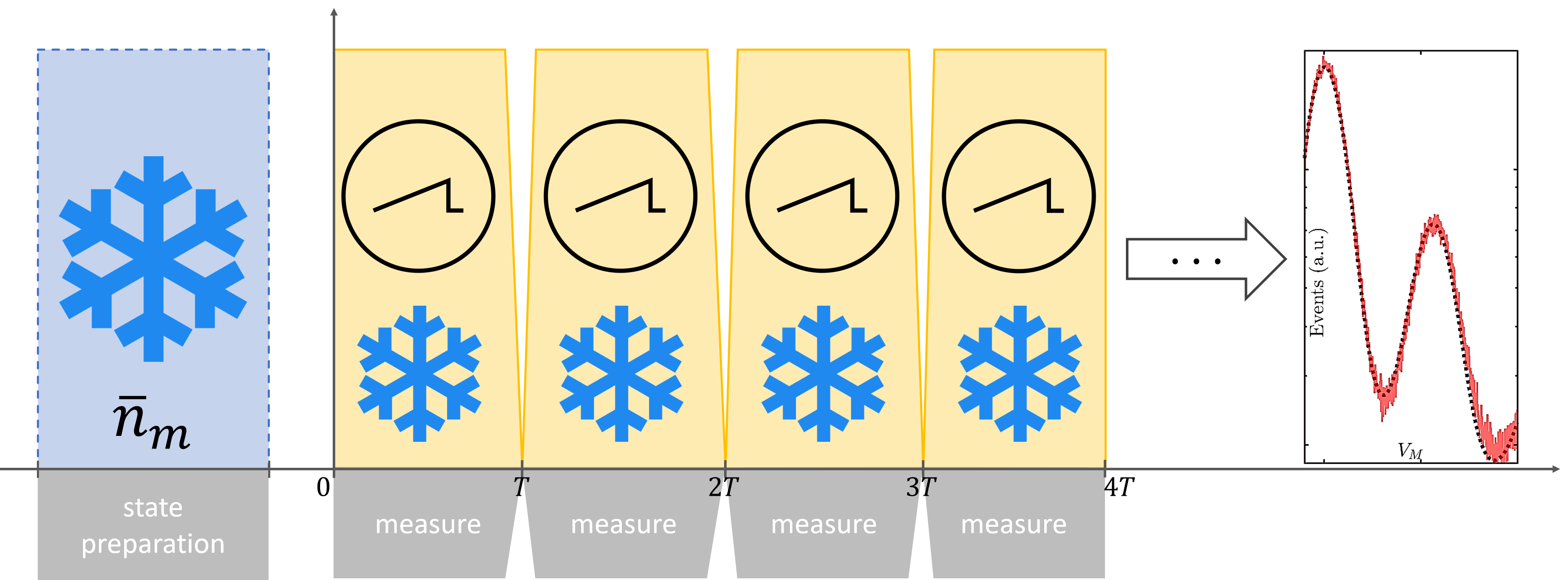}
	\caption[FigS8]{Considered measurement sequence. The mechanical state is cooled down to some average thermal phonon number $\bar{n}_{m}$, and then measured several times, while keeping the cooling active.}
	\label{Fig:FigS8}
\end{figure}

The parameter $\lambda$ specifies how suitable a specific experimental setup is for carrying out the QND detection. The outcome of an experimental run will, however, depend on the procedure used in the experiment. In this subsection, we consider the protocol presented in the main text and illustrated in Fig. \ref{Fig:FigS8}. We assume that the mechanical oscillator is prepared and continuously cooled to some average phonon number $\bar{n}_{m}$ during the entire measurement sequence. We determine the best experimental parameter $\Delta n_{b}$ to be used to maximize the visibility $\xi$ for a given value of $\lambda$. Finally, we compare the analytical model with numerical simulations.
%%%
%%%

First, we develop a model for the protocol presented in Fig. \ref{Fig:FigS8}, where we only allow for a single jump during the measurement time $T$. After the state preparation, the membrane is in a mixture of the Fock states $\hat{\rho}_{in} = \sum_{i} p_{i} \lvert i \rangle \langle i \rvert$, where $p_{i} = \bar{n}_{m}^{i}/\left[1+\bar{n}_{m}\right]^{i+1}$ is the probability to be in the $i$--th state. When we start probing the system, the measurement outcome $V_{M}$ follows a probability distribution that depends on three parameters: $\bar{n}_{m}$, $\Delta n_{b}$, and $D^{2}$. $\bar{n}_{m}$ determines the initial thermal state of the membrane, and fixes the probabilities $p_{i}$. $\Delta n_{b}$ determines the rate at which the membrane jumps out the ground state, and can be used to calculate the probability, for each Fock state in $\hat{\rho}_{in}$, to jump up or down. In particular, the ground state $\lvert 0 \rangle$ has a probability $p_{0} \left( 1 - \exp\left[ -\Delta n_{b} \right]\right)$ to jump up, while any other Fock state $\lvert i \rangle$ has probabilities $ p_{i}\left(1 - \exp\left[ -(i+1) \Delta n_{b} \right] \right) $ and $ p_{i}\left(1 - \exp\left[ -i \Delta n_{b}  (\bar{n}^{-1}_{m}+1) \right] \right) $ to jump up or down, respectively. Finally, $D^{2}$ sets the distance between measurement outcomes with different phonon occupations. Therefore, whenever the membrane remains in the same state during the whole duration $T$, the outcome $V_{M}$ is a Gaussian distributed random number with average $n_{b} d$ and variance $\sigma^{2} = d^{2}/D^{2}$. If, on the other side, the mechanical state changes during the measurement, then $V_{M}$ is again a Gaussian distributed random number with variance $\sigma^{2}$, but with an average value given by $T^{-1}\left[n^{(i)}_{b} T_{j} d + n^{(f)}_{b} (T - T_{j}) d\right]$. Here, $T_{j}$ is the (uniformly distributed) random time at which the jump happens, and $n^{(i)}_{b}$ and $n^{(f)}_{b}$ are the phonon numbers before and after the jump. Therefore, knowing $\Delta n_{b}$, $D^{2}$ and $\bar{n}_{m}$, the probability distribution function $PDF$ of the outcomes $V_{M}$ can be determined:
\begin{equation}
PDF(V_{M})=\lim\limits_{N_{p}\rightarrow\infty}\left\lbrace \sum_{i=0}^{N_{p}}\frac{P_{p}(i)}{\sqrt{2\pi}\sigma}e^{-\frac{(V_{M}-i d)^{2}}{2\sigma^{2}}} + \sum_{i=1}^{N_{p}}\int_{(i-1)d}^{i d}\frac{P_{r}(i-1;1)}{\sqrt{2\pi} d \sigma}e^{-\frac{(V_{M}-V)^{2}}{2\sigma^{2}}}dV \right\rbrace, \label{Eq:PDFvM}
\end{equation}
where, we recall, $D=d/\sigma$, and $\sigma^{2}=(\Delta V_{M})^{2}$ is a scaling factor. In this last equation, $P_{p}(i)$ is the probability for the mechanical state to remain in $\lvert i \rangle$ for the whole measurement, while $P_{r}(i-1;i)$ is the likelihood that either $\lvert i-1 \rangle$ jumps up or $\lvert i \rangle$ jumps down. The parameter $N_{p}$ is the size of the Hilbert space of the mechanical subsystem. Since we are interested in the peaks relative to $n_{b}=0$ and $n_{b}=1$, we can assume $N_{p}=1$ and thus rewrite \eqref{Eq:PDFvM} in the following form:
%\begin{equation}
%\begin{split}
%PDF(V_{M})\simeq& \sqrt{\frac{2}{\pi}}\frac{e^{-\frac{V_{M}^{2}}{2}}(1+\bar{n}_{m})(1-\Delta n_{b}) + e^{-\frac{(V_{M}-D)^{2}}{2}}\left[ \Delta n_{b}+\bar{n}_{m}(1-3 \Delta n_{b}) \right]}{2(1+\bar{n}_{m})^{2}}\\ & + \frac{\Delta n_{b}(3+5\bar{n}_{m})\left[ \text{Erf}\left(\frac{V}{\sqrt{2}}\right)-\text{Erf}\left(\frac{V-D}{\sqrt{2}}\right) \right]}{4D(1+\bar{n}_{m})^{2}},
%\end{split}
%\end{equation}
\begin{equation}
\begin{split}
PDF(V_{M})\simeq& \sqrt{\frac{1}{2\pi\sigma^{2}}}\left(\frac{e^{-\frac{V_{M}^{2}}{2\sigma^{2}}-\Delta n_{b}}}{1+\bar{n}_{m}} + \frac{e^{-\frac{(V_{M}-d)^{2}}{2\sigma^{2}}-\Delta n_{b}\left(3+\bar{n}_{m}^{-1}\right)}\bar{n}_{m}}{(1+\bar{n}_{m})^{2}} \right) \\& + \left[ (1+\bar{n}_{m})(1-e^{-\Delta n_{b}}) + \bar{n}_{m}\frac{1-e^{-\Delta n_{b}(3+\bar{n}_{m}^{-1})}}{1+e^{-\Delta n_{b}(1-\bar{n}_{m}^{-1})}} \right] \frac{\text{Erf}\left( \frac{V_{M}}{\sqrt{2}\sigma} \right)-\text{Erf}\left( \frac{V_{M}-d}{\sqrt{2}\sigma} \right)}{2d(1+\bar{n}_{m})^{2}}, \label{Eq:PDFvMa}
\end{split}
\end{equation}
where $\text{Erf}(\cdot)$ denotes the error function. From Eq. \eqref{Eq:PDFvMa}, we can derive an analytical expression for the visibility that depends on the mentioned parameters: $\xi \left( \bar{n}_{m},D^{2},\Delta n_{b} \right)$. Importantly, $D^{2}$ is a dummy parameter, since $\lambda = D^{2}/\Delta n_{b}$. It follows that, for given $\lambda$ and $\bar{n}_{m}$, we can maximise $\xi \left( \bar{n}_{m},\lambda \Delta n_{b},\Delta n_{b} \right)$ by tuning $\Delta n_{b}$. Using this model, we obtain the analytical  expression for the optimal $\xi$ in the limit of $\lambda \gg 1$, presented in Eq. (5) of the main text. Note that tuning $\Delta n_{b}$ can be done by adjusting the probe power and choosing an appropriate measurement time $T$. 

\begin{figure}[htbp]
	\centering
	\includegraphics[width=15 cm]{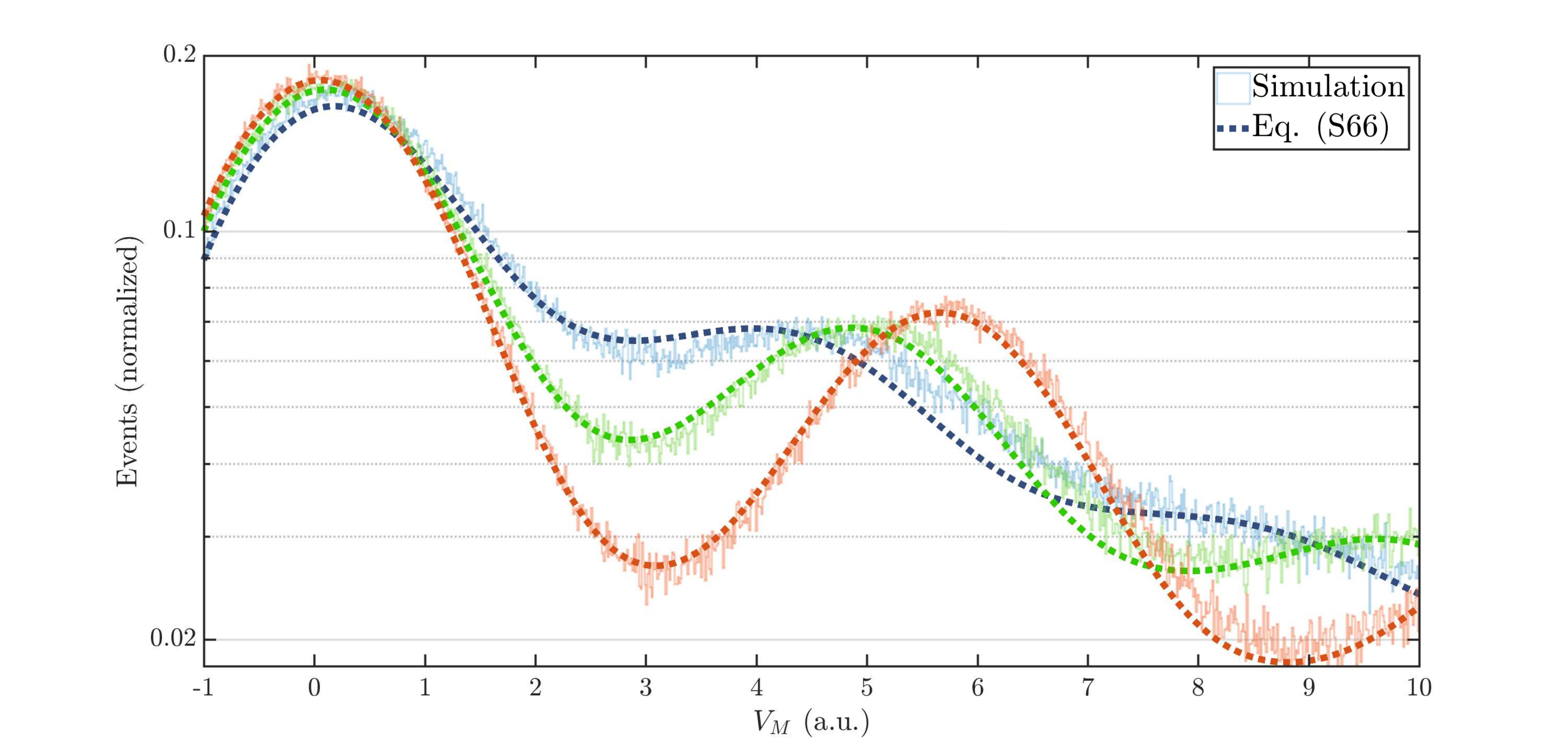}
	\caption[FigS9]{Analytical (dotted curves, Eq. \eqref{Eq:PDFvMa}) and numerical (solid, noisy curves) probability distribution function for the experimental outcome $V_{M}$. We choose $\bar{n}_{m}=1$ and $\lambda$ equal to $20$ (blue), $100$ (green) and $200$ (red). The values of $\Delta n_{b}$ that have been used were determined using the analytical maximization of the visibility $\xi$, with the cited $\lambda$ and $\bar{n}_{m}$ as free parameters.}
	\label{Fig:FigS9}
\end{figure}

\begin{figure}[htbp]
	\centering
	\includegraphics[width=15 cm]{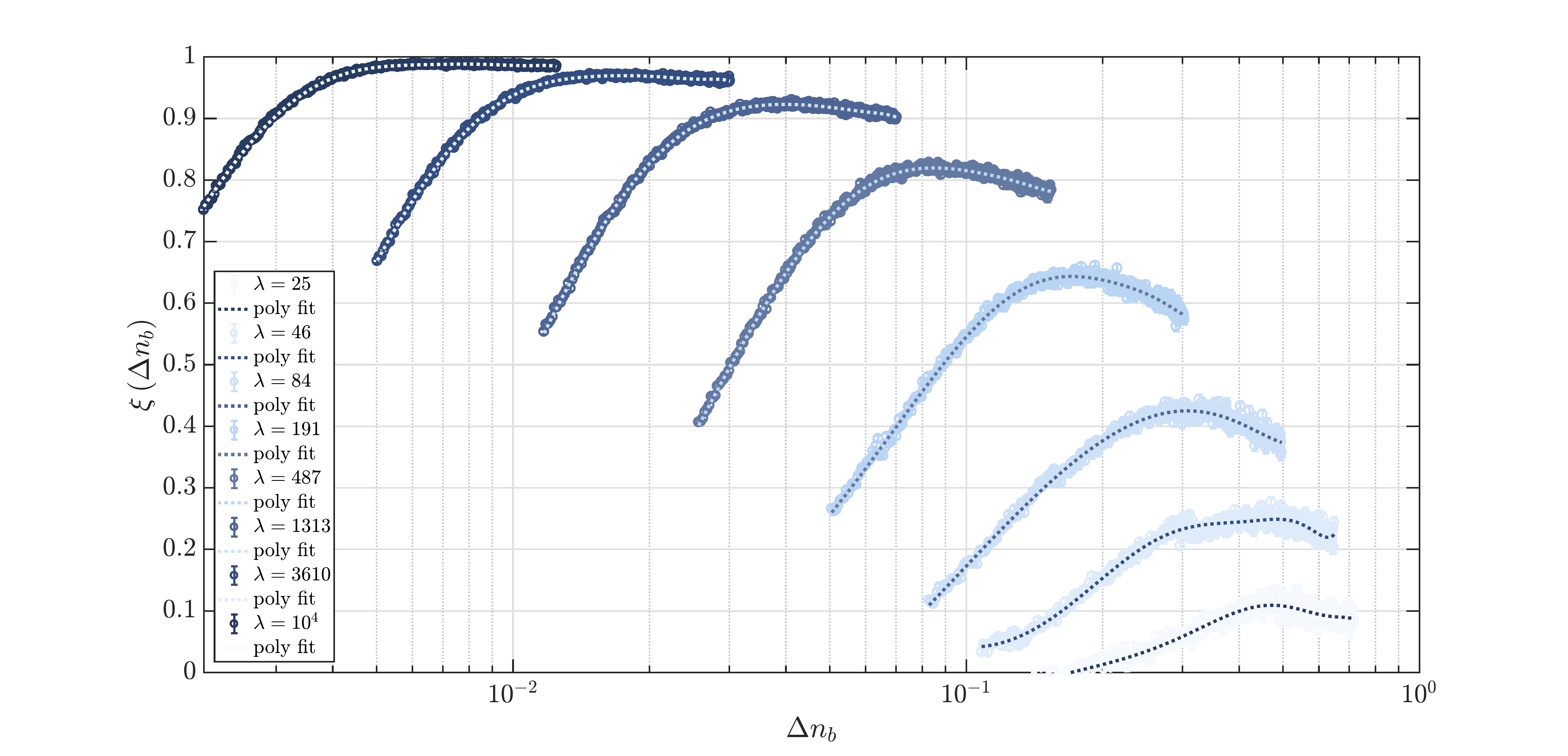}
	\caption[FigS10]{Visibility as a function of the parameter $\Delta n_{b}$, for different values of $\lambda$. The points are derived numerically using the Monte--Carlo method, and the dotted lines are polynomial fit. Notice that error bars are included, and have been determined assuming Poissonian statistics in each bin of the histogram collecting the outcomes $V_{m}$.}
	\label{Fig:FigS10}
\end{figure}

To investigate the validity of the single jump approximation used above, we simulate the probability distribution function for the outcome $V_{M}$ in a Monte--Carlo simulation. The results are presented in Fig. \ref{Fig:FigS9}. In the simulation, we allow multiple jumps to happen by dividing the measurement time $T$ into smaller segments, during which the mechanics is allowed to change. Similar to above, the outcome of the measurement can then be sampled from a Gaussian distribution of variance $\sigma^{2}$ and a mean determined by the average phonon number during the measurement. The hardly visible deviation between the analytical prediction and numerics in Fig. \ref{Fig:FigS9} comes from the single jump restriction, and can be eliminated by including the two--jumps events in the model. Note that, since the optimal $\Delta n_{b}$ decreases for higher values of $\lambda$, the single jump approximation becomes more and more accurate with increasing $\lambda$.
%%%
%%%
Secondly, for given $\lambda$ and $\bar{n}_{m}$, we can compare the analytical with the numerical maximum visibility $\xi$. The latter is determined using repeated Monte--Carlo simulations with different values for $\Delta n_{b}$, as shown in Fig. \ref{Fig:FigS10}. Each point in such plot corresponds to a single simulation with given parameters $\lambda$, $\bar{n}_{m}$ and $\Delta n_{b}$. The error bars are derived assuming Poissonian statistics in each bin of the histogram collecting the outcomes $V_{M}$. A polynomial fit is then used for determining the maximum visibility $\xi$, that is compared with the analytical prediction in Fig. 3 of the main text (2D plot; analytical corresponds to the red dotted line, numerical results are represented by blue circles).

As a final remark, it is important to say that the results of this section can easily be adapted to other experimental schemes \cite{FollowUp}. For instance, it is possible to first cool down the mechanical motion to the ground state, and then let it thermalize while measuring several times. An advantage of this approach is that it reduces the probability of jumping down from the excited Fock states, since the cooling is absent. On the other hand, operating the experiment in a pulsed regime may add an extra degree of complexity, e.g., transient effects associated with the change in equilibrium position of the membrane when the fields are turned on. 
%%%

\section{Realistic experimental parameters}\label{sec:ExptEst}

Here we study in more detail a possible experimental setup for our proposal, and the potential challenges that may arise pursuing the QND measurement of the phonon number. First, we present a simple derivation for determining the linear and quadratic electromechanical couplings $g_{1}$ and $g_{2}$ \cite{EmilThesis}. Then, we discuss the presence of a stray capacitance $C_{s}$, the main effect of which is to reduce these couplings. Finally, we analyse the feasibility of an experiment, considering aspects such as the intracavity photon number, the mechanical quality factor, the measurement time, and probe power. The parameters employed are the same as introduced in the main text: $\omega_{s} = (2\pi)7$ GHz, $\omega_{m} = (2\pi)80$ MHz, $\gamma_{r}\simeq\gamma_{t} = (2\pi)150$ kHz, $R=Z_{out}/10$, $\bar{n}_{e}=0$, $g_{r} = 10^{-2} g_{1}$ and $C_{s} = 100 C_{0}$. The membrane is assumed to be $1$ $\mu$m long and $0.3$ $\mu$m wide, with a quality factor $Q=10^6$ \cite{Settima}. We discuss the average occupation $\bar{n}_{m}$ of the mechanical bath in the following.

\begin{figure}[htbp]
	\centering
	\includegraphics[width=15 cm]{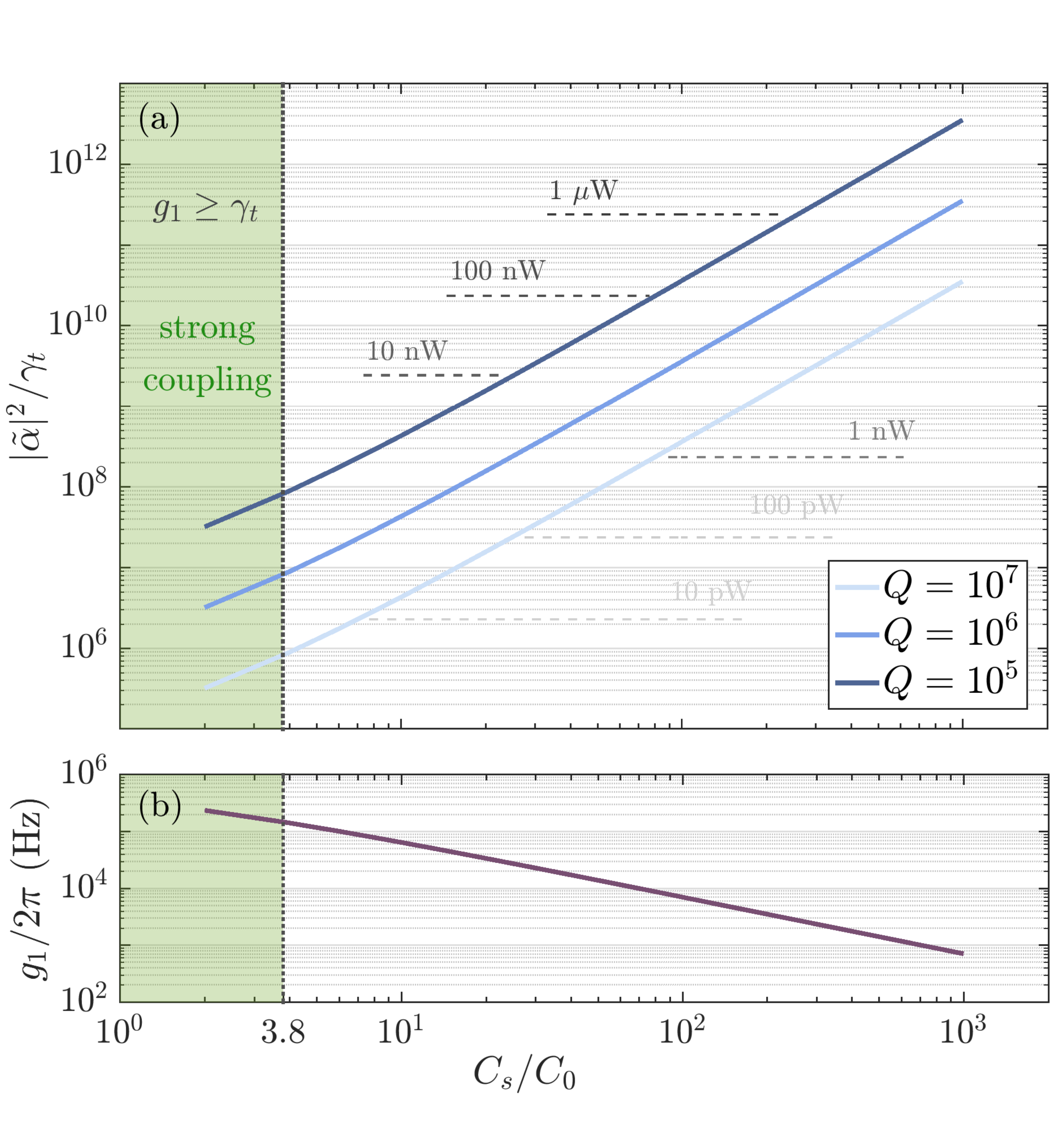}
	\caption[FigS11]{\textbf{Above}: average intracavity photons $\tilde{\alpha}^{2}/\gamma_{t}$ as a function of the (normalized) stray capacitance. The three lines correspond to different values of the mechanical quality factor, as indicated in the legend. We assume $\bar{N}_{e} = 3$, and equal heating contribution from the mechanical and electrically induced reservoirs, with $\bar{n}_{m} = 3$. \textbf{Below}: linear coupling $g_{1}$ as a function of the relative value of the stray capacitance $C_{s}/C_{0}$. For both figures, in the shadowed region the strong coupling $g_{1}\geq \gamma_{t}$ is achieved.}
	\label{Fig:FigS11}
\end{figure}

The mechanical membrane is fixed along all its boundaries, such as in Fig. 1(a) of the main text. A basis of modes describing its motion is thus $\left\lbrace u_{i,j}(x,y)  \right\rbrace $, where
\begin{equation}
u_{i,j}(x,y)= \epsilon \sin\left( i\frac{\pi}{L}x \right)\sin\left( j\frac{\pi}{W}y \right).
\end{equation}
Here, $L$ and $W$ are the length and width of the membrane, respectively. The mode of interest has indices $i=2$ and $j=1$, and the constant $\epsilon$ fixes an effective gauge for the mass. By setting $\int_{0}^{L}\int_{0}^{W}\left\lvert u_{i,j}(x,y)  \right\rvert^{2}dx dy=1 $, we choose the gauge in which the so--called effective mass is the physical mass, and $\epsilon=2$. Recalling that
\begin{subequations}\label{Eq:Couplings}
	\begin{align}
	g_{1} = & \frac{x_{0}\omega_{s}}{C_{0}} \frac{\partial C}{\partial \beta}\Big\rvert_{x=0}\label{Eq:LinCoupling},\\
	g_{2} = & \frac{x_{0}^{2}\omega_{s}}{2C_{0}} \frac{\partial^{2} C}{\partial \beta^{2}}\Big\rvert_{x=0}	\label{Eq:QuadCoupling},
	\end{align}
\end{subequations}
the couplings can be determined once the derivatives are found. As discussed in Ref. \cite{EmilThesis}, $\beta$ describes the amplitude of the considered mode, and can be viewed as a canonical position. An approximate value of the derivatives in Eq. \eqref{Eq:Couplings} is then 
\begin{equation}\label{Eq:CouplingDerivative}
\frac{\partial^{k} C}{\partial \beta^{k}}\Big\rvert_{x=0} \simeq (-1)^{k}k! \frac{C_{0}}{d^{k}} \frac{2}{L W}\int_{0}^{L/2}\int_{0}^{W} u_{2,1}^{k}(x,y) dx dy,
\end{equation}
where $d$ is the distance separating the two capacitor's plates, and the integral is taken over \textit{half} the membrane, for the reasons discussed in section \ref{sec:CompleteCirc}. Using Eqs. \eqref{Eq:Couplings} and \eqref{Eq:CouplingDerivative}, we can finally determine the values of $g_{1}$ and $g_{2}$ to be
\begin{subequations}\label{Eq:CouplingsValue}
	\begin{align}
	g_{1} = & \frac{8}{\pi^{2}} \frac{x_{0}\omega_{s}}{d} \label{Eq:LinCouplingVal},\\
	g_{2} = & \frac{x_{0}^{2}\omega_{s}}{d^{2}} \label{Eq:QuadCouplingVal}.
	\end{align}
\end{subequations}
For the parameters introduced in the main text, we find $g_{1}\simeq (2\pi) 715$ kHz and $g_{2}\simeq (2\pi) 111$ Hz. Considering that we assumed an electrical damping $\gamma_{t} = (2\pi) 150$ kHz, it could seem that we reach the strong coupling regime $g_{1}>\gamma_{t}$. However, we have so far ignored stray capacitances, that are the main reason for which the strong coupling regime (and thus the phonon QND measurement) has never been accomplished in electromechanics.
%In this regime, schemes for phonon QND measurement are already available, based on an effective quadratic coupling originated from the linear one $(g_{2} \simeq g_{1}^{2}/\omega_{s})$ \cite{Marquardt,Clerk2016}. The problem arises when stray capacitances are considered. 
For such small geometries, the stray capacitance $C_{s}$ always exceed the intrinsic capacitance $C_{0}$ by up to several orders of magnitude \cite{ExpNumb4}, and severely limits the attainable values of the linear and quadratic couplings [see Fig. \ref{Fig:FigS11}(b)]. Looking at Eqs. \eqref{Eq:Couplings}, and considering that the stray capacitance does not affect the derivatives $\partial^{k} C/\partial \beta^{k}$, we replace the coupling constants by
\begin{subequations}\label{Eq:CouplingsReal}
	\begin{align}
	g_{1} \rightarrow & \frac{C_{0}}{C_{0}+C_{s}} g_{1} \label{Eq:LinCouplingReal},\\
	g_{2} \rightarrow & \frac{C_{0}}{C_{0}+C_{s}} g_{2}	\label{Eq:QuadCouplingReal}.
	\end{align}
\end{subequations}
For $C_{s}=100 C_{0}$, we find $g_{1} \simeq (2\pi) 7$ kHz and $g_{2} \simeq (2\pi) 1$ Hz. This value of $C_{s}$ is optimistic for geometries similar to the ones in Ref. \cite{ExpNumb4,ExpNumb5}, where graphene sheets are laid on a substrate in such a way that the contact area between the two interfaces is very large. However, for a setup like the one described in Ref. \cite{ExpNumb1}, membranes of similar dimensions as the ones conjectured here are assembled onto small localized gates, dramatically reducing the stray capacitance. With this fabrication technique, a stray capacitance of $50$ $f$F is obtained, for a membrane that is about two and a half times the size of the one conjectured here. In our case, this value for the stray capacitance would correspond to $376$ times $C_{0}$, where $C_{0} \simeq 0.13$ $f$F. Due to the smaller size of our membrane, we assume a reduction in the stray capacitance, such that we consider $C_{s} = 100 C_{0}$. 

In our settings, reaching the strong coupling would require $C_{s} < 3.8 C_{0}$. In this regime, phonon QND detection could also be performed with the strategy proposed in \cite{Marquardt}. With a large stray capacitance, however, we cannot accomplish such requirement, and investigate different approaches to QND detection. With the experimental parameter described above, we find $\lambda_{b} = 105 \times Z_{out}/R$ and $\lambda_{p} = 0.014(g_{1}/g_{r})^{2}$, regardless of the value of $C_{s}$. The highest quoted value $\lambda = 122$ is found assuming $g_{r} = 10^{-2} g_{1}$ and $Z_{out} = 10 R$.

In the remainder of this section, we study the conditions under which the QND detection could be implemented, focusing on the incident power and the measurement time. First, we recall that the membrane's heating has several contributions. Two of them, denoted  $\Delta n_{b}^{(b)}$ and $\Delta n_{b}^{(p)}$, are the ones identified in Eqs. \eqref{Eq:DeltaNbCompl} (rhs, second term) and \eqref{Eq:HeatingPar}, respectively. They describe the feedback of the electrical system on the mechanical motion, and are the denominators of the parameters $\lambda_{b}$ and $\lambda_{p}$ defined in the main text. The third contribution comes from the mechanical reservoir. Indicated with $\Delta n_{b}^{(m)}$, this is given in Eq. \eqref{Eq:DeltaNbCompl} (rhs, first term), and is independent of the strength of the probing field. For determining the parameter $\lambda$, we have so far ignored $\Delta n_{b}^{(m)}$, assuming that the measurement is fast enough. This is an excellent approximation, as far as $\Delta n_{b}^{(m)} \ll \Delta n_{b}^{(b)}+\Delta n_{b}^{(p)}$. Below, we discuss the case in which there is a seizable contribution from the mechanical bath, and describe its effect on the QND detection of the phonon number. 

Above we have seen that, for the considered experimental parameters, $\lambda = 122$, for which we find the optimal $\Delta n_{b} = 0.21$. For $C_{s} = 100 C_{0}$, the total number of photons that we need to send within the measurement is $\alpha^{2}\simeq 4.5 \cdot 10^{11}$ [see Eqs. \eqref{Eq:DeltaNbCompl} and \eqref{Eq:HeatingPar}]. By sending these photons within a sufficiently short time, we can neglect the influence of the mechanical reservoir. To derive precise conditions for this, we define the effective temperature $\bar{N}_{e}$ of the electrically induced reservoir:
\begin{equation}\label{Eq:AvgOccElR}
\bar{N}_{e} = \frac{\Delta n_{b}^{(b)}+\Delta n_{b}^{(p)}}{T(\gamma_{b} + \Gamma_{b})}.
\end{equation}
This last equation is derived from Eq. \eqref{Eq:PhononEvCompl}, by sending the time $t$ to infinity, and recalling the definitions of $\Delta n_{b}^{(b)}$ and $\Delta n_{b}^{(p)}$. $\bar{N}_{e}$ is the average phonon number at which the membrane stabilizes in the absence of the mechanical reservoir. The condition $\Delta n_{b}^{(m)} \ll \Delta n_{b}^{(b)}+\Delta n_{b}^{(p)}$, under which the mechanical bath can be neglected, can be rewritten as $\bar{n}_{m}\ll \bar{N}_{e}$. The measurement time $T$ is thus a knob that allows us to adjust the relative weights of the mechanical reservoir and the electrical feedback on the membrane. For short measurement times, $\bar{N}_{e}$ is increased, and this condition is easier to satisfy. As an example, if we choose $T=0.4$ ms and $Q=10^{6}$, we get $\bar{N}_{e}=1$, implying that the average occupation of the membrane's reservoir needs to be less than unity to neglect the mechanical reservoir. For our setup, this requires the temperature of the cryostat to be lower than $6$ mK, a challenging task for current technology. On the other side, choosing $T=0.04$ ms fixes $\bar{N}_{e} = 10$, and the temperature for which $\bar{n}_{m}$ becomes negligible is $40$ mK, that has been already achieved in experiments involving graphene resonators \cite{ExpNumb4}. %For high average occupations, the time that the experiment has to run to acquire enough statistics for the lowest Fock states may become long. To avoid this problem, it is possible to introduce an additional electrical cooling of the mechanical resonator. This cooling, red detuned of $\omega_{m} \gg \gamma_{t}$, would not interfere with the QND probe, and would effectively increase the probability to find the membrane in its ground state. Finally, 
Once the measurement time is chosen, and assuming the photon flux $\lvert\tilde{\alpha}\rvert^{2} = \lvert\alpha\rvert^{2}/T$ to be constant, one can determine the probing power $P_{in}$ and the average intracavity photon number $\lvert\tilde{\alpha}\rvert^{2}/\gamma_{t}$. As an example, with $T=0.4$ ms, $P_{in}$ becomes $5.3$ nW and $\lvert\tilde{\alpha}\rvert^{2}/\gamma_{t} = 1.2 \cdot 10^{9}$. These parameters are highly dependent on the mechanical quality factor $Q$, and the stray capacitance, as investigated below in Fig.\ref{Fig:FigS11}(a). %This high power may be worrisome in a cryogenic scenario, where the temperature of the fridge can be raised by the electrical dissipation \cite{TemperatureRising}. The process of thermalization of the mechanical bath is yet to be understood, and hard to be exactly quantified for our setup. Two qualitative arguments for which, we believe, we are not concerned too much are the following. First, $T$ is a parameter that can be freely chosen, and tuned to optimize the experiment. As an example, by taking a shorter measurement time, we would indeed increase the net temperature rise in the cryostat, but we also make the system less susceptible to heating mechanisms (see above). Second, by going into a pulsed regime, we can let the system cool down in between different pulses. Thus, given that the measurement time $T$ is generally very short, the raise of temperature in the cryostat remains negligible. 

Finally, we discuss the case in which the electrical heating is in equilibrium with the heating and damping of the mechanical reservoir. In that case, we need to renormalize $\lambda$ by the total heating:
\begin{equation}\label{Eq:LambdaRen}
\lambda' = \frac{\Delta n_{b}^{(b)}+\Delta n_{b}^{(p)}}{\Delta n_{b}^{(m)}+\Delta n_{b}^{(b)}+\Delta n_{b}^{(p)}}\lambda,
\end{equation}
where $\lambda$ contains both the terms $\lambda_{b}$ and $\lambda_{p}$. 

Importantly, we can always make $\lambda' \simeq \lambda$ by increasing the driving strength and reducing the measurement time (thus decreasing $\Delta n_{b}^{(m)} = \bar{n}_{m} \gamma_{b} T$). The parameter $\lambda'$ substitutes $\lambda$ in the description of the system. The mechanical bath thus reduces the quality of the phonon QND measurement, but the analysis above still applies, with $\lambda'$ instead of $\lambda$. As an example, assume that the temperature is $14$ mK \cite{ExpNumb4}. Then, the average occupation of the membrane's reservoir is $\bar{n}_{m} = 3$. We choose $T$ such that the electrical heating is equal to the mechanical: $\bar{N}_{e} = \bar{n}_{m}$. Then $\lambda' = \lambda/2 = 61$ with the previous parameters. This value is well in the regime of good visibility $\xi$, and for this $\lambda'$, the optimal value of the total heating is $\Delta n_{b} = \Delta n_{b}^{(m)} + \Delta n_{b}^{(b)} + \Delta n_{b}^{(p)} = 0.3$. Depending on the value of the mechanical quality factor $Q$ and stray capacitance $C_{s}$, we can then determine the required incident power and the intracavity photon number, as shown in Fig. \ref{Fig:FigS11}. Since we assume $\Delta n_{b}^{(m)} = \Delta n_{b}/2$, the measurement time is $T = \Delta n_{b}/(2 \gamma_{b} \bar{n}_{m}) \simeq 0.05 Q/\omega_{m}$. For the parameters in the figure and a quality factor $Q$ varying between $10^5$ and $10^7$, we find $T \in \left[ 10^{-5}, 10^{-3}\right]$ s.